\documentclass[preprint,3p,times,longtitle, twocolumn]{elsarticle}
\usepackage{mathtools}
\usepackage{textcomp}
\usepackage{amssymb}
\usepackage{lineno}
\usepackage{subcaption}

\begin{document}

\begin{frontmatter}



\title{Characterization of the Atmospheric Muon Flux in IceCube}


%\author{Ice$^{3}$}

%\iffalse

\author[Adelaide]{M.~G.~Aartsen}
\author[Munich]{K.~Abraham}
\author[Zeuthen]{M.~Ackermann}
\author[Christchurch]{J.~Adams}
\author[BrusselsLibre]{J.~A.~Aguilar}
\author[MadisonPAC]{M.~Ahlers}
\author[StockholmOKC]{M.~Ahrens}
\author[Erlangen]{D.~Altmann}
\author[PennPhys]{T.~Anderson}
\author[Mainz]{M.~Archinger}
% ...
\address[Adelaide]{School of Chemistry \& Physics, University of Adelaide, Adelaide SA, 5005 Australia}
% ...
\address[BrusselsLibre]{Universit\'e Libre de Bruxelles, Science Faculty CP230, B-1050 Brussels, Belgium}
% ...
\address[Christchurch]{Dept.~of Physics and Astronomy, University of Canterbury, Private Bag 4800, Christchurch, New Zealand}
% ...
\address[Erlangen]{Erlangen Centre for Astroparticle Physics, Friedrich-Alexander-Universit\"at Erlangen-N\"urnberg, D-91058 Erlangen, Germany}
% ...
\address[MadisonPAC]{Dept.~of Physics and Wisconsin IceCube Particle Astrophysics Center, University of Wisconsin, Madison, WI 53706, USA}
\address[Mainz]{Institute of Physics, University of Mainz, Staudinger Weg 7, D-55099 Mainz, Germany}
% ...
\address[Munich]{Technische Universit\"at M\"unchen, D-85748 Garching, Germany}
% ...
\address[StockholmOKC]{Oskar Klein Centre and Dept.~of Physics, Stockholm University, SE-10691 Stockholm, Sweden}
% ...
\address[PennPhys]{Dept.~of Physics, Pennsylvania State University, University Park, PA 16802, USA}
% ...
\address[Zeuthen]{DESY, D-15735 Zeuthen, Germany}
\fntext[Tokyofn]{Earthquake Research Institute, University of Tokyo, Bunkyo, Tokyo 113-0032, Japan}
\fntext[Goddard]{NASA Goddard Space Flight Center, Greenbelt, MD 20771, USA}
\cortext[cor1]{Corresponding Author: berghaus@icecube.wisc.edu}
%\fi

\begin{abstract}

Muons produced in atmospheric cosmic ray showers account for the by far dominant part of the event yield in large-volume underground particle detectors. The IceCube detector, with an instrumented volume of about a cubic kilometer, has the potential to conduct unique investigations on atmospheric muons by exploiting the large collection area and the possibility to track particles over a long distance. Through detailed reconstruction of energy deposition along the tracks, the characteristics of muon bundles can be quantified, and individual particles of exceptionally high energy identified. The data can then be used to constrain the cosmic ray primary flux and the contribution to atmospheric lepton fluxes from prompt decays of short-lived hadrons.

In this paper, techniques for the extraction of physical measurements from atmospheric muon events are described and first results are presented. The multiplicity spectrum of TeV muons in cosmic ray air showers for primaries in the energy range from the knee to the ankle is derived and found to be consistent with recent results from surface detectors. The single muon energy spectrum is determined up to PeV energies and shows a clear indication for the emergence of a distinct spectral component from prompt decays of short-lived hadrons. The magnitude of the prompt flux, which should include a substantial contribution from light vector meson di-muon decays, is consistent with current theoretical predictions.

The variety of measurements and high event statistics can also be exploited for the evaluation of systematic effects. In the course of this study, internal inconsistencies in the zenith angle distribution of events were found which indicate the presence of an unexplained effect outside the currently applied range of detector systematics. The underlying cause could be related to the hadronic interaction models used to describe muon production in air showers.

\end{abstract}


\end{frontmatter}

%\clearpage

%\linenumbers


\section{Introduction}

IceCube is a particle detector with an instrumented volume of about one cubic kilometer, located at the geographic South Pole \cite{Karle:2014bta}. The experimental setup consists of 86 cables (``strings''), each supporting 60 digital optical modules (``DOMs''). Every DOM contains a photomultiplier tube and the electronics required to handle data acquisition, digitization and transmission. The main active part of the detector is deployed at a depth of 1450 to 2450 meters below the surface of the ice, which in turn lies at an altitude of approximately 2830 meters above sea level. The volume detector is supplemented by the surface array IceTop, formed by 81 pairs of tanks filled with - due to ambient conditions solidified - water.

The main scientific target of IceCube is the search for astrophysical neutrinos. At the time of design, the most likely path to discovery was expected to be the detection of upward-going tracks caused by Earth-penetrating muon neutrinos interacting shortly before the detector volume. All DOMs were consequently oriented in the downward direction, such that Cherenkov light emission from charged particles along muon tracks can be registered after minimal scattering in the surrounding ice.

The first indication for a neutrino signal exceeding the expected background from cosmic ray-induced atmospheric fluxes came in the form of two particle showers with a total visible energy of approximately 1 PeV \cite{Aartsen:2013bka}. Detailed analysis of their directionality strongly indicated an origin from above the horizon. The result strengthened the case for the astrophysical nature of the events, since no accompanying muons were seen, as would be expected for neutrinos produced in air showers. This serendipitous detection motivated a dedicated search for high-energy neutrinos interacting within the detector volume, which led first to a strong indication \cite{Aartsen:2013jdh} and later, after evaluating data taken during three full years of detector operation, to the first discovery of an astrophysical neutrino flux \cite{Aartsen:2014gkd}. In each case, the decisive contribution to the event sample were particle showers pointing downward.

Despite the large amount of overhead material, the deep IceCube detector is triggered at a rate of approximately 3000 $\textrm{s}^{-1}$ by muons produced in cosmic ray-induced air showers. Formerly regarded simply as an irksome form of background, these have since proved to be an indispensable tool to tag and exclude atmospheric neutrino events in the astrophysical discovery region \cite{Gaisser:2014bja}.

Apart from their application in neutrino searches, muons can be used for detector verification and a wide range of physics analyses. Examples are the measurement of cosmic ray composition and flux in coincidence with IceTop \cite{IceCube:2012vv}, the first detection of an anisotropy in the cosmic ray arrival direction in the southern hemisphere \cite{Abbasi:2010mf, Abbasi:2011ai, Abbasi:2011zka}, investigation of QCD processes producing high-$p_{\rm{t}}$ muons \cite{Abbasi:2012kza} and the evaluation of track reconstruction accuracy by taking advantage of the shadowing of cosmic rays by the moon \cite{Aartsen:2013zka}.

Remaining to be demonstrated is the possibility to develop a comprehensive and consistent picture of atmospheric muon physics in IceCube. The goal of this paper is to outline how this could be accomplished, illustrate the scientific potential and discuss consequences of the actual measurement for the understanding of detector systematics.

\section{Physics}

\subsection{Cosmic Rays in the IceCube Energy Range}

\begin{figure}
  \centering
  \includegraphics[width=220pt]{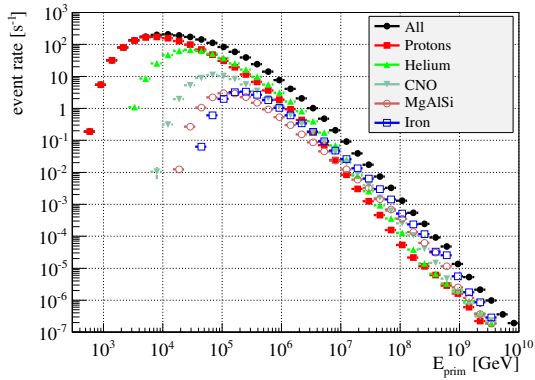}
  \caption{Atmospheric muon event yield in IceCube in dependence of primary type simulated with CORSIKA \cite{corsika}. The cosmic ray flux was weighted according to the H3a model \cite{Gaisser:2013bla}.}
  \label{fig-allev}
\end{figure}

The energy range of cosmic ray primaries producing atmospheric muons in IceCube is limited by the minimum muon energy required to penetrate the ice at the low, and the cosmic ray flux rate at the high end. Predicted event yields are shown in Fig. \ref{fig-allev}. Since the muon energy is related to the energy per nucleon $E_{\rm{prim}}/A$, threshold energies increase in proportion to the mass of the primary nucleus.

The energy range of atmospheric muon events in IceCube covers more than six orders of magnitude. Neutrinos, not attenuated by the material surrounding the detector, can reach even lower. With a ratio between lepton and parent nucleon energy of about one order of magnitude \cite{tomsbook}, the lowest primary energies relevant for neutrinos in IceCube fall in the region around 100 GeV.

Coverage of this vast range of energies by specialized detectors varies considerably, and overlapping measurements are not always consistent. At energies well below 1 TeV, important for production of atmospheric neutrinos in oscillation measurements \cite{Aartsen:2013jza}, both PAMELA \cite{Adriani:2011cu} and AMS-02 \cite{Aguilar:2015ooa} find a clear break in the proton spectrum at about 200 GeV. The exact behavior of the primary spectrum should be an important factor in upcoming precision measurements of oscillation parameters by the planned IceCube sub-array PINGU \cite{Aartsen:2013aaa}.

In the energy region where the bulk of atmospheric muons triggering the IceCube detector are produced, the most recent measurement was performed by the balloon-borne CREAM detector \cite{Ahn:2010gv}. In the range from 3-200 TeV, proton and helium spectra are found to be consistent with power laws of the form $E^{-\gamma}$. The proton spectrum with $\gamma_{\rm{p}}=2.66\pm0.02$ is somewhat softer than that of helium with $\gamma_{\rm{He}}=2.58\pm0.02$. The cross-over between the two fluxes lies at approximately 10 TeV.

Between a few hundred GeV and 3 TeV, and again from 100 TeV to 1 PeV, there are large gaps where experimental measurements of individual primary fluxes are sparse and contain substantial uncertainties \cite{Kochanov:2008pt}. Especially the second region is of high importance to IceCube physics, because it corresponds to neutrino energies of tens of TeV where indications for astrophysical fluxes start to become visible.

The situation improves around the ``knee'' located at about 4 PeV, which has long been a major focus of cosmic ray physics. The well constrained overall primary flux has been resolved into its individual components by the KASCADE array \cite{Antoni:2005wq}, although the result depends strongly on the model used to describe nuclear interactions within the air shower. There is a general consensus that the primary composition changes towards heavier elements in the range between the knee and 100 PeV, confirmed by various measurements \cite{Bluemer:2009zf}, including IceCube \cite{IceCube:2012vv}. An exact characterization of the all-nucleon spectrum around the knee is necessary to constrain the contribution to atmospheric lepton fluxes from prompt hadron decays and accurately describe backgrounds in diffuse astrophysical neutrino searches.

%GAMMA with bump\cite{Garyaka:2008gs}. 
Between 100 PeV and approximately 1 EeV lies another region with sparse coverage, which has only recently begun to be filled. In the past, data taken near the threshold of very large surface arrays indicated a ``second knee'' at about 300 PeV \cite{Bergman:2007kn}. Approaching from the other side, KASCADE-Grande found evidence for a knee-like structure closer to 100 PeV, along with a hardening of the all-primary spectrum around 15 PeV \cite{Apel:2012rm}. This result confirms earlier tentative indications from the Tien-Shan detector using data taken before 2001, but published only in 2009 \cite{Shaulov:2009zzd} and is supported by subsequent measurements using the TUNKA-133 \cite{Prosin:2014dxa} detector. The currently most precise spectrum in terms of statistical accuracy and hadronic model dependence was derived from data taken by the IceTop surface array \cite{Aartsen:2013wda}. KASCADE-Grande later extended the original result by indications for a light element ankle \cite{Apel:2013ura}, a heavy element knee \cite{Apel:2011mi} and separate spectra for elemental groups \cite{Apel:2013dga}.

The emergent picture has yet to be theoretically interpreted in a comprehensive manner. The data indicate that several discrete components are present in the cosmic ray flux, and that the behavior of individual nuclei closely corresponds to a power law followed by a spectral cutoff at an energy proportional to their magnetic rigidity $R=E_{\rm{prim}}/Z$. This explanation was first proposed by Peters in 1961 \cite{Peters:1961} and later elaborated by, among others, Ter-Antonyan and Haroyan \cite{TerAntonyan:2000hh} as well as H\"orandel \cite{Hoerandel:2002yg}. Exactly how many components there are, where they originate, and the precise values and functional dependence of their transition energies are still open questions. A well-known proposal by Hillas postulates two galactic sources, one accounting for the knee, the other for the presumptive knee-like feature at 300 PeV \cite{Hillas:2005cs}. Another model, by Zatsepin and Sokolskaya, identifies three distinct types of galactic sources to account for the flux up to 100 PeV \cite{Zatsepin:2006ci}. The hardening of the spectrum around the ``ankle'' at several EeV can be described elegantly by a pure protonic flux and its interaction with CMB radiation \cite{Berezinsky:2005cq} or, more in line with recent experimental results, in terms of separate light and heavy components \cite{Aloisio:2009sj}. The consensus is in either case that the origin of the highest-energy cosmic rays is extragalactic. 

This paper, like other IceCube analyses, relies for purposes of model testing mainly on the parametrizations by Gaisser, Stanev and Tilav \cite{Gaisser:2013bla}. These incorporate various basic features of the models described above, while updating numerical values to conform with the latest available measurements. Specifically, the ``Global Fit'' (GF) parametrization introduces a second distinct population of cosmic rays before the knee with a transition energy of 120 TeV. The knee itself, and the feature at 100 PeV, are interpreted as helium and iron components with a common rigidity-dependent cutoff, eliminating the need for an intermediate galactic flux component as in the H(illas) 3a and 4a parametrizations. The difference between H3a and H4a lies in the composition of the highest-energy component which becomes dominant at energies beyond 1 EeV, which is mixed in the former, and purely protonic in the latter case. In the region around the knee, the two models are for practical purposes indistinguishable.

\subsection{Muons vs. Neutrinos}

The flux of atmospheric neutrinos in IceCube is modeled using extrapolated parametrizations based on a Monte Carlo simulation for energies up to 10 TeV \cite{Honda:2006qj}. To account for the influence of uncertainties of the cosmic ray nucleon flux, the energy spectrum is adjusted by a correction factor \cite{Gaisser:2013ira}. The result can be demonstrated to agree reasonably well with full air shower simulations \cite{Fedynitch:2012fs}, but necessarily contains inaccuracies, for example by neglecting variations in the atmospheric density profile at the site and time of production.

Atmospheric muon events on the other hand are simulated through detailed modeling of individual cosmic ray-induced air showers. In standard simulation packages such as CORSIKA \cite{corsika}, specific local conditions like the direction of the magnetic field and the profile of the atmosphere including seasonal variations can be fully taken into account. Energy spectra for each type of primary nucleus are separately adjustable. Hadronic interaction models can be varied and their influence quantified in terms of a systematic uncertainty.

In contrast to neutrinos, astrophysical fluxes, flavor-changing effects and hypothetical exotic phenomena do not affect muons. All observations can be directly related to the primary cosmic ray flux and the detailed mechanisms of hadron collisions. Due to the close relation between neutrino and charged lepton production, high-statistics measurements using muon data are therefore invaluable to constrain atmospheric neutrino fluxes.

Perhaps most importantly, atmospheric muons represent a high-quality test beam for the verification of detector performance, because the variety of possible measurements along with high event statistics permit detailed consistency checks. A particular advantage in the case of IceCube is that muons probe the region above the horizon for which the down-looking detector configuration is not ideal, but where contrary to original expectation the bulk of astrophysical detections has taken place.

\subsection{Primary Flux and Atmospheric Muon Characteristics}\label{sec:prim-muchar}

\begin{figure}[ht!]
  \centering
  \begin{subfigure}{220pt}
    \includegraphics[width=220pt]{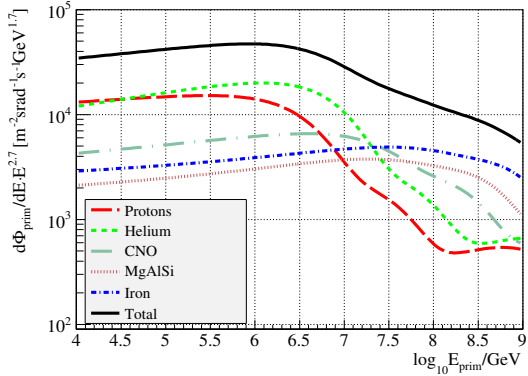}
    \caption{Primary Energy $E_{\rm{prim}}$}
  \end{subfigure}
  \begin{subfigure}{220pt}
    \includegraphics[width=220pt]{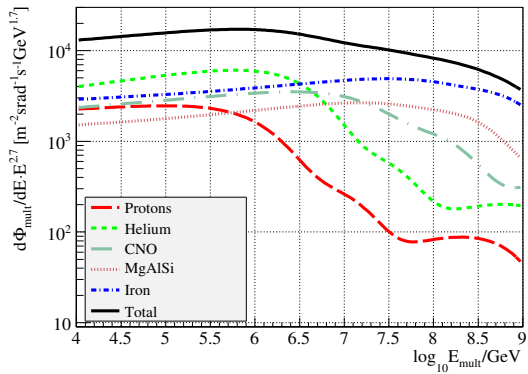}
    \caption{Rescaled Muon Multiplicity $E_{\rm{mult}}$}
  \end{subfigure}
  \begin{subfigure}{220pt}
    \includegraphics[width=220pt]{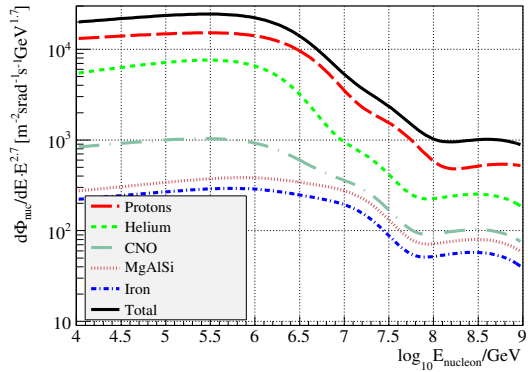}
    \caption{Nucleon Energy $E_{\rm{nuc}}$}
  \end{subfigure}
  \caption{Contribution of individual elemental components to overall flux spectra relevant for atmospheric muon measurements, here shown for the Gaisser/Hillas model with mixed-composition extragalactic component (H3a) \cite{Gaisser:2013bla}. For definition of $E_{\rm{mult}}$, see Section \ref{sec:prim-muchar}.}
  \label{fig-h3aspecs}
\end{figure}

The connection between the measurable quantities of atmospheric muon events and the properties of the primary cosmic ray flux is illustrated in Fig. \ref{fig-h3aspecs}. The relation of muon multiplicity to primary type and energy is expressed in terms of the parameter $E_{\rm{mult}}$, defined such that $E_{\rm{mult}}=E_{\rm{prim}}$ for iron primaries. The average number of muons in a bundle can then be expressed as $<N_{\mu}> = \kappa \cdot E_{\rm{mult}}$, where the proportionality factor $\kappa$ depends on the specific experimental circumstances. Due to fluctuations in the atmospheric depth of shower development and the total amount of hadrons produced in nuclear collisions, the variation in the number of muons is slightly wider than a Poissonian distribution \cite{Boziev:1991xw}.

Since the muon multiplicity itself is a function of zenith angle, atmospheric conditions, detector depth and surrounding material, it is convenient to re-scale it such that the derived quantity is directly related to primary mass and energy. This study uses the parameter

\begin{equation}\label{emultdef}
E_{\rm{mult}}\equiv E_{\rm{prim}}\cdot (A/56)^{\frac{1-\alpha}{\alpha}}.
\end{equation}

The definition was chosen such that $E_{\rm{mult}}$ is equal to $E_{\rm{prim}}$ in the case of iron primaries  with atomic mass number $A=56$, which will in practice dominate the multiplicity spectrum above a few PeV, as shown in Fig. \ref{fig-h3aspecs} (b). Exact definition and construction of $E_{\rm{mult}}$ are discussed in Section \ref{sec:emult}.

As the ratio of muons to electromagnetic particles in an air shower increases with the primary mass, the contribution of light elements to the multiplicity spectrum is suppressed. For a power law spectrum of the form $E^{-\gamma}$, the contribution of individual elements to the muon multiplicity, here expressed in terms of a flux $\Phi_{\rm{mult}}$, scales as:

\begin{equation}\label{emult_eprimrel}
\frac{\Phi_{\rm{mult}}(A)}{\Phi_{\rm{mult}}(1)}\cdot\frac{\Phi_{\rm{prim}}(1)}{\Phi_{\rm{prim}}(A)}\simeq A^{\frac{1-\alpha}{\alpha}\cdot(\gamma-1)},
\end{equation}
where $\alpha \approx 0.79$ is an empirical parameter derived from simulation \cite{tomsbook}.

Single-particle atmospheric lepton fluxes, on the other hand, are related to the nucleon spectrum. Under the same assumptions as above, the relation between all-nucleon and primary flux is:

\begin{equation}\label{enuc_eprimrel}
 \frac{\Phi_{\rm{nuc}}(A)}{\Phi_{\rm{nuc}}(1)}\cdot\frac{\Phi_{\rm{prim}}(1)}{\Phi_{\rm{prim}}(A)} = A^{2-\gamma}.
\end{equation}
For a power law with an index of approximately -2.6 to -2.7, such as the cosmic ray spectrum before the knee, the nucleon spectrum therefore becomes strongly dominated by light elements.

\subsection{Prompt Muon Production}

A particular difficulty in the description of atmospheric lepton fluxes is the emergence at high energies of a component originating from prompt hadron decays. The reason is the harder spectrum compared to the light meson contribution, which is the consequence of the lack of re-interactions implicit in the definition.

An important source of prompt atmospheric lepton fluxes is the decay of charmed hadrons. While it is possible to estimate their production cross section using theoretical calculations based on perturbative QCD, substantial contributions from non-perturbative mechanisms cannot be excluded. The problem can therefore at the moment only be resolved experimentally \cite{Lipari:2013taa}. One major open question currently under investigation \cite{Bednyakov:2014pqa} is whether nucleons contain ``Intrinsic Charm'' quarks, which might considerably increase charmed hadron production \cite{Brodsky:1980pb}. 

Inclusive charm production cross sections were measured during recent LHC runs by the collider detectors LHCb \cite{Britsch:2013lca}, ATLAS \cite{Mountricha:2011zz}, and ALICE \cite{ALICE:2011aa,Abelev:2012vra}, and previously by the RHIC collaborations PHENIX \cite{Adare:2006hc} and STAR \cite{Adams:2004fc}. Data points are consistently located at the upper end of the theoretical uncertainty, which covers about an order of magnitude \cite{Cacciari:2012ny}. On a qualitative level, the new results suggest that charm-induced atmospheric neutrino fluxes could be somewhat stronger than previously assumed. A straightforward translation is however difficult. Although collider measurements probe similar center-of-mass energies, they are for technical reasons restricted to central rapidities of approximately $|y| \leq 1$. For lepton production in cosmic ray interactions, forward production is much more important.

A variety of descriptions for the flux of atmospheric leptons from charm have been proposed in the past \cite{Costa:2000jw}. In recent years, the model by Enberg, Reno and Sarcevic \cite{Enberg:2008te} has become the standard, especially within the IceCube collaboration, which usually expresses prompt fluxes in ``ERS units''. For muons, electromagnetic decays of unflavored vector mesons make a significant additional contribution not present in neutrinos \cite{Illana:2010gh}, and at the very highest energies di-muon pairs are produced by Drell-Yan processes \cite{Illana:2009qv}. Especially the first process should lead to a substantial enhancement of the prompt muon flux compared to neutrinos \cite{Fedynitch:2015zma}. A detailed discussion can be found in \ref{sec:simple-prompt}.

It has long been suggested to use large-volume neutrino detectors to constrain the prompt component of the atmospheric muon flux directly  \cite{Gelmini:2002sw}. Apart from the aspect of particle physics, the approximate equivalence between prompt muon and neutrino fluxes would help to constrain atmospheric background in the energy region critical for astrophysical searches.

Past measurements of the muon energy spectrum in volume detectors were not able to identify the prompt component. Usually based on the zenith angle distribution alone, the upper end of their energy range fell one order of magnitude or more below the region where the prompt flux is expected to become measurable \cite{Kochanov:2009rn}. The LVD collaboration, by exploiting azimuthal variations in the density of the surrounding material, was able to set a weak limit \cite{Aglietta:1999ic}. The Baksan Underground Scintillation Telescope reported a significant excess above even the most optimistic predictions \cite{Bogdanov:2009ny}, but the result has not yet been confirmed independently.

\section{Data Samples}

\subsection{Experimental Data}

The data used in this study were taken during two years of detector operation from 2010 to 2012. Originally the analysis was developed for the first year only, but problems related to simulation methods as discussed in Section \ref{sec:simulation} made it necessary to base the high-energy muon measurement on the subsequent year instead.

\begin{table}[h!]
  \footnotesize
  \begin{center}
    \begin{tabular}{|c|c|c|}
      \hline
      Time Period & Detector Configuration & Livetime \\
      \hline
      05-31-2010 - 05-13-2011 & 79 Strings (IC79) & 313.3 days \\
      05-13-2011 - 05-15-2012 & 86 Strings (IC86) & 332.1 days \\
      \hline
    \end{tabular}	
    \caption{Experimental Data Sets.}
    \label{dataset_livetime}
  \end{center}
\end{table}

\vspace{-0.5cm}

The main IceCube trigger requires four or more pairs of neighboring or next-to-neighboring DOMs to register a signal within a time of 5 \textmu s. Full event information is read out for a window extending from 10 \textmu s before to 22 \textmu s after the moment at which the condition was fulfilled. Including events triggered by the surface array IceTop and the low-energy extension DeepCore, for which special conditions are implemented, this results in a total event rate of approximately 3000 $\textrm{s}^{-1}$ for the full 86-string detector configuration.

As data transfer from the South Pole is constrained by bandwidth limitations, only specific subsets are available for offline analyses. The main requirement in the studies presented here was an unbiased base sample. The main physics analyses therefore use the filter stream containing all events with a total of more than 1,000 photo-electrons. Additionaly, minimum bias data corresponding to every 600th trigger were applied to evaluate detector systematics.

Reconstruction of track direction and quality parameters followed the standard IceCube procedure for muon candidate events \cite{Aartsen:2014cva}, based on multiple photo-electron information and including isolated DOMs registering a signal. In addition, various specific energy reconstruction algorithms were applied. For all data, the differential energy deposition was calculated using the deterministic method discussed in \ref{sec:ddddr}, and the track energy was estimated by a truncation method \cite{Abbasi:2012wht}. Likelihood-based energy reconstructions \cite{Aartsen:2013vja} were applied to the first year of data only, primarily for evaluation purposes.

\subsection{Simulation}\label{sec:simulation}

The standard method used for simulation of cosmic ray-induced air showers in IceCube is the CORSIKA software package \cite{corsika}, in which the physics of hadronic interactions are implemented via externally developed and freely interchangeable modules. In this study, as in all IceCube analyses, mass air shower simulation production was based on SIBYLL 2.1 \cite{Ahn:2009wx}. To investigate systematic variations, smaller sets of simulated data were produced using the QGSJET-II \cite{Ostapchenko:2010vb} and EPOS 1.99 \cite{Werner:2009zzc} models.

In the current version of CORSIKA (7.4), the contribution from prompt decays of charmed hadrons and short-lived vector mesons to the muon flux is usually neglected. An accurate simulation would in any case be difficult due to strong uncertainies on production and re-interaction cross sections. For this study, the prompt component of the atmospheric muon flux was modeled by re-weighting events produced in decays of light mesons. The exact procedure is described in \ref{sec:simple-prompt}.

\begin{figure}
  \centering
  \includegraphics[width=220pt]{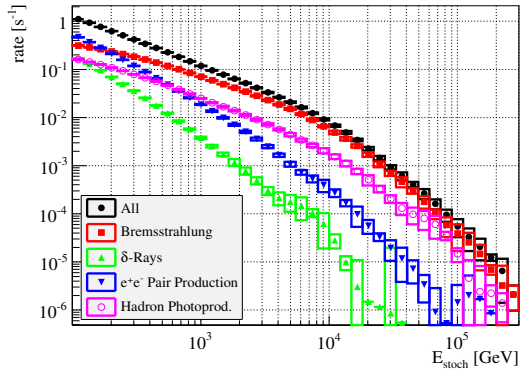}
  \caption{Energy spectra of discrete stochastic energy losses along muon tracks simulated using the mmc code \cite{Chirkin:2004hz}. The data sample corresponds to events with more than 1,000 registered photo-electrons in the IceCube detector. For demonstration purposes, the primary cosmic ray spectrum is modeled as an unbroken $E^{-2.7}$ power law.}
  \label{fig-losspec}
\end{figure}

High-energy muons passing through matter lose their energy through a variety of specific processes \cite{Koehne:2013gpa}, which in IceCube are modeled by a dedicated simulation code \cite{Chirkin:2004hz}. The energy spectra of discrete catastrophic losses along atmospheric muon tracks predicted to occur within the IceCube detector volume are shown in Fig. \ref{fig-losspec}.

For all energy loss processes, the corresponding Cherenkov photon emission is calculated. Every photon is then tracked through the detector medium until it is either lost due to absorption or intersects with an optical module \cite{Chirkin:2013tma}. This detailed procedure is necessary to account for geometrically complex variations in the optical properties of the ice, but has the disadvantage of being computationally intensive, limiting the amount of simulated data especially for bright events. 

The variations between direct photon propagation and the tabular method previously used in IceCube simulations were evaluated for each of the studies presented in this paper. It was found that in the case of high-multiplicity bundles the difference can be accounted for by a simple correction factor, while for high-energy tracks the distortion was so severe that simulations produced with the obsolete method were unusable. Simulation mass production based on direct photon propagation is only available for the 86-string detector configuration, requiring the use of a corresponding experimental data set. In order to reduce computational requirements, the measurement of bundle multiplicity was not duplicated and instead solely relies on data from the 79-string configuration.

The low cosmic ray flux rate at the highest primary energies means that even relatively few events correspond to large amounts of equivalent livetime. Accordingly, for the measurement of the bundle multiplicity spectrum simulation statistics are not a limiting factor. In the region before and at the knee, where the dominant part of high-energy muons are produced, far more showers need to be simulated. For this reason, the statistical accuracy of the single muon energy spectrum measurement is limited by the amount of simulated livetime, generally corresponding to substantially less than one year.

The calculation of detector acceptance and conversion of muon fluxes from South Pole to standard conditions for high energy muons as described in Section \ref{sec-hemu-espec} made use of an external simulated data set produced for a dedicated study on the effect of hadronic interaction models on atmospheric lepton fluxes \cite{Fedynitch:2012fs}.

%\begin{figure}
%\centering
%\includegraphics[width=220pt]{Lead_muensurf_3primtypes.png}
%\caption{Lead\_muen\_surf}
%\label{fig-leadmuons}
%\end{figure}

\section{Low-Energy Muons}\label{sec:le-muons}

\subsection{Observables}\label{sec:lolev-obs}

A comprehensive verification of detector performance requires the demonstration that atmospheric muon data are understood at a basic level. Sufficient statistics for this purpose are in IceCube provided by the minimum bias sample, consisting of every 600th event triggering the detector.

Two simple parameters were used in the evaluation. These are the zenith angle $\theta_{\rm{zen}}$ of the reconstructed track, with $\theta_{\rm{zen}} = 0$ for vertically down-going muons from zenith, and the total number of photo-electrons $Q_{\rm{tot}}$ registered in the event. 

\begin{figure}[ht]
  \centering
  \begin{subfigure}{220pt}
    \includegraphics[width=220pt]{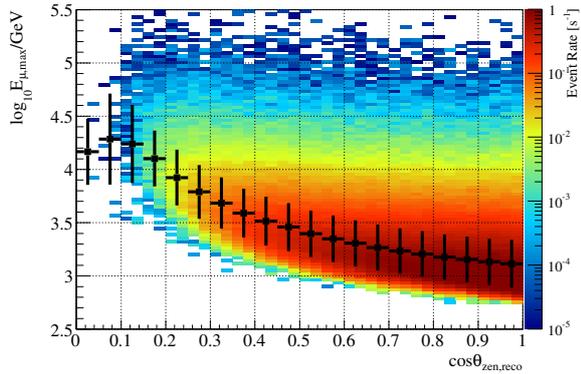}
    \caption {Muon Surface Energy}
  \end{subfigure}
  \begin{subfigure}{220pt}
    \includegraphics[width=220pt]{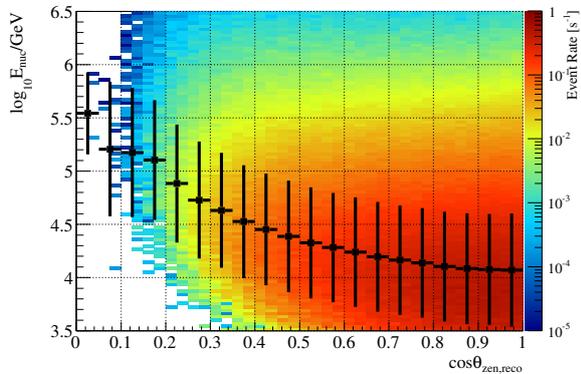}
    \caption {Parent Nucleon Energy}
  \end{subfigure}
  \caption{Relation between reconstructed zenith angle and energy for simulated muon showers triggering the IceCube detector. The distributions correspond to minimum bias data after track quality selection described in Sec. \ref{sec:lolev-result}. Superimposed are mean and spread of the distribution.}
  \label{fig-zenang-emu}
\end{figure}

%The respective result from IceCube is shown in Fig. \ref{fig-pelusita}. 
The angular dependence of the muon flux can be directly related to the energy spectrum in the TeV range, because the threshold increases as a function of the amount of matter that a muon has to traverse before reaching the detector. The limiting factors near the horizon are the rapid increase of the mean surface energy approximately proportional to $\exp(\sec\theta_{\rm{zen}})$, the corresponding decrease in flux, and eventually the irreducible background from atmospheric muon neutrinos. Purely angular-based muon energy spectra therefore only reach up to energies of 20-30 TeV, depending on the depth of the detector and the type of surrounding material. For the specific case of IceCube, the relation of zenith angle to muon and primary nucleon energy is shown in Fig. \ref{fig-zenang-emu}.

%\begin{figure}
%\centering
%\includegraphics[width=220pt]{pelusita.jpg}
%\caption{A bunny.}
%\label{fig-pelusita}
%\end{figure}

\begin{figure}[t]
  \centering
  \includegraphics[width=220pt]{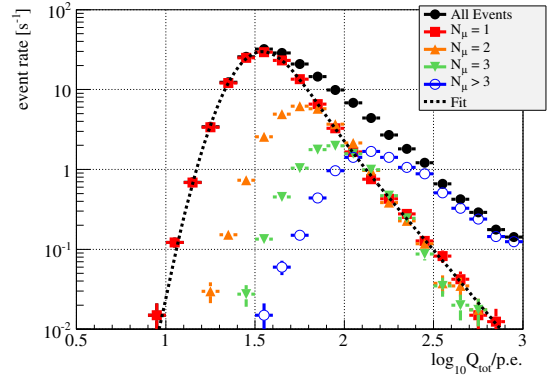}
  \includegraphics[width=220pt]{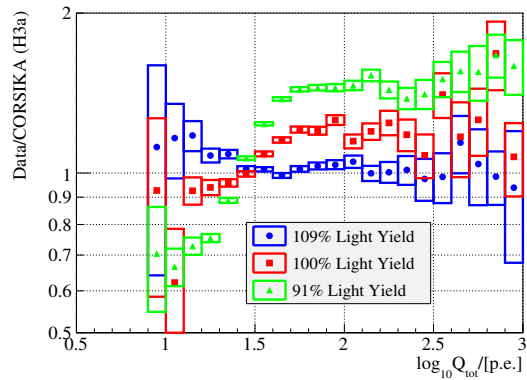}
  \caption{Top: Simulated distributions of total number of photo-electrons in event, separated in dependence of number of muons in bundle at closest approach to the center of the IceCube detector. The functional dependence of the fit is described in the text. Bottom: Change of data/simulation ratio for different assumptions about the light yield, effectively corresponding to the relation between energy deposition and number of registered photo-electrons. The simulation was weighted according to the H3a primary flux model.}
  \label{fig-12many}
\end{figure}

The total number of photo-electrons (``brightness'') of atmospheric muon events is closely related to the muon multiplicity, as demonstrated in Fig. \ref{fig-12many}, where events with photons registered by the DeepCore array were excluded to avoid minor biases at the very low end of the distribution. In the experimental measurements described below, all events were included. The emitted Cherenkov light is in good approximation proportional to the total energy loss, and the multiplicity spectrum can therefore be measured even at low energies, although its interpretation is difficult because of the varying threshold for the individual components of the cosmic ray flux.

The distribution for a fixed number of muons can be described by a transition from a Gaussian distribution to an exponential in terms of the parameter $q\equiv \log_{\rm{10}}(Q_{\rm{tot}}/\textrm{p.e.})$:

\begin{equation}
  \label{singmufit_param}
  \frac{\Delta n_{\rm{event}}}{\Delta q} = N\cdot\exp\left(\frac{-\frac{1}{2\sigma^{2}}(q-q_{\rm{peak}})^2}{1+\exp^{a(q-q_{\rm{peak}})}}+\frac{\beta_{\mu}(q-q_{\rm{peak}})}{1+\exp^{-a(q-q_{\rm{peak}})}}\right)
\end{equation}

\begin{table}[h!]
  \small
  \begin{center}
    \begin{tabular}{|c|c|c|}
      \hline
      Fit Parameter & Value & Interpretation \\
      \hline
      $q_{\rm{peak}}$ & $1.615\pm0.002$ & 42.2 p.e \\
      a & $5.35\pm0.34$ & Transition Smoothness \\
      $\sigma$ & $0.160\pm0.004$ & Width of Gaussian \\
      $\beta_{\mu}$ & $-6.23\pm0.07$ & Power Law Index \\
      N & arbitrary & normalization \\
      \hline
    \end{tabular}	
    \caption{Parameters and values for the fit to the single muon distribution shown in Fig. \ref{fig-12many}. The $\chi^{2}/$dof of the fit is 26.75/16, where the main deviation from the fit is found in the first three bins of the histogram.}
    \label{onemuon_table}
  \end{center}
\end{table}

The free fit parameters for the case of single muon events are described in Table \ref{onemuon_table}. While all values depend on the exact detector setup and event sample and have no profound physical meaning, the description nevertheless provides valuable insights. For example, the peak position corresponds to the average number of photo-electrons detected from a minimum ionizing track crossing the full length of the detector, and represents an approximate calorimetric scale from which the response to a given energy deposition can be estimated.  

The lower, Gaussian half of the one-muon distribution only depends on the experimental setup and shows minimal sensitivity to physics effects in simulations. In particular, the peak value $q_{\rm{peak}}$ varies as a function of the optical efficiency, a scalar parameter which expresses the effects of a wide variety of underlying phenomena \cite{Aartsen:2013eka}. As shown in the lower panel of Fig. \ref{fig-12many}, above a certain threshold only the flux level, not the shape of the distribution is affected by detector systematics. This is a common observation for energy-related observables and a simple consequence of the effect of a slight offset on a power law function. Note that the measured distribution is fully consistent with expectation within the 10\% light yield variation usually assumed as systematic uncertainty in IceCube. 

\subsection{Connection to Primary Flux}\label{sec:primflux-conn}

\begin{figure}[ht]
  \centering
  \includegraphics[width=220pt]{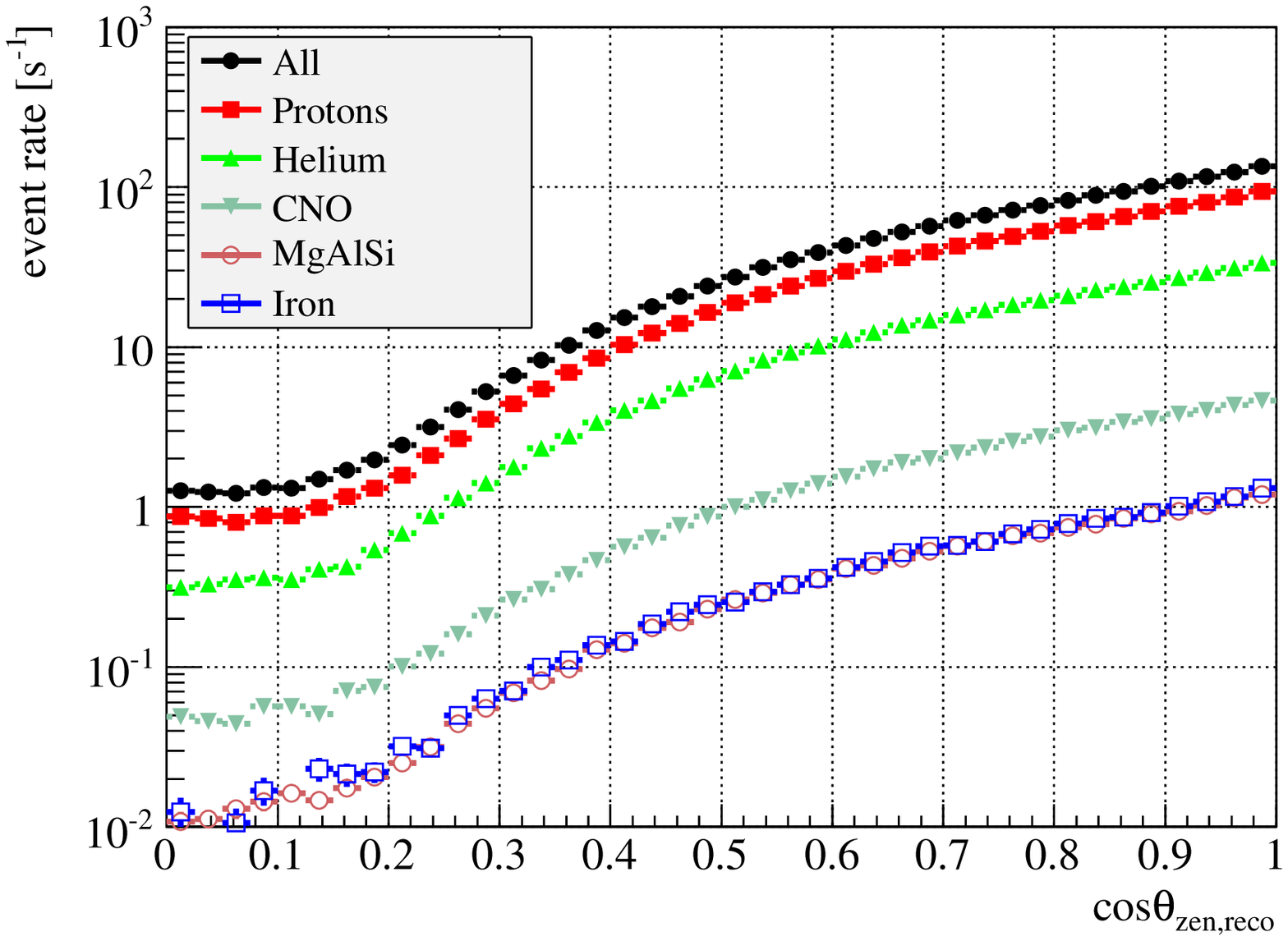}
  \includegraphics[width=220pt]{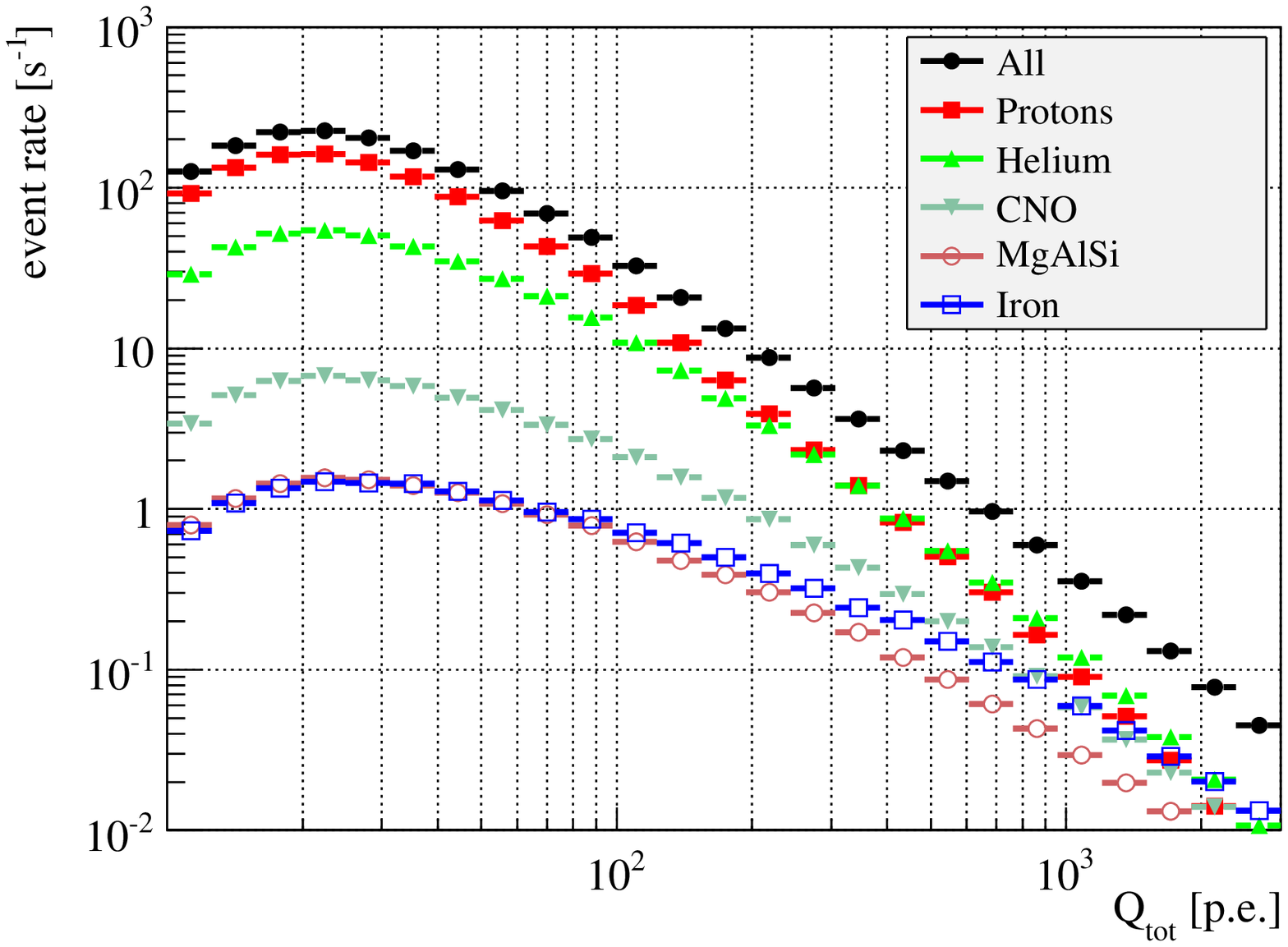}
  \caption{Low-level observables for IceCube atmospheric muon events at trigger level, separated by cosmic ray primary type. The simulated data were generated with CORSIKA \cite{corsika} and weighted according to the H3a model \cite{Gaisser:2013bla}.}
  \label{fig-lolev-primtype}
\end{figure}

The consistency of measurements on separate observables can be checked by relating them to the primary cosmic ray flux. Assuming that the current understanding of muon production in air showers is correct, there should be a model which describes both energy and multiplicity spectra of atmospheric muons. 

Figure \ref{fig-lolev-primtype} shows the two proxy variables described in the previous section, separated by elemental type of the cosmic ray primary. At all angles, the muon flux is strongly dominated by proton primaries. This is a simple consequence of the connection between muon energy and energy per nucleon of the primary particle, and does not depend strongly on the specific cosmic ray flux model \cite{Gaisser:2012zz}. Likewise, the multiplicity-related brightness distribution is for low values dominated by light primaries, a consequence of the varying threshold energies shown in Fig. \ref{fig-allev}.

The cosmic ray flux models best reproducing the latest direct measurement in the relevant energy region from 10 to 100 TeV \cite{Ahn:2010gv} are GST-GF and H3a \cite{Gaisser:2013bla}. For the comparisons between data and simulation in the following section, they are used as benchmark models representing the best prediction at the current time. In addition, toy models corresponding to straight power law spectra are discussed to illustrate the effect of variations in the primary nucleon index. In these, elemental composition and absolute flux levels at 10 TeV primary energy correspond to the rigidity-dependent poly-gonato model \cite{Hoerandel:2002yg}, used as default setting for the production of IceCube atmospheric muon systematics data sets.

\subsection{Experimental Result}\label{sec:lolev-result}

\begin{figure}[t]
  \centering
  \includegraphics[width=220pt]{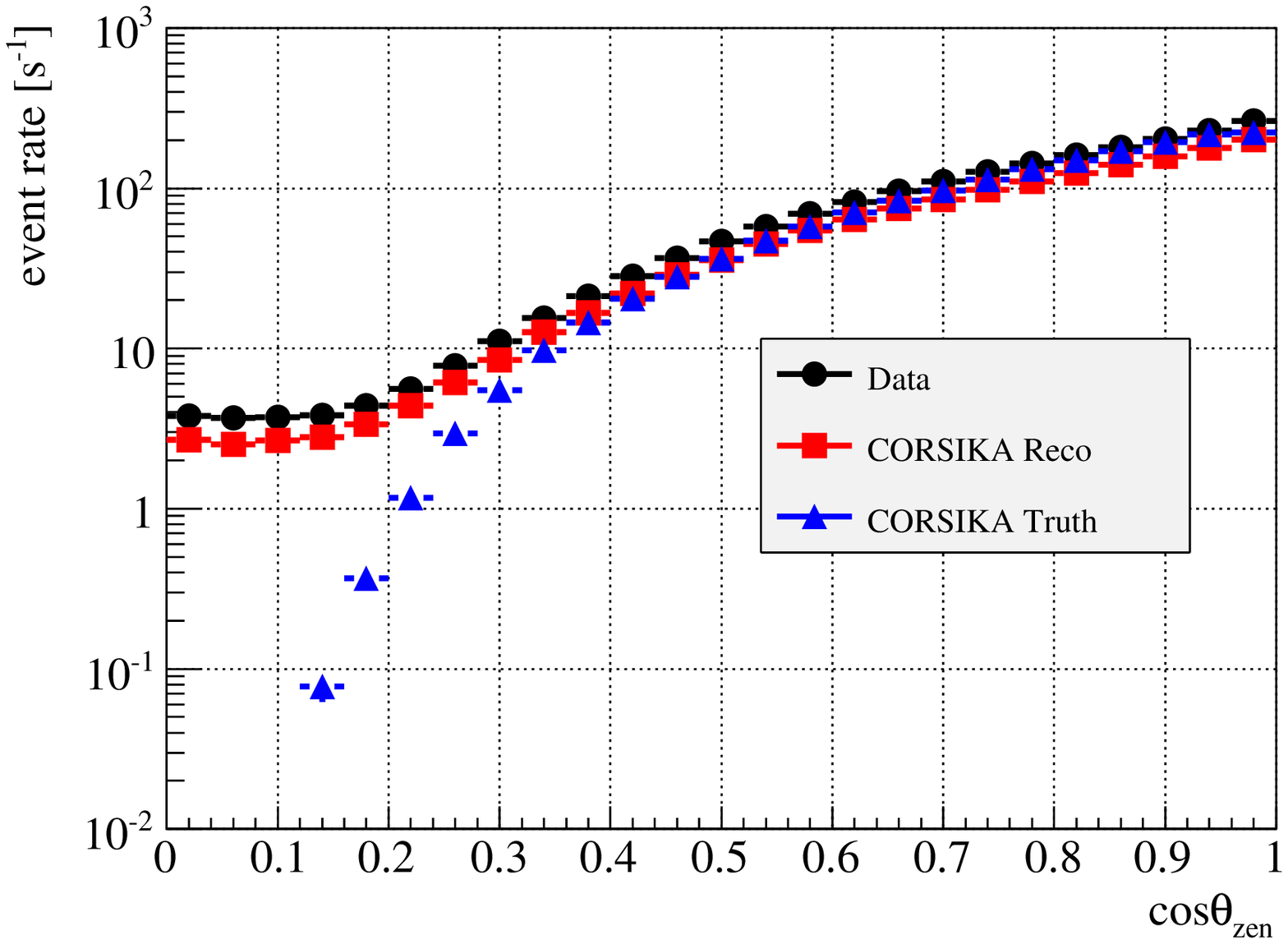}
  \includegraphics[width=220pt]{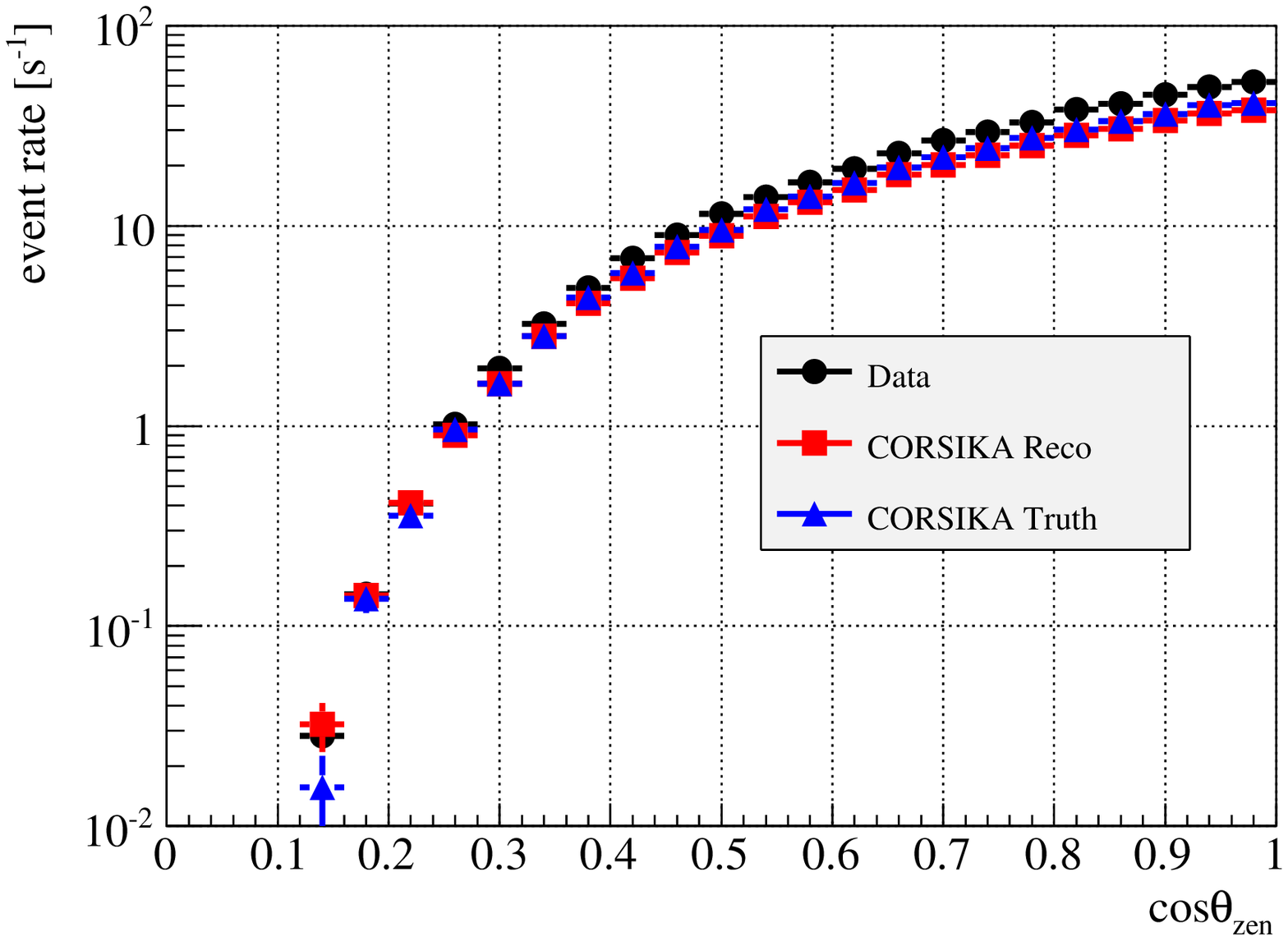}
  \caption{Angular distribution of true and reconstructed atmospheric muon tracks in simulation compared to experimental data. Top: Trigger Level, Bottom: High-Quality Selection. The event sample corresponds to minimum bias data encompassing all trigger types. The ratio of experimental data to simulation is shown in Figs. \ref{fig-minbias-paramrat} (a) and (c).}
  \label{fig-angdest-true-reco}
\end{figure}

For the study presented in this section, minimum bias data and simulation were compared at trigger level and for a sample of high-quality tracks requiring:

\begin{itemize}
\item Reconstructed track length within the detector exceeding 600 meters.
\item $llh_{\rm{reco}}/(N_{\rm{DOM}}-2.5) < 7.5$, where $llh_{\rm{reco}}$ corresponds to the likelihood value of the track reconstruction and $N_{\rm{DOM}}$ to the number of optical modules registering a signal.
\end{itemize}

The stringency of the quality selection is slightly weaker than in typical neutrino analyses. For tracks reconstructed as originating from below the horizon, the contribution from mis-reconstructed atmospheric muon events amounts to about 50\%.

Simulated and experimental zenith angle distributions are shown in Fig. \ref{fig-angdest-true-reco}. Even at trigger level, the influence of mis-reconstructed tracks can be neglected in the region above 30 degrees from the horizon ($\cos\textrm{ }\theta_{\rm{zen}} = 0.5$).
%This fact was exploited in the measurement of the cosmic ray anisotropy \cite{Abbasi:2011zka}, where due to the reduced data format background reduction by imposition of quality criteria was not possible. 
For the high-quality data set, true and reconstructed distributions are approximately equal down to angles of $\cos\textrm{ }\theta_{\rm{zen}} = 0.15$, or 80 degrees from zenith. 

%Consistency checks can therefore be performed by comparing the measured angular distribution for the full data sample and after application of track quality criteria. If detector systematics are correctly understood, the results should agree within the expected error margin.

\begin{figure*}[!th]
  \centering
  \begin{subfigure}{220pt}
    \includegraphics[width=220pt]{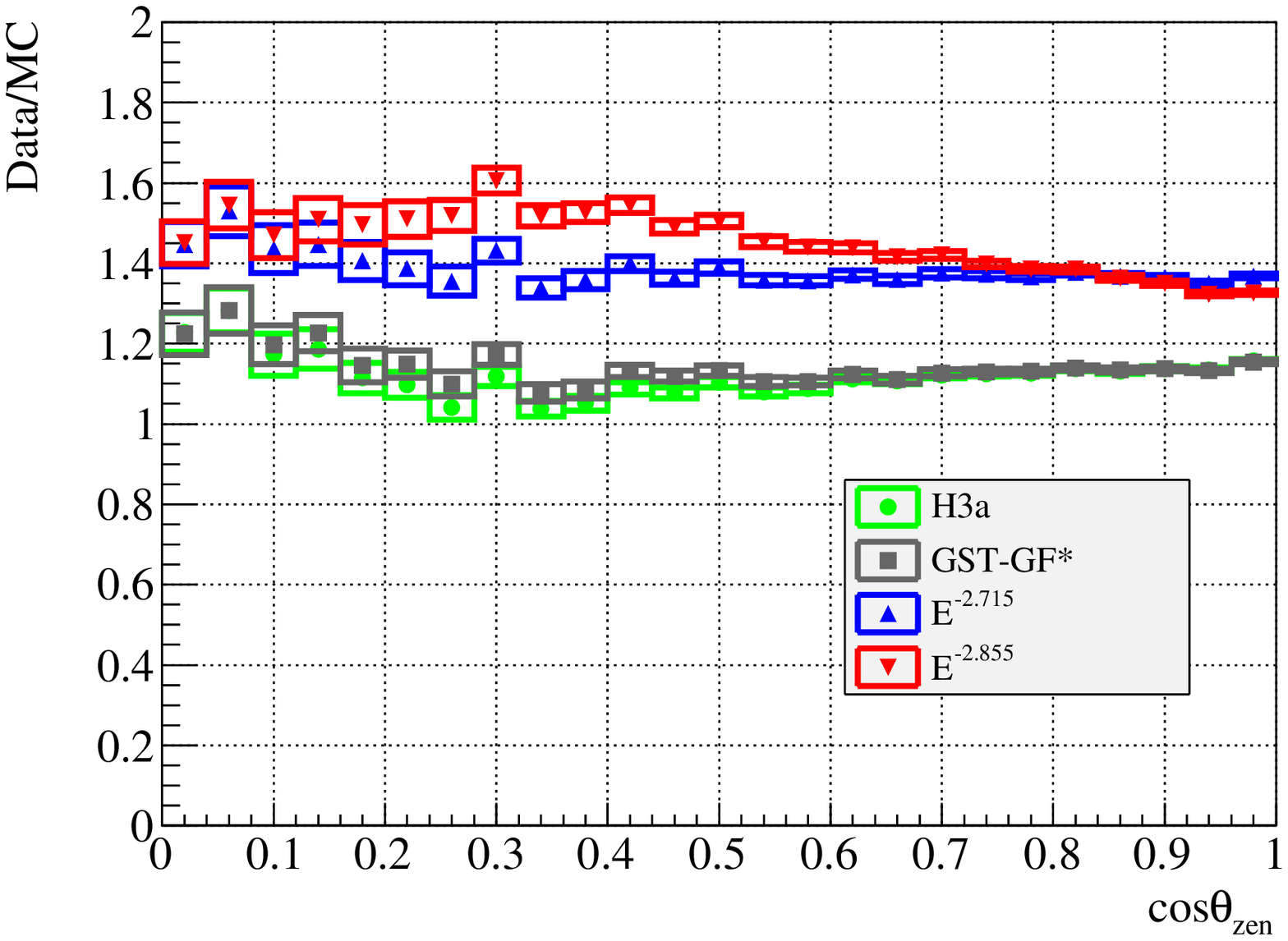}
    \caption {$\cos\textrm{ }\theta_{\rm{zen}}$, Trigger Level}
  \end{subfigure}
  \begin{subfigure}{220pt}
    \includegraphics[width=220pt]{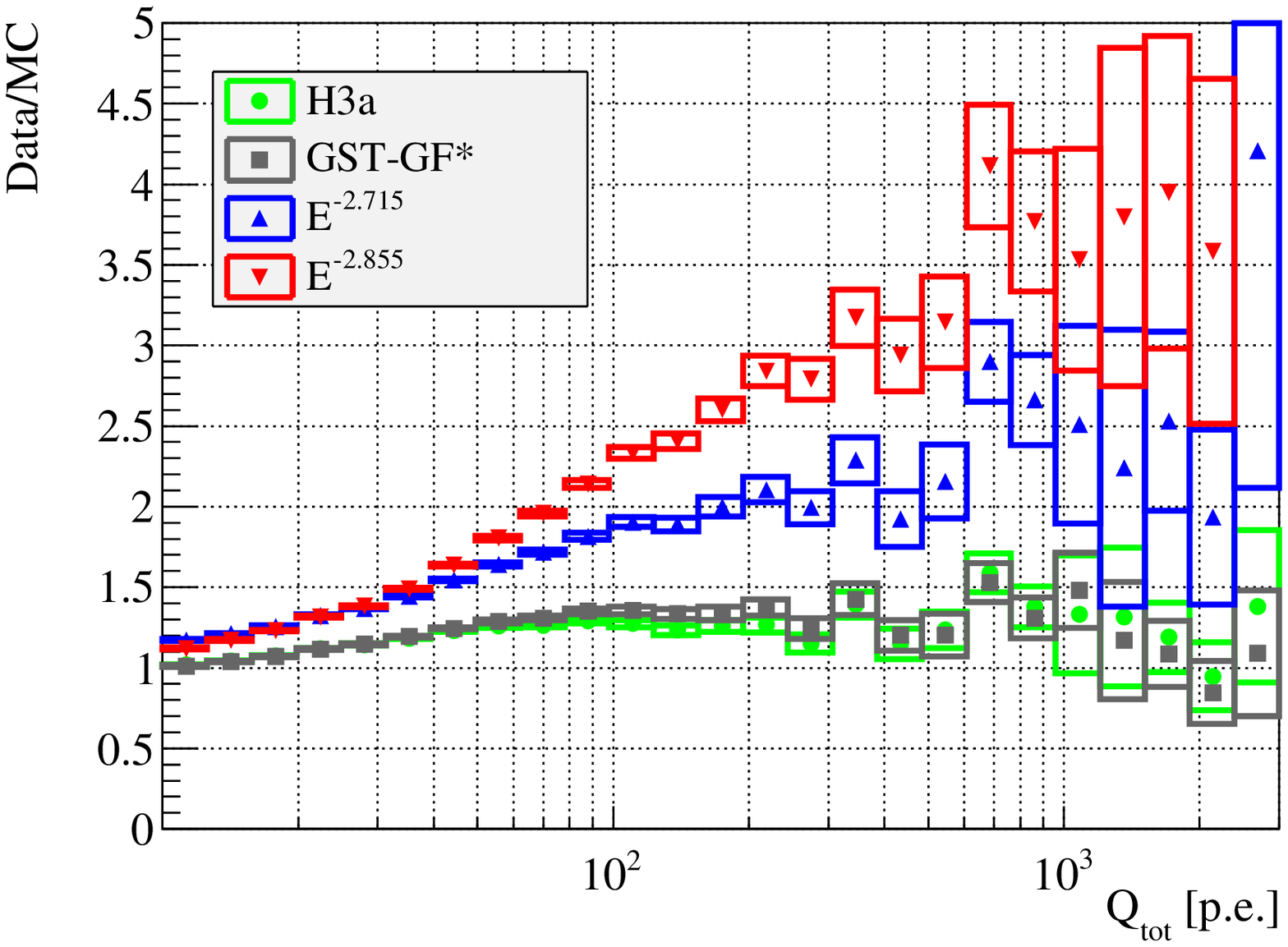}
    \caption {$Q_{\rm{tot}}$, Trigger Level}
  \end{subfigure}
  \begin{subfigure}{220pt}
    \includegraphics[width=220pt]{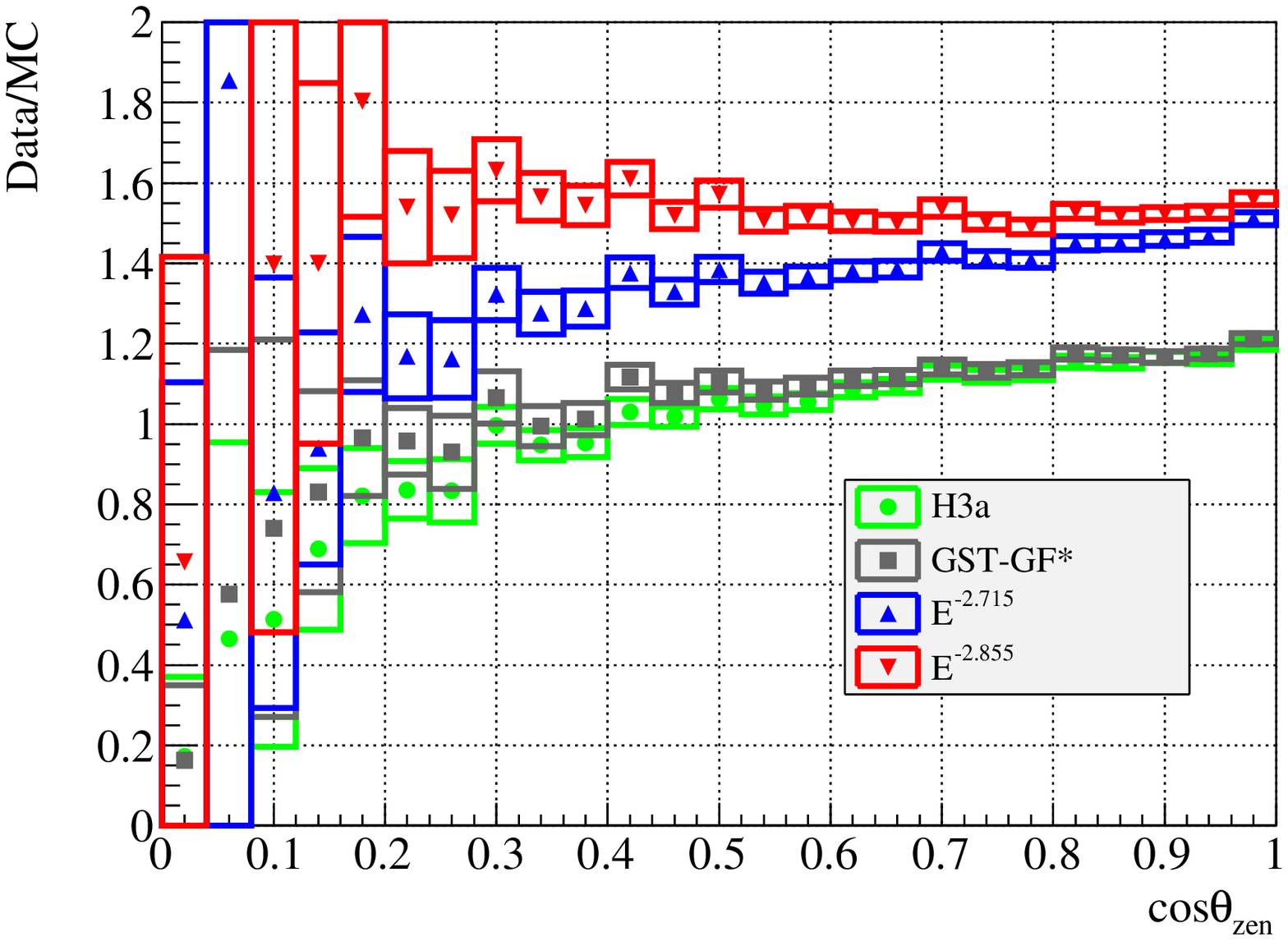}
    \caption {$\cos\textrm{ }\theta_{\rm{zen}}$, High-Quality Tracks}
  \end{subfigure}
  \begin{subfigure}{220pt}
    \includegraphics[width=220pt]{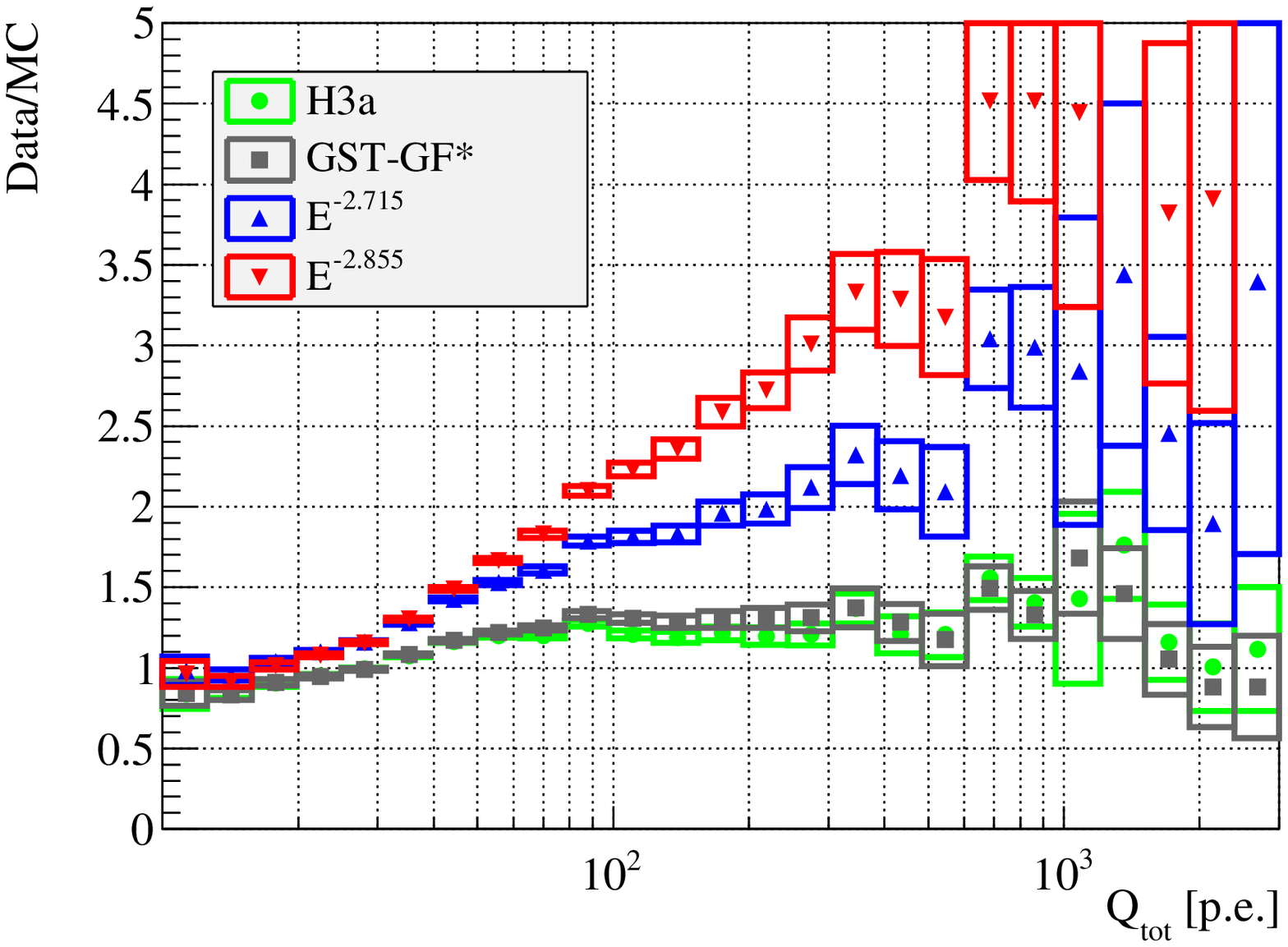}
    \caption {$Q_{\rm{tot}}$, High-Quality Tracks}
  \end{subfigure}
  \caption{Ratio of experimental data to simulation in terms of reconstructed zenith angle $\theta_{\rm{zen}}$ and total amount of registered photo-electrons $Q_{\rm{tot}}$. The primary flux models used in this comparison are discussed in Section \ref{sec:primflux-conn}.}
  \label{fig-minbias-paramrat}
\end{figure*}

\begin{table*}[ht!]
  \begin{center}
%    \small
    \begin{tabular}{|c|c|c|c|c|}
      \hline
      Type  & Variation & $\gamma_{\rm{CR,Trigger}}$ & $\gamma_{\rm{CR,High-Q}}$ & $\Delta\gamma_{\rm{CR}}$ \\
      \hline
      Hole Ice Scattering & 30cm/100cm & $\pm 0.03$ & $+0.03/-0.05$ & $+0.01/-0.02$ \\
      Bulk Ice Absorption & $\pm 10\%$ & $\pm 0.03$ & $\pm 0.02$ & $\pm 0.05$ \\
      Bulk Ice Scattering & $\pm 10\%$ & $<0.01$ & $\pm 0.01$ & $<0.015$ \\
      Primary Composition & p/He & $<0.01$ & $+0.03/-0.10$ & $-0.03/+0.10$ \\
      Hadronic Model & QGSJET-II/EPOS~1.99 & $+0.02/<0.01$ & $+0.03/<0.02$ & $<0.02$\\
      DOM Efficiency & $\pm 10\%$ & $<0.02$ & $+<0.02/-0.04$ & $+0.02/-<0.02$\\
      \textbf{Experimental Value} & Statistical Error & $2.715 \pm 0.003$ & $2.855 \pm 0.007$ & $0.140 \pm 0.008$\\
      \hline
    \end{tabular}	
    \caption{Cosmic ray nucleon spectrum measurement and influence of systematic uncertainties. The goodness of the experimental fit is $\chi^{2}$/dof = 13.0/11 at trigger and 12.6/11 at high-quality level.}
    \label{angsyst_table}
  \end{center}
\end{table*}

 Figure \ref{fig-minbias-paramrat} shows comparisons between data and simulation weighted according to several primary flux predictions. The total number of photo-electrons is described reasonably well by the simulation weighted according to the H3a model. Application of quality criteria does not lead to any visible distortion. The angular distribution, on the other hand, shows substantial inconsistencies. At trigger level, the spectrum is clearly harder than for the high-quality sample. The discrepancy does not depend on the particular track quality parameters used in the selection.

It is important to note that the absolute level of the ratio is not a relevant quantity for the evaluation. Consistency between measurement and expectation within the range of systematic uncertainties on the photon yield was demonstrated for the brightness distribution in Section \ref{sec:lolev-obs}. Also, absolute primary flux levels derived from direct measurements are typically constrained no better than to several tens of percents. For the toy models, the normalization was consciously chosen to produce a clear separation from the realistic curves in the interest of clarity. 

The trigger-level angular distribution in the region near the horizon becomes dominated by mis-reconstructed events consisting of two separate showers crossing the detector in close succession. The frequency of these ``coincident'' events scales with the square of the overall shower rate, leading to a spurious distortion of the ratio between data and simulation in cases where the absolute normalization is not exactly equal. This effect is visible in Fig. \ref{fig-minbias-paramrat} (a) at values below 0.3.

To quantify the discrepancy between trigger and high-quality level and investigate the influence of systematic uncertainties, the toy model simulation was fitted to data for $1>\cos\textrm{ }\theta_{\rm{zen}}>0.5$. In this region, influences of mis-reconstructed tracks are negligible even at trigger level, as demonstrated in Fig. \ref{fig-angdest-true-reco}. From Fig. \ref{fig-zenang-emu} it can be seen that this corresponds to a relatively small energy range for muons and parent nuclei, over which the power law index of the cosmic ray all-nucleon spectrum can be assumed to be approximately constant and used as sole fit parameter. As the normalization was left free, the best result simply corresponds to a flat curve for the ratio between data and simulation. Possible effects of variations in the primary elemental composition can be taken into account as a systematic error.

The numerical results of the fit to the angular distribution is shown in Table \ref{angsyst_table}.  Note that for cases where the statistical error due to limited simulated data exceeds the absolute value of the variation, only an upper limit is given. The best fit results for the spectral index at trigger and high-quality level, 2.715 and 2.855, are illustrated by the toy model curves in Fig. \ref{fig-minbias-paramrat}. Both measurements are softer than those of the realistic models, in which $\gamma_{\rm{nucleon}}\approx2.64$.

\subsection{Interpretation}

For the strong discrepancy between the measurements at trigger and high-quality level of $\Delta\gamma_{\rm{CR}} = 0.140\pm0.008 \textrm{(stat.)}$, the following explanations can be proposed:

\begin{itemize}
\item A global adjustment to the bulk ice absorption length of more than 20\%. This explanation would imply a major flaw in the method used to derive the optical ice properties \cite{Aartsen:2013rt}, and is strongly disfavored by the good agreement between the effective attenuation length in data and simulation demonstrated in \ref{sec:ddddr}.
\item A substantial change of the primary cosmic ray composition towards heavier elements. In an event sample entirely excluding proton primaries, the observed effect can be approximately reproduced. However, the increased threshold energy would require the overall primary flux to be more than three times higher than in the default assumption to produce the observed event rate. An explanation based purely on a heavier cosmic ray composition therefore appears highly unlikely.
\item A major inaccuracy of hadronic interaction simulations common to SIBYLL, QGSJET-II and EPOS. While this explanation seems improbable, especially given the almost perfect agreement between SIBYLL and EPOS, it should be noted that the models used in the IceCube CORSIKA simulation were developed before LHC data became available. Improved models are in preparation \cite{Pierog:2013ria} and it should be possible to evaluate them in the near future.
\item An unsimulated detector effect with a significant influence on the behavior of track quality parameters. Detectors using naturally grown ice are inherently difficult to model in simulations. The optical properties of the medium are inhomogeneous and photon scattering has a substantial influence on the data. The situation is complicated further by the placement of the active elements in re-frozen ``hole ice'' columns containg sizable amounts of air bubbles. Studies on possible error sources are ongoing at the time of writing, but currently there is no indication for a major oversight.
\end{itemize}

While the presence of an inconsistency is clear, from IceCube data alone there is no strict way to conclude whether the brightness or the angular measurement is more reliable. However, the evidence strongly points to an unrecognized angular-dependent effect introduced by track quality-related observables. The reasons are:

\begin{itemize}
\item The brightness distributions are consistent both between the two data samples and with direct measurements of the cosmic ray flux.
\item At trigger level, there are no major contradictions between brightness and zenith angle distributions.
\item The angular spectrum for the high-quality data set is significantly steeper than both the neutrino-derived result \cite{Aartsen:2013eka} and direct measurements. In comparisons to the latter, the error from the variation in primary composition does not apply, as proton and helium fluxes are constrained individually. The total systematic uncertainty on the all-nucleon power law index would in this case be reduced to about $\pm 0.06$, whereas the difference in measurement is larger than 0.2. On the other hand, it is interesting to note that the LVD detector found a value of $\gamma_{\rm{cr}} = 2.78 \pm 0.05$ \cite{Aglietta:1998nx}, closer to the IceCube high-quality sample result.
\end{itemize}

Even though angular distributions of atmospheric muons have been published by practically all large-volume neutrino detectors and prototypes \cite{Babson:1989yy, Bakatanov:1992gp, Ambrosio:1995cx, Belolaptikov:1997ry, Andres:1999hm, Aggouras:2005bg, Aiello:2009uh, Aguilar:2010kg}, none of the measurements is accurate enough to provide a strict external constraint. For the time being, there is no other choice than to note the effect and continue to investigate possible explanations. In the main physics analyses described in the subsequent sections, the possible presence of an angular distortion was taken into account as a systematic error on the result.

\section{Physics Analyses}\label{sec:phys-mu}

\begin{figure}[t]
  \centering
  \includegraphics[width=220pt]{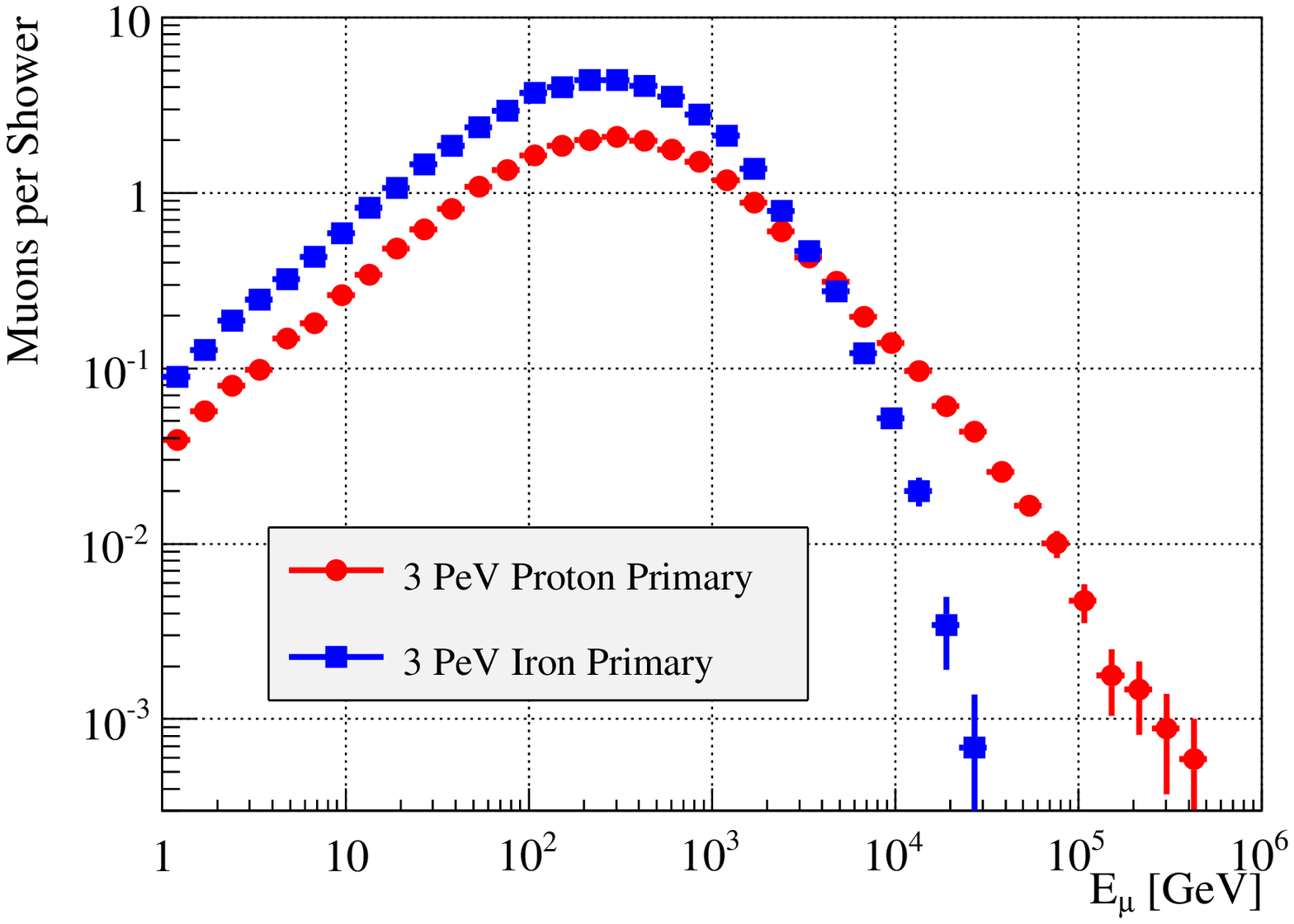}
  \includegraphics[width=220pt]{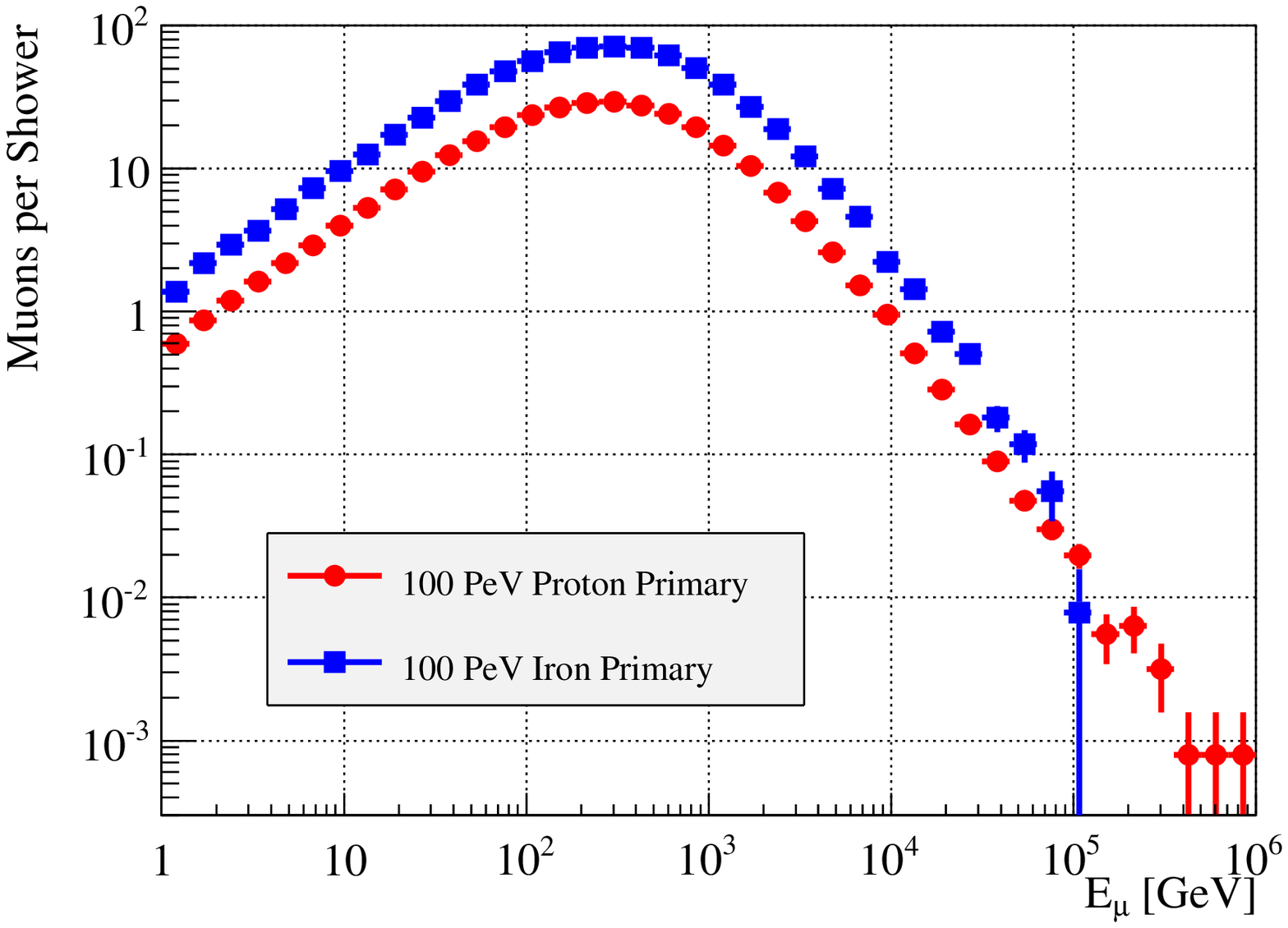}
  \caption{Distribution of muon energies in individual air showers at the IceCube detector depth simulated with CORSIKA/SIBYLL \cite{corsika, Ahn:2009wx}, averaged over all angles. Top: $E_{\rm{prim}}$ = 3 PeV. Bottom: $E_{\rm{prim}}$ = 100 PeV. The threshold effect visible at high muon energies in the top plot is due to the lower energy per nucleon in iron primaries. As the total energy increases, this effect becomes less and less visible and the spectra are identical except for a scaling factor.}
  \label{fig-emupdf}
\end{figure}

\begin{figure}[t]
  \centering
  \includegraphics[width=220pt]{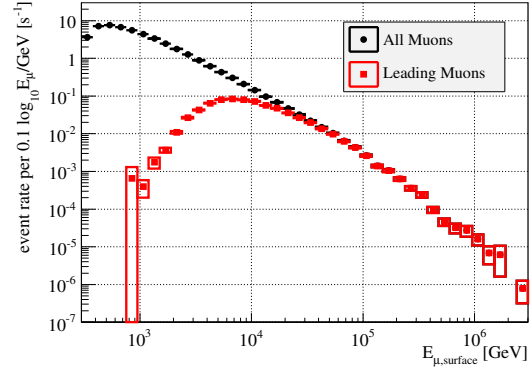}
  \caption{Surface energy distribution for all and most energetic (``leading'') muons in simulated events with a total of more than 1,000 registered photo-electrons in IceCube.}
  \label{fig-leadmuons}
\end{figure}

\begin{figure}[ht]
  \centering
  \includegraphics[width=220pt]{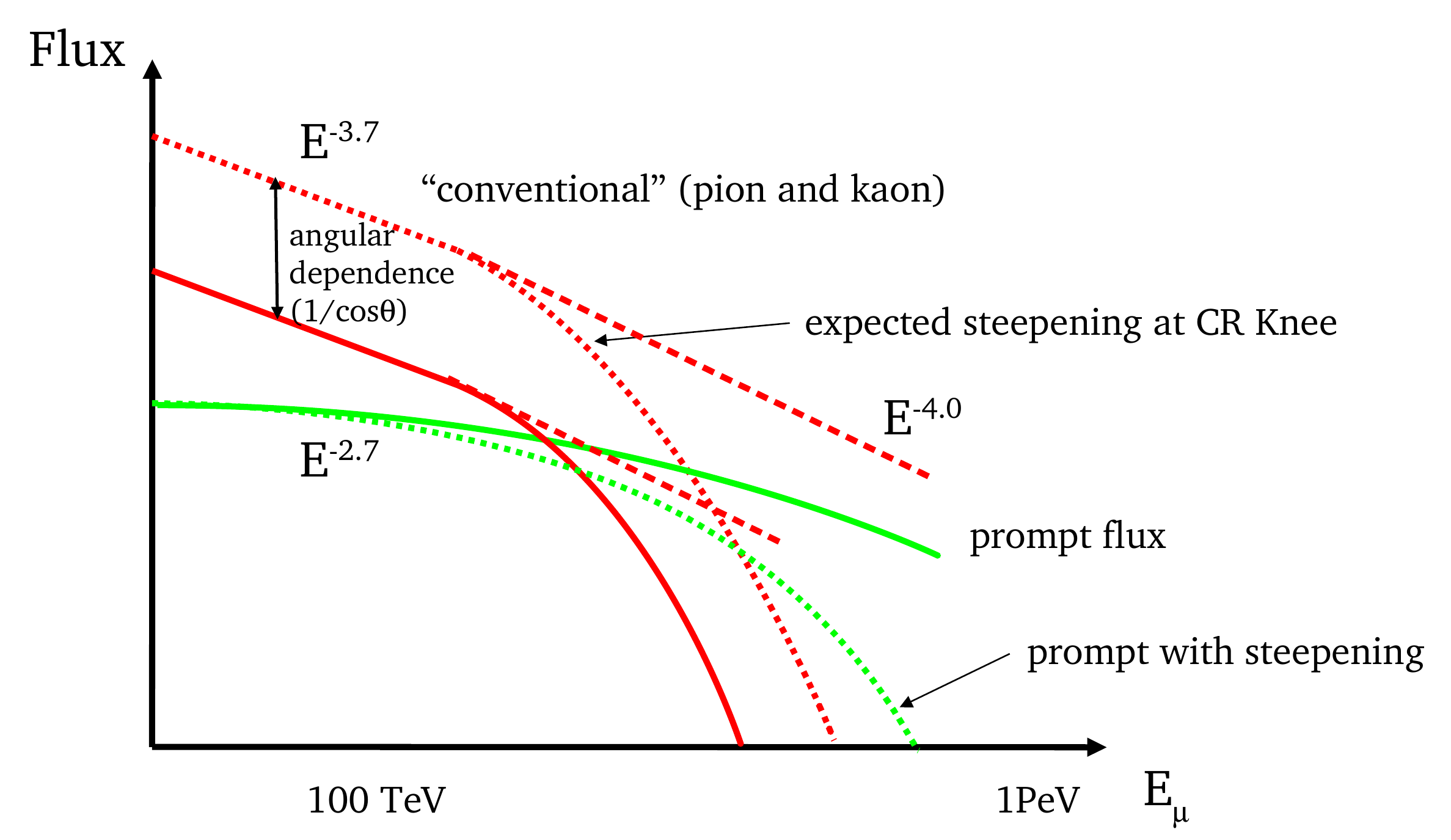}
  \caption{Sketch illustrating the contribution to the single muon spectrum at energies beyond 100 TeV. The ``conventional'' component from light mesons is sensitive to atmospheric density and varies as a function of the zenith angle \cite{Illana:2010gh}, that from prompt decays of short-lived hadrons is isotropic. Re-interactions cause the non-prompt spectrum to be steeper. The exact spectral shape depends on the all-nucleon cosmic ray flux, with a significant steepening expected due to the cutoff at the ``knee''.}
  \label{fig-promptsketch}
\end{figure}

While the study of low-energy atmospheric muons is instructive for detector verification and the evaluation of systematic uncertainties, the main physics potential lies in the measurement of events at higher energies. Here it is necessary to distinguish two main categories:

\begin{itemize}
\item \textbf{High-Multiplicity Bundles}, in which muons conform to typical energy distributions as shown in Fig. \ref{fig-emupdf}. The total energy $\sum{E_{\mu}}$ contained in the bundle is approximately proportional to the number of muons $N_{\mu}$, and related to primary mass $A$ and energy $E_{\rm{prim}}$ as 

\begin{equation}
\sum{E_{\mu}}\propto N_{\mu}\propto E_{\rm{prim}}^{\alpha}\cdot A^{1-\alpha},
\end{equation}
with $\alpha\approx0.79$. The dependence of the muon multiplicity on the mass of the cosmic ray primary is the main principle underlying composition analyses using deep detector and surface array in coincidence \cite{IceCube:2012vv}. Low-energy muons lose their energy smoothly, and fluctuations in the energy deposition are usually negligible.

\item \textbf{High-Energy Muons} with energies significantly exceeding the main bundle distribution. Their production is dominated by exceptionally quick decays of pions and kaons at an early stage in the development of the air shower. Figure \ref{fig-leadmuons} shows that showers with more than one muon with an energy above several tens of TeV are very rare. Any muon with an energy of 30 TeV or more will therefore very likely be the leading one in the shower, although this does not exclude the presence of other muons at lower energies.
The primary nucleus can in this case be approximated as a superposition of individual nucleons, each carrying an energy of $E_{\rm{nucleon}}=E_{\rm{prim}}/A$. High-energy lepton spectra are therefore a function of the primary nucleon flux.

Hadronic models, cosmic ray spectrum and composition all have a significant influence on TeV muons \cite{Lipari:1993ty}. In addition, at muon energies approaching 1 PeV prompt decays of short-lived hadrons play a significant role. The result is a complex picture with substantial uncertainties, as neither the exact behavior of the nucleon spectrum at the knee nor the production of heavy quarks in air showers is fully understood. A schematic illustration of the muon flux above 100 TeV is given in Fig. \ref{fig-promptsketch}. 

Charged leptons and neutrinos are usually produced in the same hadron decay. The energy spectrum of single muons is therefore the quantity most relevant for the constraint of atmospheric neutrino fluxes. Since the stochasticity of energy losses in matter increases with the muon energy, the signal registered in the detector can vary substantially, as in the case of neutrino-induced muons. 
\end{itemize}

The transition between the two atmospheric muon event types is gradual. 
%High-energy events rarely consist of single particles, and the characteristics of the accompanying bundle of low-energy muons can in some cases be determined and used to extract additional information about the primary nucleus, as discussed in \ref{sec:he-bundle}. 
High-energy events rarely consist of single particles, and the characteristics of the accompanying bundle of low-energy muons could in principle for some cases be determined and used to extract additional information about the primary nucleus.
At low energies the distinction becomes meaningless, as events are usually caused by single or very few muons with energies below 1 TeV.

\begin{figure}[ht]
  \centering
  \includegraphics[width=220pt]{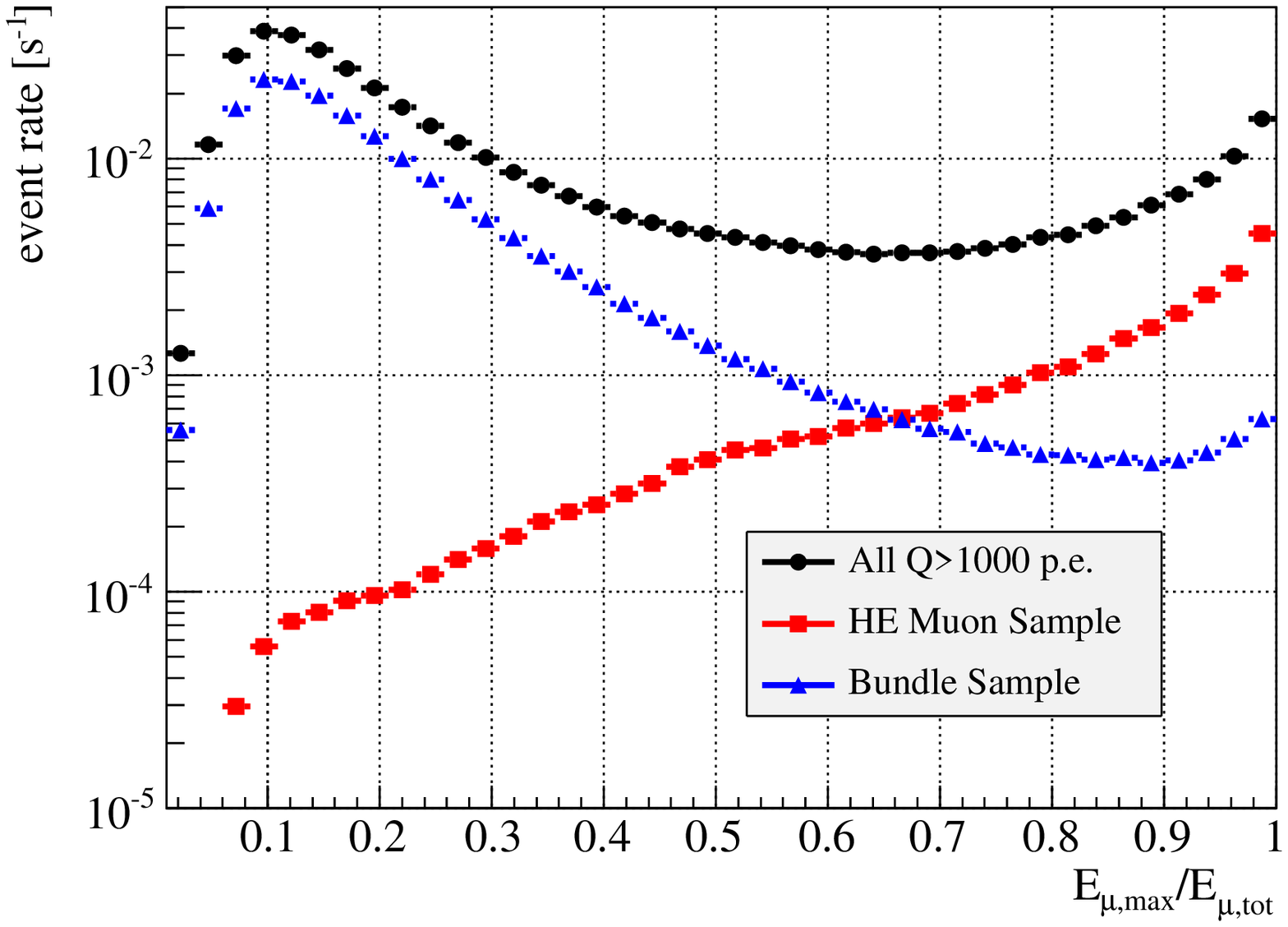}
  \includegraphics[width=220pt]{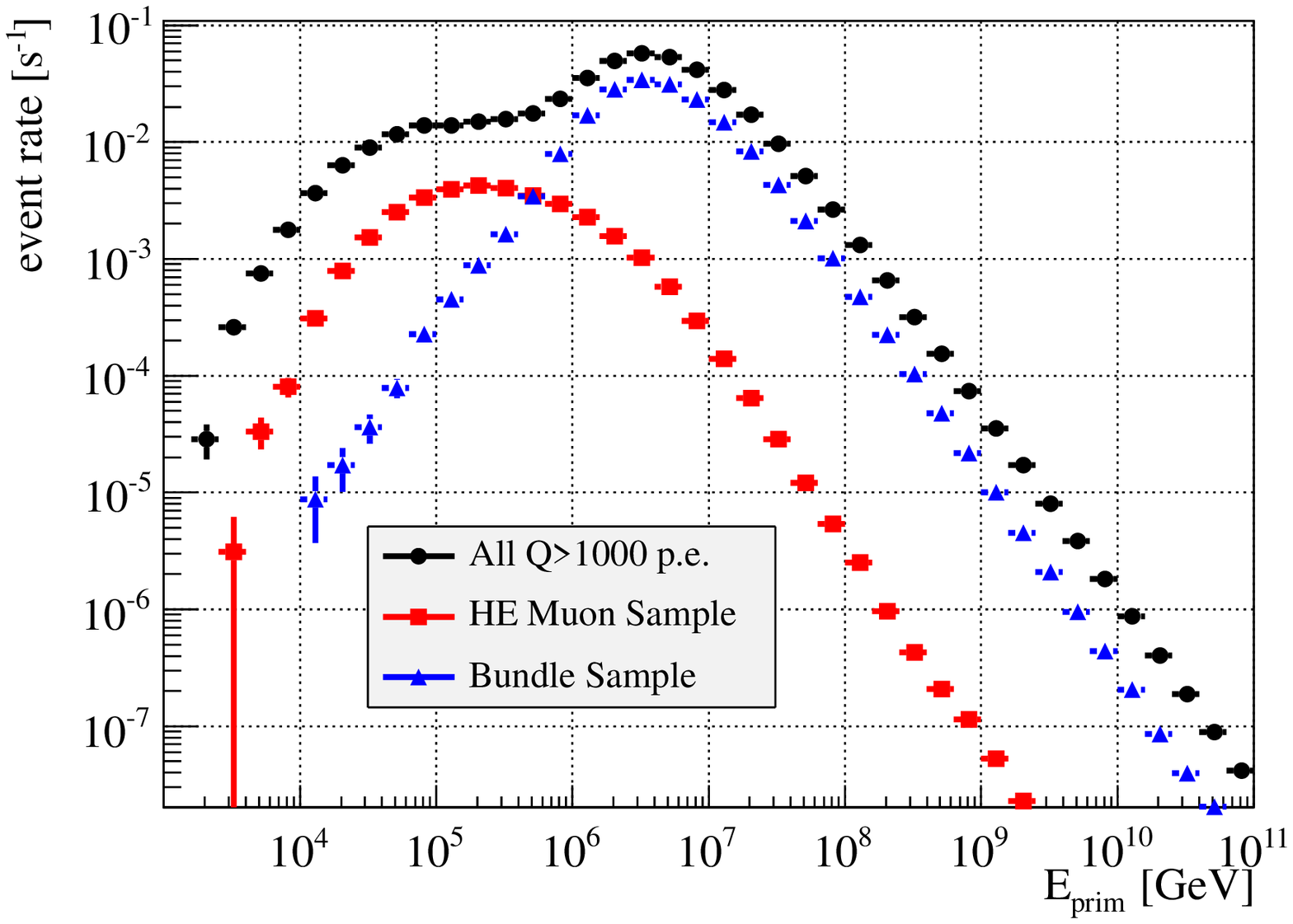}
  \caption{Event samples used for the measurements described in Sec. \ref{sec:bundles} and \ref{sec:hemu}. Shown are true parameter distributions for simulated data with more than 1,000 registered photo-electrons. Top: Fraction of total bundle energy carried by the leading muon. Bottom: Energy of CR primary. The bimodal shape of the distributions becomes more pronounced with increasing brightness.}
  \label{fig-evsamp-trupar}
\end{figure}

Two separate analysis samples were extracted from the data, corresponding to high-energy muon and high-multiplcity bundle event types. Figure \ref{fig-evsamp-trupar} illustrates their characteristics in terms of true event parameters derived from Monte-Carlo simulations. High-energy events, in which the total muon energy is dominated by the leading particle, are outnumbered by a factor of approximately ten. The corresponding need for more rigorous background suppression leads to a lower selection efficiency than in the case of large bundles. The details of the selection methods are described in the following sections.

\section{Muon Bundle Multiplicity Spectrum}\label{sec:bundles}

\subsection{Principle}\label{sec:emult}

The altitude of air shower development, and with it the fraction of primary energy going to muons, decreases as a function of parent energy $E_{\rm{prim}}$, but increases with the nuclear mass $A$. The relation between the energy of the cosmic ray primary and the number of muons above a given energy $E_{\rm{\mu,min}}$ is therefore not linear. A good approximation is given by the ``Elbert formula'':

\begin{equation}\label{elbert}
\small
N_{\mu}(E>E_{\rm{\mu,min}})=A\cdot\frac{E_{0}}{E_{\rm{\mu,min}}\cos\textrm{ }\theta}\cdot\left(\frac{E_{\rm{prim}}}{AE_{\mu}}\right)^{\alpha}\cdot\left(1-\frac{AE_{\mu}}{E_{\rm{prim}}}\right)^{\beta},
\end{equation}
where $\cos\textrm{ }\theta$ is the incident angle of the primary particle, and $\alpha$, $\beta$ and $E_{0}$ are empirical parameters that need to be determined by a numerical simulation \cite{tomsbook}. The index $\beta$ describes the cutoff near the production threshold, and $E_{\rm{0}}$ is a proportionality factor applicable to the number of muons at the surface. In this analysis, only the parameter $\alpha$, describing the increase of muon multiplicity as a function of primary energy and mass, is relevant. For energies not too close to the production threshold $E_{\rm{prim}}/A$, the relation can be simplified to:

\begin{equation}\label{elbsimp}
N_{\mu}\propto A^{1-\alpha}\cdot E_{\rm{prim}}^{\alpha}
\end{equation}

For deep underground detectors, $E_{\rm{\mu,min}}$ corresponds to the threshold energy for muons penetrating the surrounding material. In the case of IceCube, this corresponds to about 400 GeV for vertical showers, increasing exponentially as a function of $\sec\theta_{\rm{zen}}$. 

%In the case of IceCube, Eq. \ref{elbsimp} is valid to about 20 percent at primary energies of 1 PeV, and rapidly converging with the exact solution as the energy increases.

Equation \ref{elbert} implies that the distribution of muon energies within a shower is independent of type and energy of the primary nucleus, except at the very highest end of the spectrum, as demonstrated in Fig. \ref{fig-emupdf}. The total energy of the muon bundle, as well as its energy loss per unit track length, is therefore in good approximation simply proportional to the muon multiplicity. After excluding rare events where the muon energy deposition is dominated by exceptional catastrophic losses, the muon multiplicity can therefore be measured simply from the total energy deposited in the detector. 

The experimental data can be directly related to any flux model expressed in terms of the parameter $E_{\rm{mult}}$ introduced in Sec. \ref{sec:prim-muchar}, as long as the measured number of muons remains proportional to the overall multiplicity in the air shower. In the case of IceCube, the corresponding threshold for iron nuclei lies at about 1 PeV. For lower primary energies, Equation \ref{emultdef} is not applicable, and the multiplicity distribution can only be used for model testing, as in Sec. \ref{sec:lolev-result}.

\subsection{Event Selection}

High-multiplicity bundles account for the dominant part of bright events in IceCube. The goal of quality selection is therefore not the isolation of a rare ``signal'', but the reduction of tails and improvement in resolution. The criteria for the high-multiplicity bundle sample are shown in Table \ref{mult_cut_table}.

\begin{table*}[ht!]
%  \footnotesize
  \begin{center}
    \begin{tabular}{|c|c|c|c|c|}
      \hline
      Selection  & Events  ($\times10^{6}$) & Rate [$s^{-1}$] & Comment & Effect\\
      \hline
      All $Q_{\rm{tot}}>1,000$ p.e. & 29.10 & 1.075 & Base Sample (79-String Configuration) & n/a\\
      $\cos\textrm{ }\theta_{\rm{zen}}>0.3$ & 28.54 & 1.054 & Track Zenith Angle & low $N_{\mu}$ \\
      $L_{\rm{dir}}>$600 m & 24.09 & 0.890 & Track Length & high $N_{\mu}$\\
      $q_{\rm{max}}/Q_{\rm{tot}}<0.3$ & 20.66 & 0.763 & Brightness dominated by single DOM & low $N_{\mu}$ \\
      $d_{\rm{mpe,cod}}<$ 425 m & 18.22 & 0.673 & Closest approach to center of detector &  high $N_{\mu}$\\
      $dE/dx$ peak/median $< 8$ & 12.34 & 0.456 & See \ref{sec:ddddr} & low $N_{\mu}$ \\
      \hline
    \end{tabular}	
    \caption{Selection criteria and passing rates for muon multiplicity measurement. The effect of each parameter corresponds to a reduction at either low or high end of the distribution shown in Fig. \ref{fig-bundlecuteffect}.}
    \label{mult_cut_table}
  \end{center}
\end{table*}

\begin{figure}[t]
\centering
\includegraphics[width=220pt]{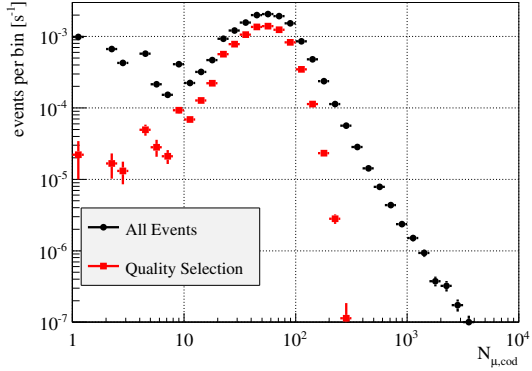}
\caption{Muon bundle multiplicity at closest approach to the center of the detector ({\it cod}) for simulated events with 3,000 to 4,000 registered photo-electrons. Distributions are shown for trigger level and final high-multiplicity bundle selection.}
\label{fig-bundlecuteffect}
\end{figure}

\begin{figure}[ht!]
\centering
\includegraphics[width=220pt]{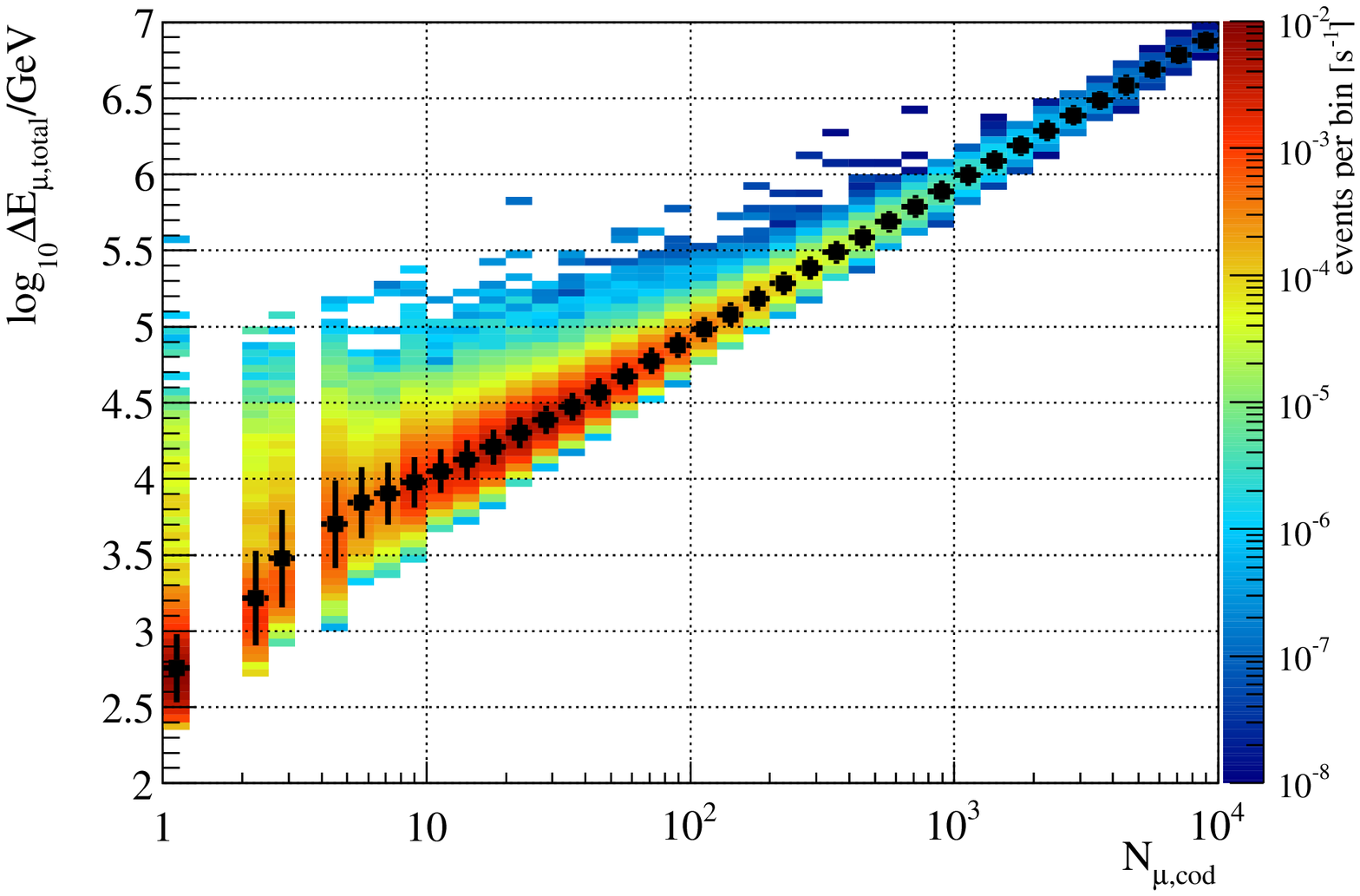}
\includegraphics[width=220pt]{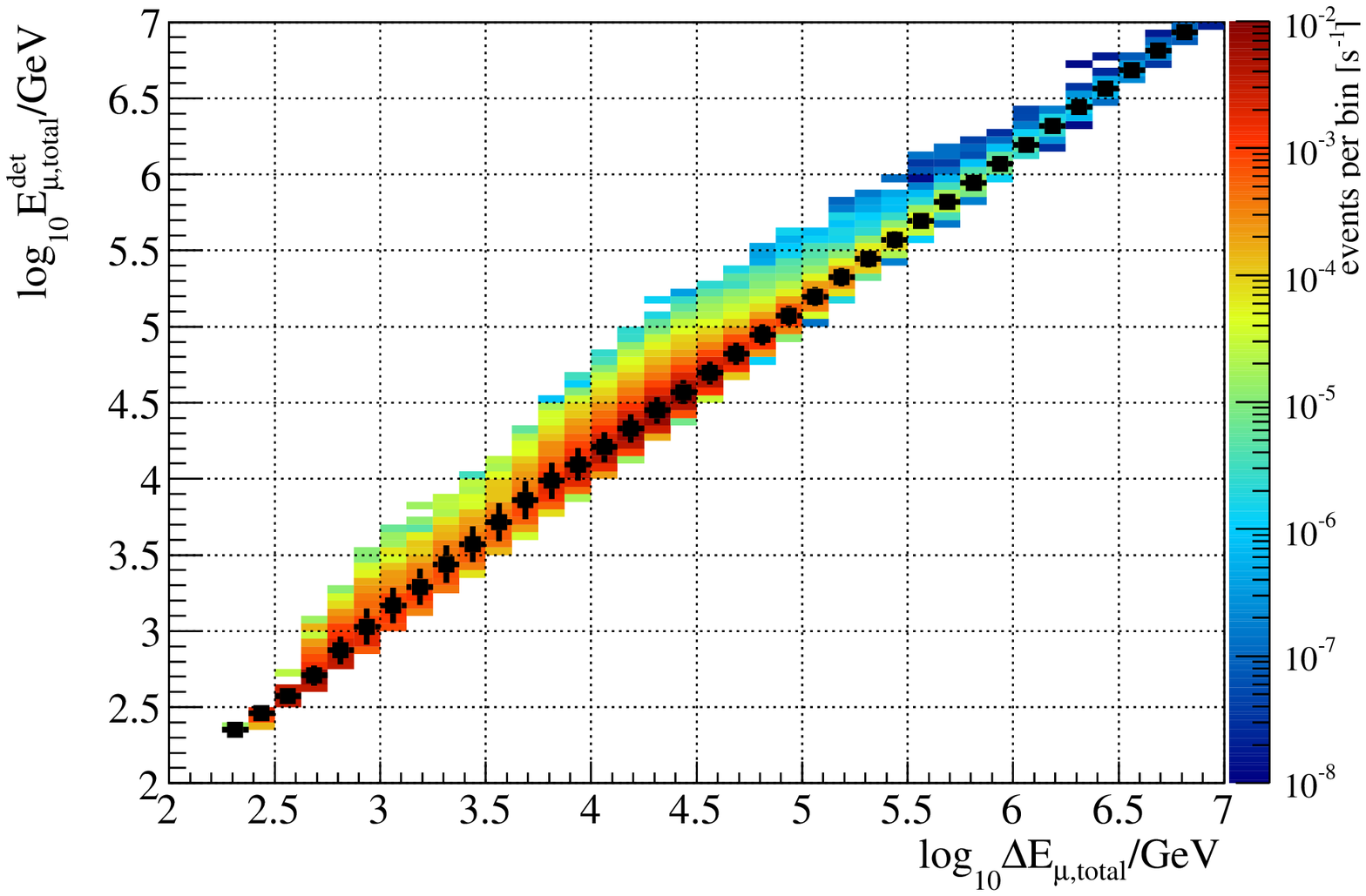}
\caption{Top: Relation between number of muons at closest approach to the center of the detector and total energy loss of muon bundle within detector volume. Bottom: Total muon energy loss vs. sum of muon energies at entry into detector volume. Data samples correspond to CORSIKA simulation after application of bundle selection quality criteria. The black curve represents a profile of the colored histogram. The error bars indicate the spread of the value.}
\label{fig-bundleparamdep}
\end{figure}

\begin{figure}[ht!]
\centering
\includegraphics[width=220pt]{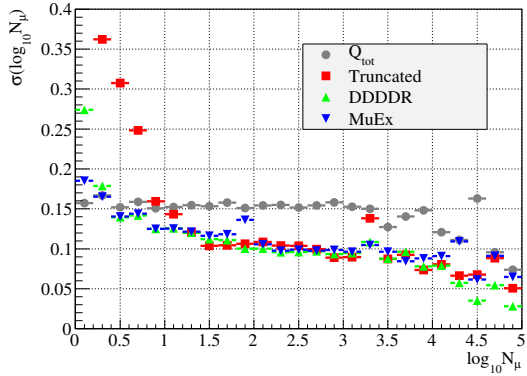}
\caption{Resolution of muon multiplicity estimators based on four different energy reconstructions. The analysis threshold of 1,000 photo-electrons corresponds to 20-30 muons.}
\label{fig-nest-res}
\end{figure}

\begin{figure}[ht!]
  \centering
  \includegraphics[width=220pt]{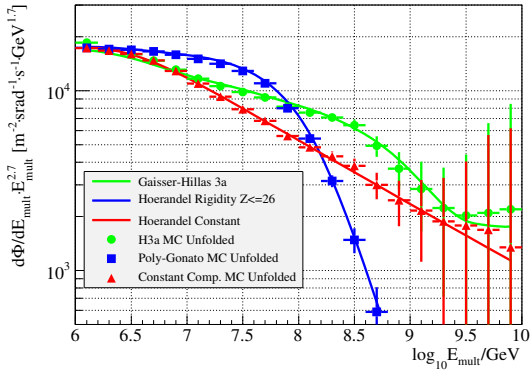}
  \caption{Unfolded spectra of simulated data compared to analytic form of spectra for three benchmark models \cite{Gaisser:2013bla, Hoerandel:2002yg}. The size of the error bars corresponds to the expected statistical uncertainty for one year of IceCube data.}
  \label{fig-fullcircle-mult}
\end{figure}

\begin{figure}[ht!]
\centering
\includegraphics[width=220pt]{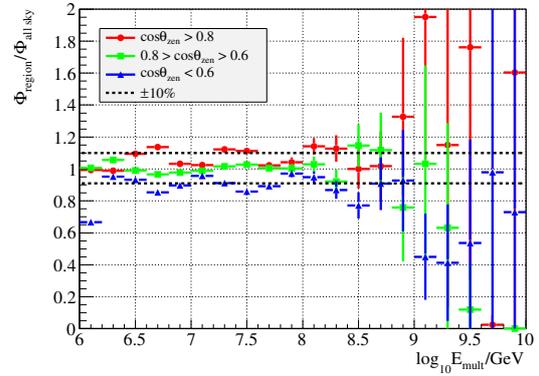}
\caption{Ratio of multiplicity spectrum unfolded separately for three zenith angle regions to all-sky result.}
\label{fig-emult-angerr}
\end{figure}

Figure \ref{fig-bundlecuteffect} shows the true simulation-derived number of muons at closest approach to the center of the detector for events with a fixed total number of registered photo-electrons. On the right hand side of the distribution, the selection criteria eliminate very energetic tracks that pass through an edge or just outside the detector. On the left, the tail of low-multiplicity tracks containing high-energy muons, which are bright mainly because of exceptional catastrophic losses, is reduced.

%The distance from the center of the detctor, here assumed to coincide with the origin of the IceCube coordinate system at $(0,0,0)$, is described by the equation:

%\begin{equation}
%\label{dtrkcod}
%d_{track,cod}=\sqrt{(z_{0}+\rho_{trk}\cdot cos\theta_{zen})^{2}+(y_{0}+\rho_{trk}\cdot sin\theta_{zen}\cdot sin\phi_{az})^{2}+(x_{0}+\rho_{trk}\cdot sin\theta_{zen}\cdot cos\phi_{az})^{2}}
%\end{equation}

%\begin{equation}
%\label{rhotrk}
%\rho_{trk} = -(sin\theta_{zen}\cdot cos\phi_{az}\cdot x_{0}+sin\theta_{zen}\cdot sin\phi_{az} \cdot y_{0}+cos\theta_{zen}\cdot z_{0})
%\end{equation}

\subsection{Derivation of Experimental Measurement}

The relation between the scaled parameter $E_{\rm{mult}}$ and the actual muon multiplicity in a specific detector $N_{\rm{\mu,det}}$ can be expressed as

\begin{equation}\label{emultrel}
E_{\rm{mult}}=g_{\rm{scale}}(\cos\textrm{ }\theta)\cdot N_{\rm{\mu,det}}^{1/\alpha},
\end{equation}
where $g_{\rm{scale}}(\cos\textrm{ }\theta)$ is a simulation-derived function accounting for angular dependence of muon production and absorption in the surrounding material. The effects of local atmospheric conditions and selection efficiency are accounted for by a separate acceptance correction term.

For the experimental measurement of the parameter $E_{\rm{mult}}$, it is first necessary to derive expressions for the terms on the right hand side of of Eq. \ref{emultrel}. The resulting parameter can then be related to the analytical form of the bundle multiplicity spectrum by spectral unfolding.

A numerical value of $0.79\pm0.02$ for the parameter $\alpha$ was determined by fitting a power law function to the relation between primary energy and muon multiplicity. The difference to the original description \cite{tomsbook}, which gives a somewhat surprisingly accurate estimate of 0.757, is likely a consequence of advances in the understanding of air shower physics during the last three decades. Recent calculations finding a lower value for $\alpha$ are only applicable in the small region of phase space of $A\cdot E_{\mu}/E_{prim} > 0.1$, where energy threshold effects become dominant \cite{Gaisser:2014bja}.

In the analysis sample, the energy loss of muons in the detector is in good approximation proportional to the number of muons $N_{\mu}$, and to the total energy of muons contained in the bundle, as illustrated in Fig. \ref{fig-bundleparamdep}. An experimental observable corresponding to the muon multiplicity can therefore be constructed through a parametrization of the detector response based on a Monte-Carlo simulation, in the simplest case using the proportionality between energy deposition and total amount of registered photo-electrons described in Sec. \ref{sec:lolev-obs}. To reduce biases and take advantage of the opportunity to investigate systematic effects, the procedure was performed for four different muon energy estimators. These are:

\begin{figure*}[ht!]
\centering
\includegraphics[width=220pt]{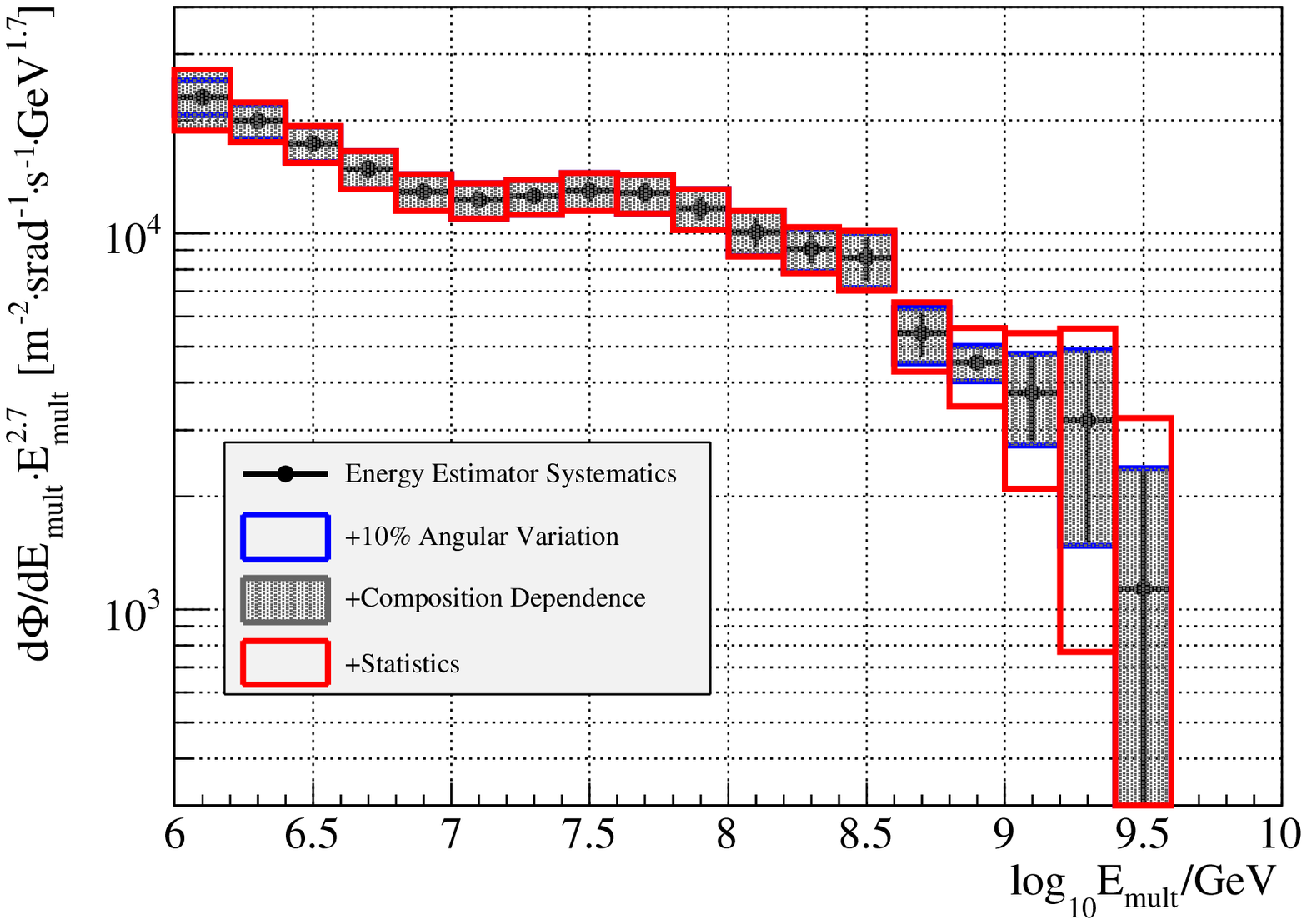}
\includegraphics[width=220pt]{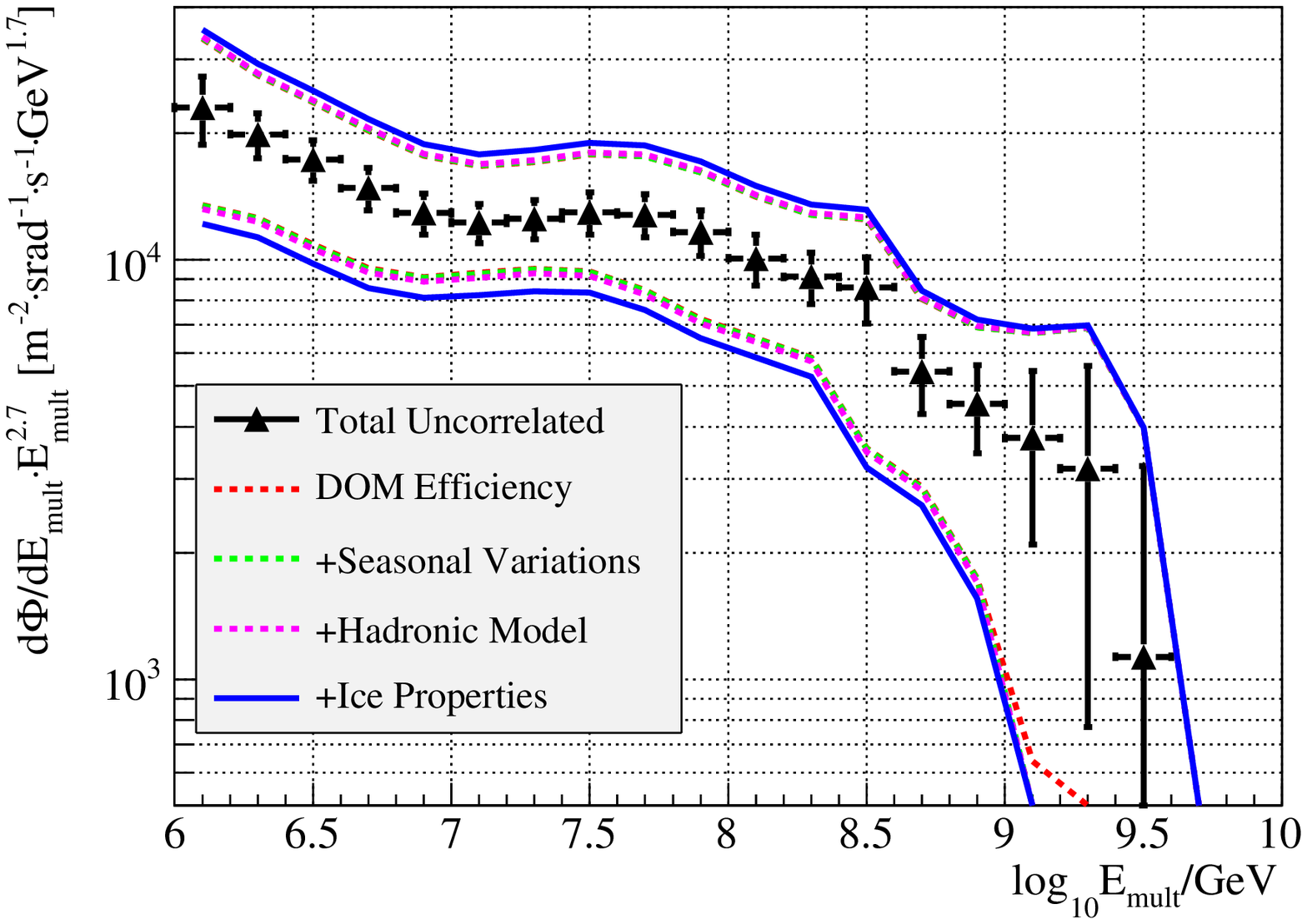}
\caption{Unfolded and acceptance-corrected experimental spectrum of rescaled muon bundle multiplcity parameter $E_{\rm{mult}}$. The influence of systematic uncertainties listed in Table \ref{mult_syst_table} is shown separately for bin-wise fluctuations (\textit{uncorrelated}, left) and overall scaling (\textit{correlated}, right).}
\label{fig-emult-errors}
\end{figure*}

\begin{table*}[ht!]
%\tiny
%\scriptsize
\footnotesize
%\small 
  \begin{center}
    \begin{tabular}{|c|c|c|c|c|}
      \hline
      Source & Type & Variation & Effect & Comment \\
      \hline
      Composition & uncorrelated & Fe, protons & variable & Residual bias near threshold \\
      Energy Estimator & uncorrelated & 4 discrete values & variable & Derived from data \\
      Angular Acceptance & uncorrelated & 3 zenith regions & $\pm 10\%$ Flux Scaling & Estimated from data \\
%      \hline
      Light Yield & correlated & $\pm 10\%$ & $\pm 13\%$ Energy Shift & Composite Scalar Factor \\
      Ice Optical & correlated & 10\% Scattering, Absorption & $ \pm 25\% $ Flux Scaling & Global variations around default model \\
      Hadronic Model & correlated & discrete & $\pm 10\%$ Flux Scaling & EPOS/QGSJET/SIBYLL \\
      Seasonal Variations & correlated & Summer vs. Winter & $\pm 5\%$ Flux Scaling & Estimated from data \\
      Muon Energy Loss & correlated & Theoretical uncertainty \cite{Koehne:2013gpa} & $\pm 1\%$ & Official IceCube Value \\
      \hline
    \end{tabular}	
    \caption{Summary of systematic uncertainties affecting the result of the bundle multiplicity spectrum measurement.}
    \label{mult_syst_table}
  \end{center}
\end{table*}

\begin{figure*}[ht!]
\centering
\includegraphics[width=220pt]{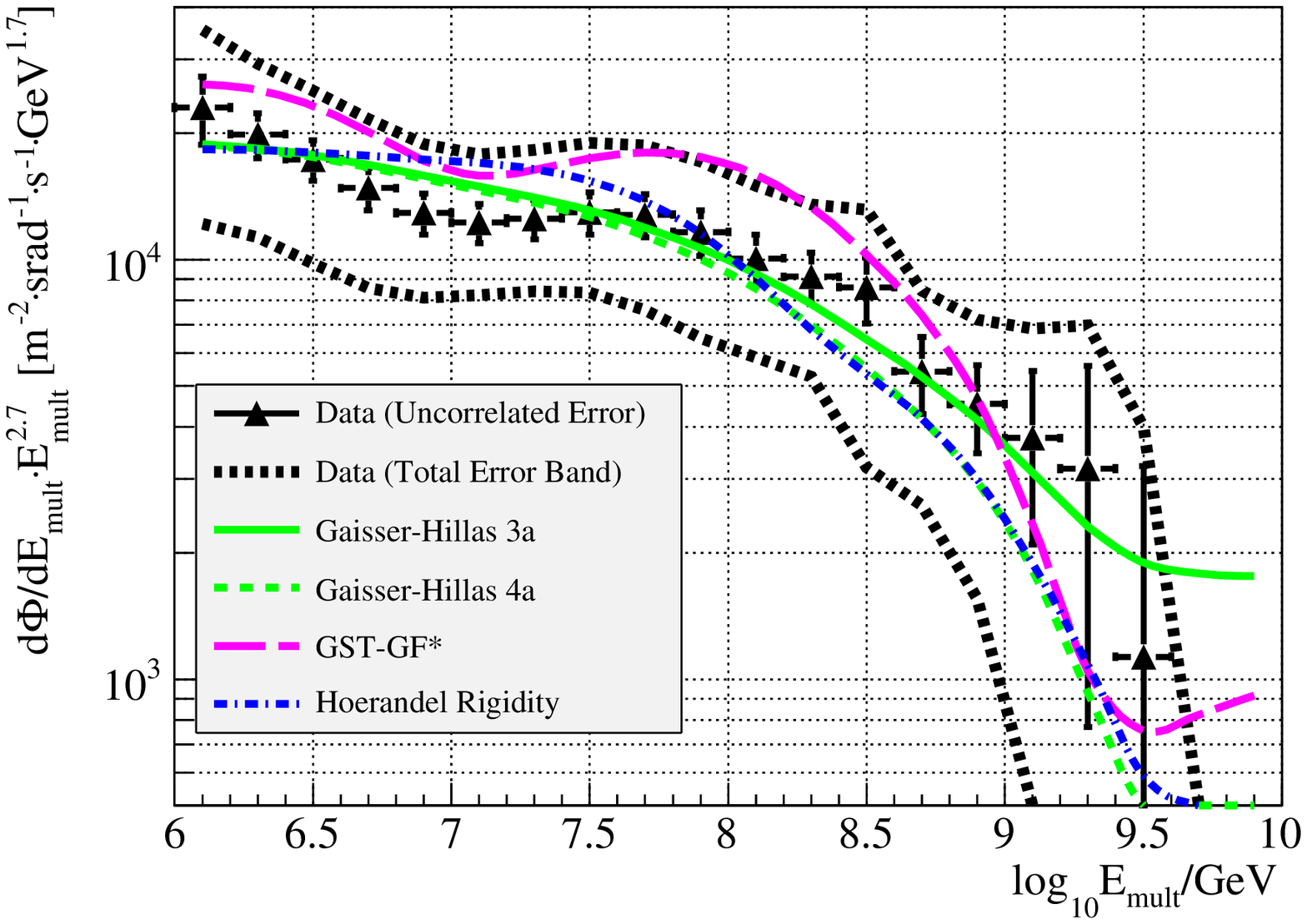}
\includegraphics[width=220pt]{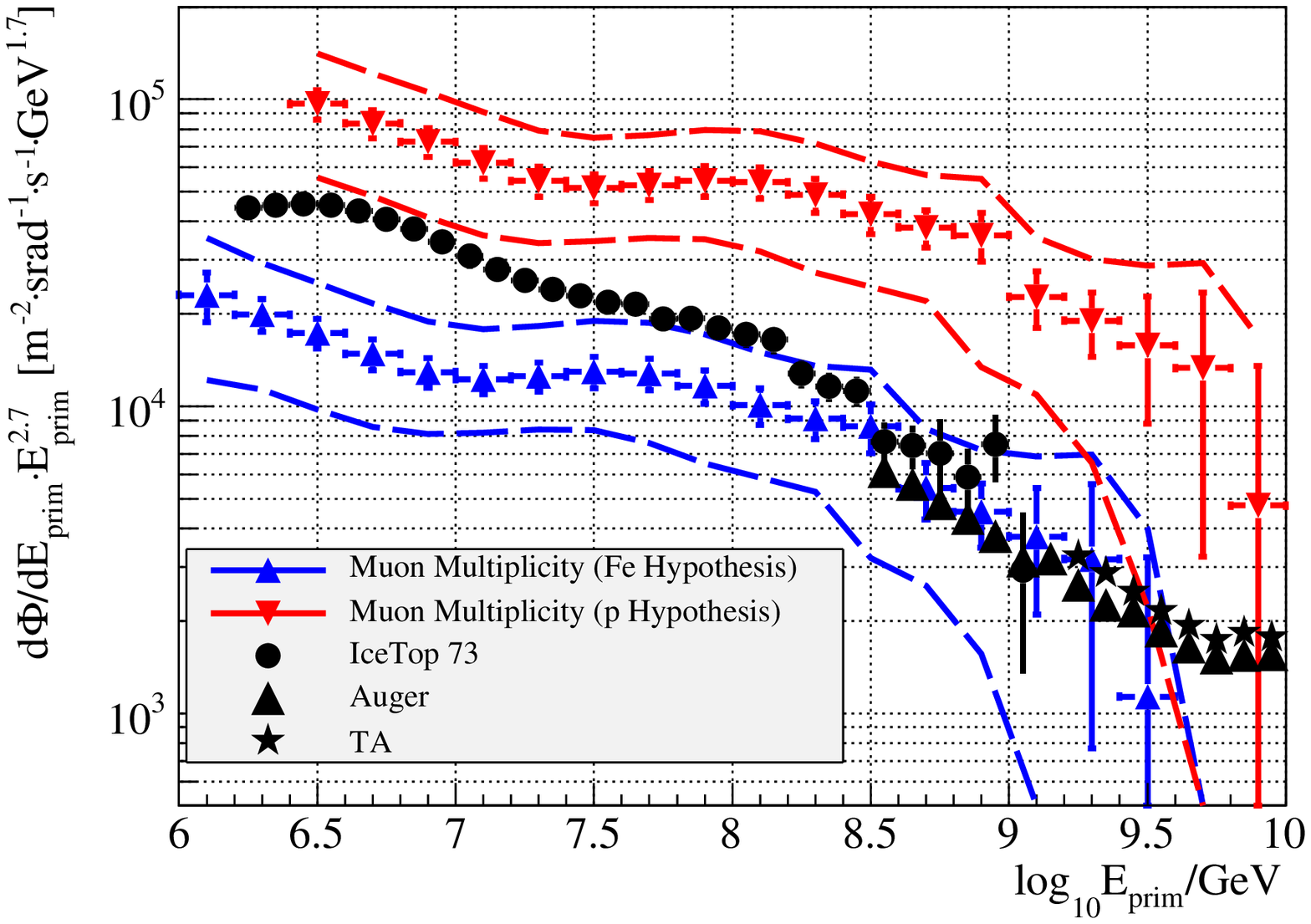}
\caption{Interpretation of muon multiplicity spectrum by comparison to specific cosmic ray models \cite{Gaisser:2013bla, Hoerandel:2002yg} (left), and by relation to all-particle flux measurements from IceCube \cite{Aartsen:2013wda} and other detectors \cite{ThePierreAuger:2013eja, Array:2013dra} (right). Note that an exact translation to average logarithmic mass is not possible.}
\label{fig-emult-interp}
\end{figure*}

\begin{itemize}
\item The total event charge $Q_{\rm{tot}}$, measured in photo-electrons. Charge registered by DeepCore was excluded to avoid biases due to closer DOM spacing and higher PMT efficiency in the sub-array.

\item The truncated mean of the muon energy loss \cite{Abbasi:2012wht}.

\item The mean energy deposition calculated with the DDDDR method described in \ref{sec:ddddr}.

\item The likelihood-based energy estimator \textit{MuEx} \cite{Aartsen:2013vja}.

\end{itemize}

The resolution of the multiplicity proxies in dependence of the true number of muons at closest approach to the center of the detector is shown in Figure \ref {fig-nest-res}. Except for the raw total number of photo-electrons, all estimators perform in a remarkably similar way in simulation. The presence of individual outliers illustrates the motivation to use more than one method to ensure stability of the result. The angular-dependent scaling function $g_{\rm{scale}}(\cos\textrm{ }\theta_{\rm{zen}})$ was parametrized based on simulated data.

%The multiplicity estimator can now be used to construct a proxy for the Multiplicity-Derived Energy Estimator $E_{mult}$:

%\begin{equation}
%\label{emultparam}
%\tilde{E}_{mult} = N_{\mu,est}^{1/\alpha_{elbert}}\cdot g_{scale}(\cos\textrm{ }\theta_{zen})
%\end{equation}

%\begin{equation}
%\label{emultparam_angscale}
%g_{scale}(cos\theta_{zen}) = 10^{3.93+0.49\cdot \cos\\textrm{ }theta_{zen}+\exp(1.082-3.0\cdot \cos \theta_{zen})}\cdot GeV
%\end{equation}

Using the RooUnfold algorithm \cite{Adye:2011gm}, a spectral unfolding was applied to the measured distribution of $E_{mult}$. The differential flux was then related to the unfolded and histogrammed experimental data as:

\begin{equation}
\label{emultparam_difflux}
\frac{d\Phi}{dE_{\rm{mult}}}=c(\Delta E_{\rm{bin}},t_{\rm{exp}})\cdot\eta(E_{\rm{mult}})\cdot\frac{\Delta N_{\rm{ev}}}{\Delta \log_{\rm{10}}E_{\rm{mult}}}
\end{equation}

Here the proportionality constant \textit{c} accounts for the effective livetime of the data sample and the bin size of the histogram. The detector acceptance $\eta(E_{\rm{mult}})$, whose exact form depends on atmospheric conditions, needs to be derived from simulation. 

%Here, it takes the functional form:

%\begin{equation}
%\label{acceptfunc}
%\eta(x) = p_{0}+\frac{1}{p_{1}\cdot(x-e_{0})+p_{2}\cdot(x-e_{0})^{2}+p_{3}\cdot(x-e_{0})^{3}}
%\end{equation}

%The exact numerical values of the parameters depend on the atmospheric conditions, as shown in Table \ref{emultpar_table}. Note that July (austral winter) and December (austral summer) atmospheres were only used to estimate higher-order systematic variations due to inaccurately modeled seasonal effects.

%\begin{table}[!h]
%  \begin{center}
%    \begin{tabular}{|c|c|c|c|c|c|}
%      \hline
%      Data Set & $e_{0}$ & $p_{0}$ & $p_{1}$ & $p_{2}$ & $p_{3}$ \\
%      \hline
%      All Seasons & 5.127 & 0.928 & 0.909 & -2.105 & 1.292 \\
%      July Atmosphere & 5.678 & 0.941 & 0.232 & -0.927 & 1.738 \\
%      December Atmosphere & 5.152 & 0.946 & 0.916 & -2.227 & 1.452 \\
%      \hline
%    \end{tabular}	
%    \caption{$E_{mult}$ Parametrization.}
%    \label{emultpar_table}
%  \end{center}
%\end{table}

The approach can be verified by a full-circle test, as shown in Fig. \ref{fig-fullcircle-mult}. Each of the benchmark models, chosen to reflect extreme assumptions about the behavior of the cosmic ray flux, can be reproduced by applying the analysis procedure to simulated data.

\subsection{Result}\label{sec:multresult}

Systematic uncertainties applying to the experimental measurement are summarized in Table \ref{mult_syst_table}. The categorization by type corresponds to bin-wise fluctuations (\textit{uncorrelated}) and overall scaling effects (\textit{correlated}). 

Of special interest is the angular variation, which dominates the total bin-wise uncertainty over a wide range. The effect is illustrated in Fig. \ref{fig-emult-angerr}. Splitting the data according to the reconstructed zenith angle into three separate event samples results in spectra that are similar in shape, but whose absolute normalization varies within a band of approximately $\pm 10\%$. As the difference appears to be not uniform, it has been conservatively assumed to lead to uncorrelated bin-wise variations in the all-sky spectrum. Notwithstanding, magnitude and direction are similar to the unexplained effect described in Section \ref{sec:le-muons}, suggesting a possible common underlying cause. The final result, after successive addition of systematic error bands in quadrature, is shown in Fig. \ref{fig-emult-errors}. 

Since the muon multiplicity is not a fundamental parameter of the cosmic ray flux, it is important to find an appropriate way for its interpretation. Two possibilities are illustrated in Fig. \ref{fig-emult-interp}. The first is by expressing cosmic ray flux models in terms of $E_{\rm{mult}}$ through application of Eq. \ref{emultdef}. Experimental result and prediction can then be directly related. The second is to translate the multiplicity distribution to an energy spectrum under a particular hypothesis for the elemental composition. By default, the scaling of $E_{\rm{mult}}$ corresponds to iron, but changing it to any other primary nucleus type is straightforward. 

\begin{figure}[b!]
  \centering
  \includegraphics[width=220pt]{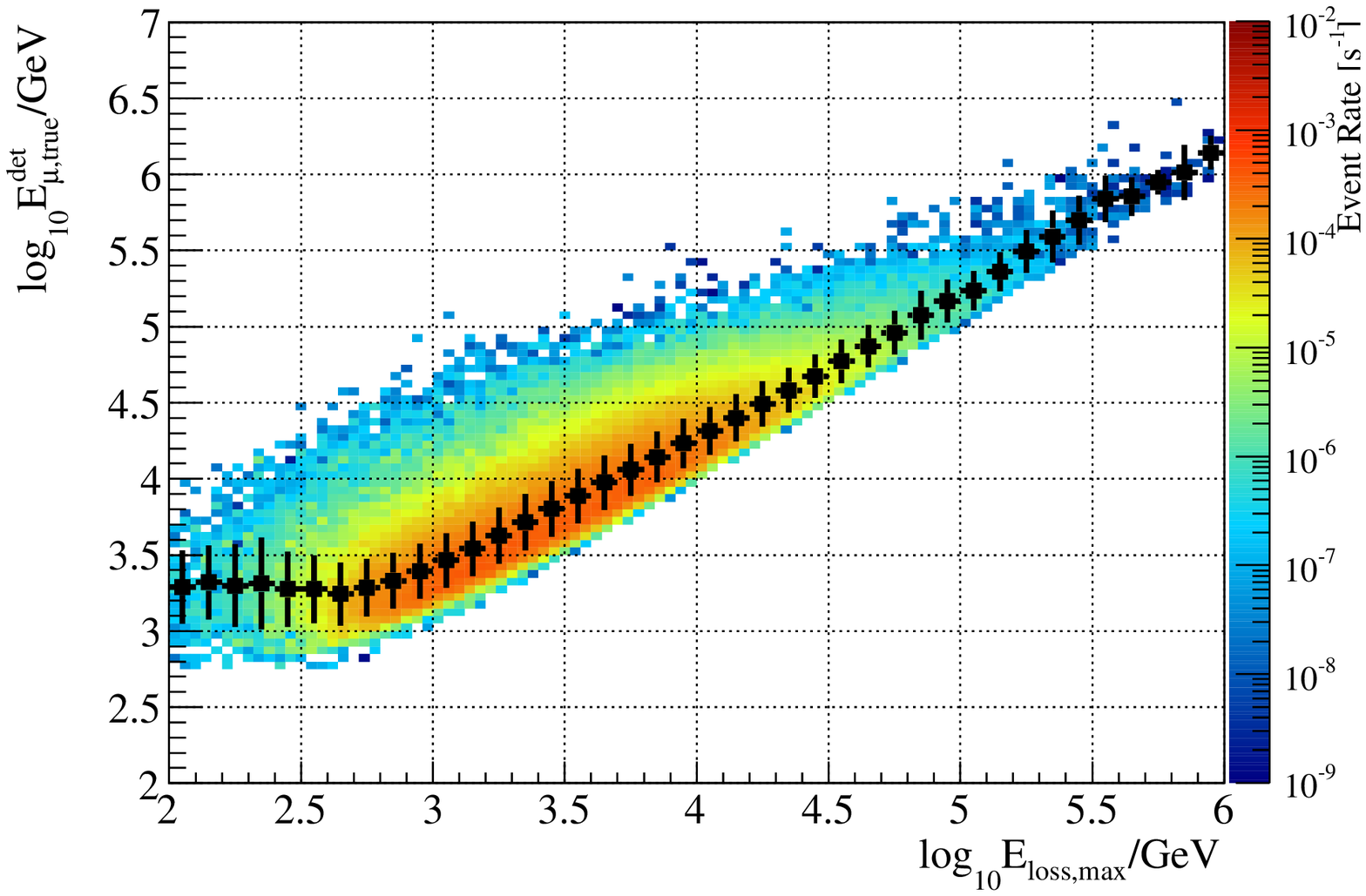}
  \includegraphics[width=220pt]{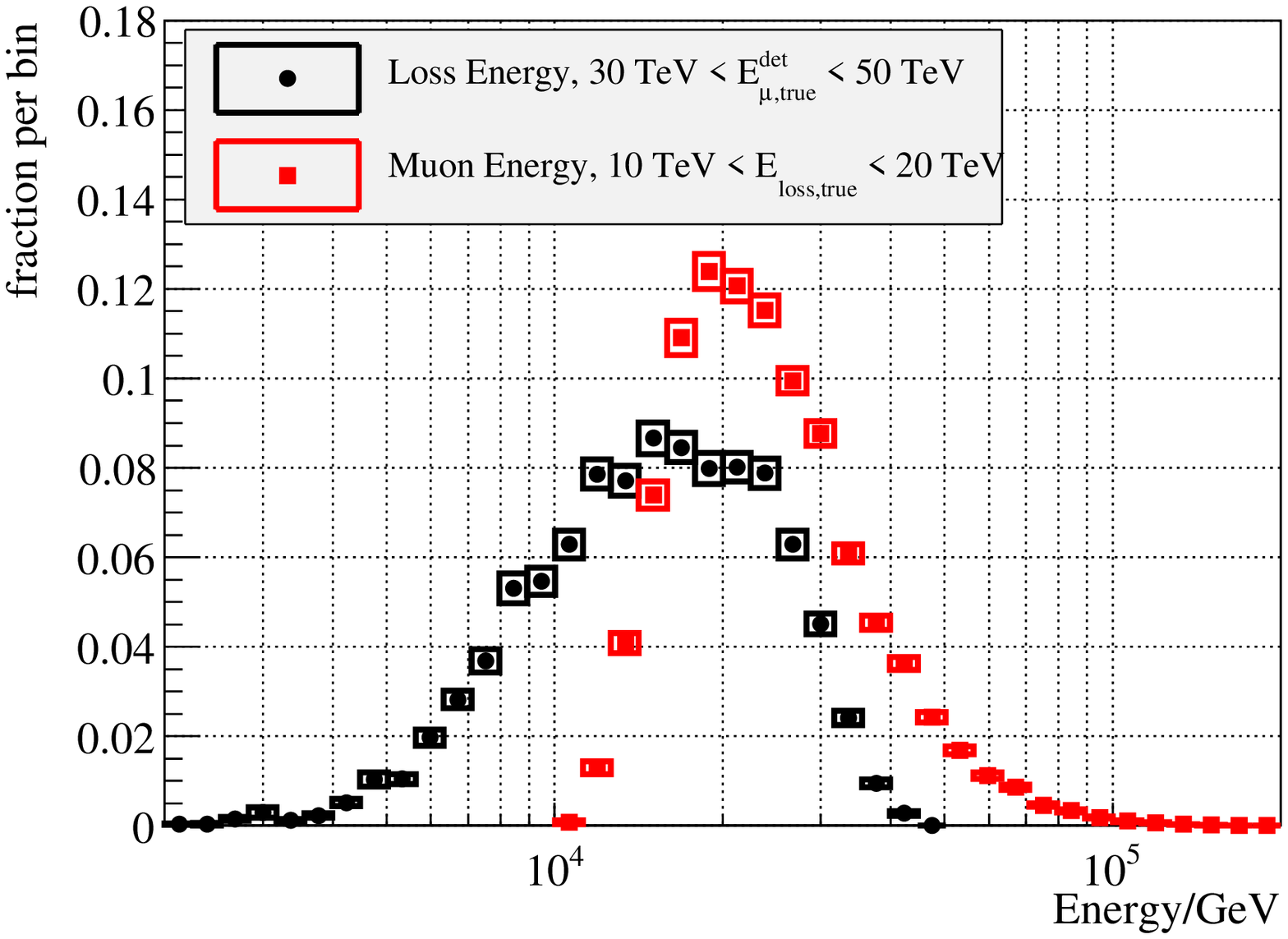}
  \includegraphics[width=220pt]{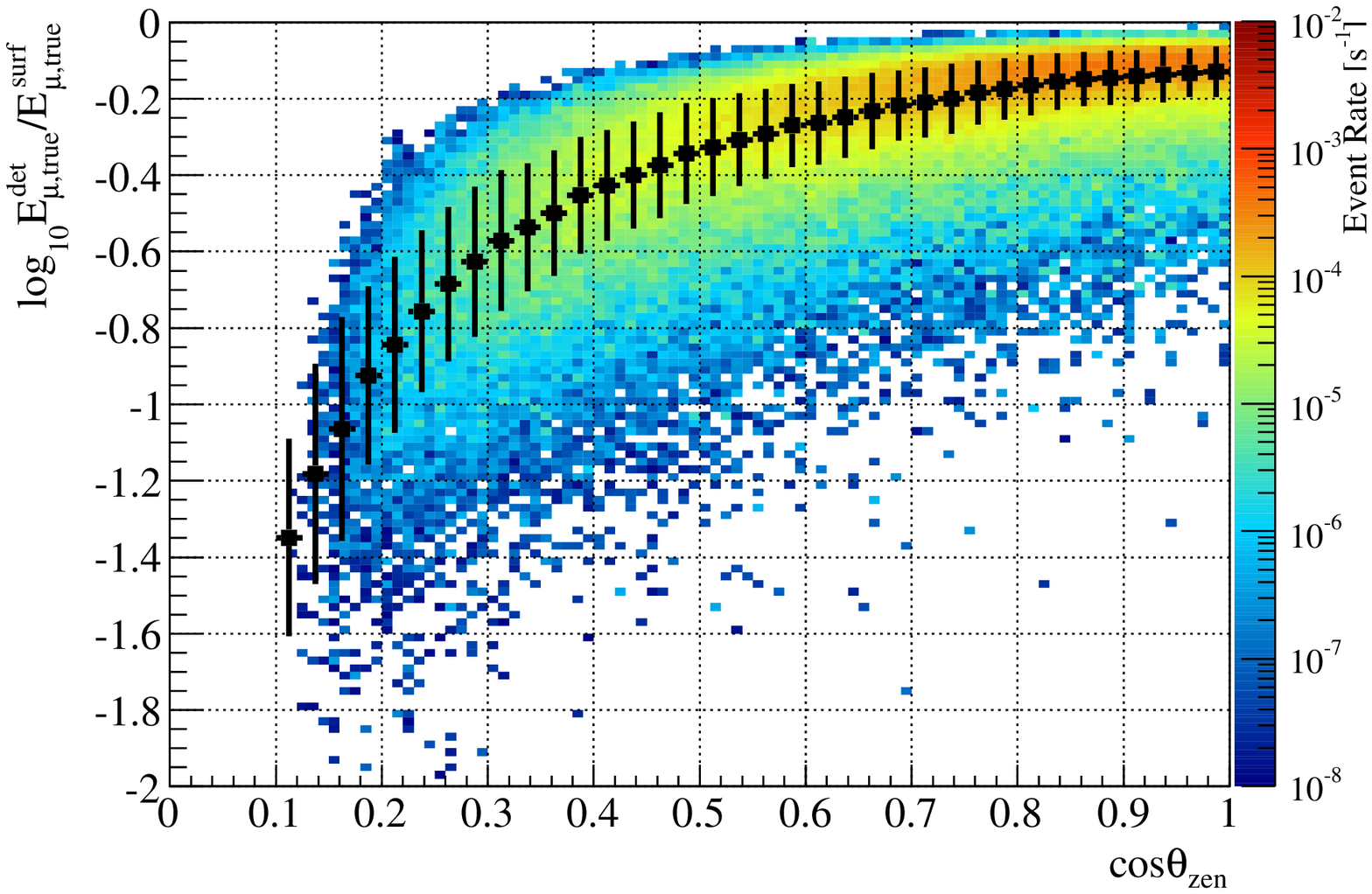}
  \caption{Top: Relation between most energetic single energy loss and leading muon energy within the IceCube detector volume. Middle: Distribution of true energy parameters for two slices in top histogram. Bottom: Fraction of muon surface energy remaining at point of entry into detector volume in dependence of zenith angle. Figures are based on simulated events with primary flux weighted to $E^{-2.7}$ power law spectrum and correspond to final analysis sample before application of minimum shower energy criterion. The black curves represent mean and spread of the distribution.}
  \label{fig-hemu_enrel}
\end{figure}

The result can then be overlayed by independent cosmic ray flux measurements. An unambiguous derivation of the average mass as a function of the primary energy is not possible due to the degeneracy between mass and energy in the multiplicity measurement. However, the qualitative variation of composition with energy is consistent with a gradual change towards heavier elements in the range between the knee and 100 PeV. If the current description of muon production in air showers is correct, and the external measurements are reliable, a purely protonic flux would be strongly disfavored up until the ankle region.

\section{High-Energy Muons}\label{sec:hemu}

\subsection{Principle}

%Events in which the total bundle energy is dominated by a single muon are relatively rare. Consequently methods must be developed to reliably identify these events and to estimate the energy of the leading particle. Here this is accomplished by making use of the stochastic nature of muon energy losses.

The presence of a single exceptionally strong catastrophic loss can be used both for tagging high-energy muons and to estimate their energy. The first part is obvious: An individual particle shower along a track can only have been caused by a parent of the same energy or above. Simulated data indicate that instances in which two or more muons in the same bundle suffer a catastrophic loss simultaneously in a way that is indistiguishable in the energy reconstruction are exceedingly rare.

The quantification is based on the close relation between the energy of the catastrophic loss used to identify the event and that of the leading muon, a consequence of the steeply falling spectrum. Once the muon energy at the point of entry into the detector volume has been determined, the most likely energy at the surface of the ice can be estimated by taking into account the zenith angle, as illustrated in Fig. \ref{fig-hemu_enrel}. This method was developed specifically for the purpose of measuring the energy spectrum of atmospheric muons. As shown in Fig. \ref{fig-evsamp-trupar}, the leading particle typically only accounts for a limited fraction of the total event energy, and the application of energy measurement techniques optimized for single neutrino-induced muon tracks could lead to substantial biases in the case of a large accompanying bundle.

Higher-order corrections are necessary to account for correlations and the effect of variations in the distance to the surface due to the vertical extension of the detector. All relations in this study were based on parametrizations using simulated events. A full multi-dimensional unfolding would be preferable, but requires a substantial increase in simulation statistics.

\subsection {Event Selection}

\begin{figure}[ht!]
  \centering
  \includegraphics[width=220pt]{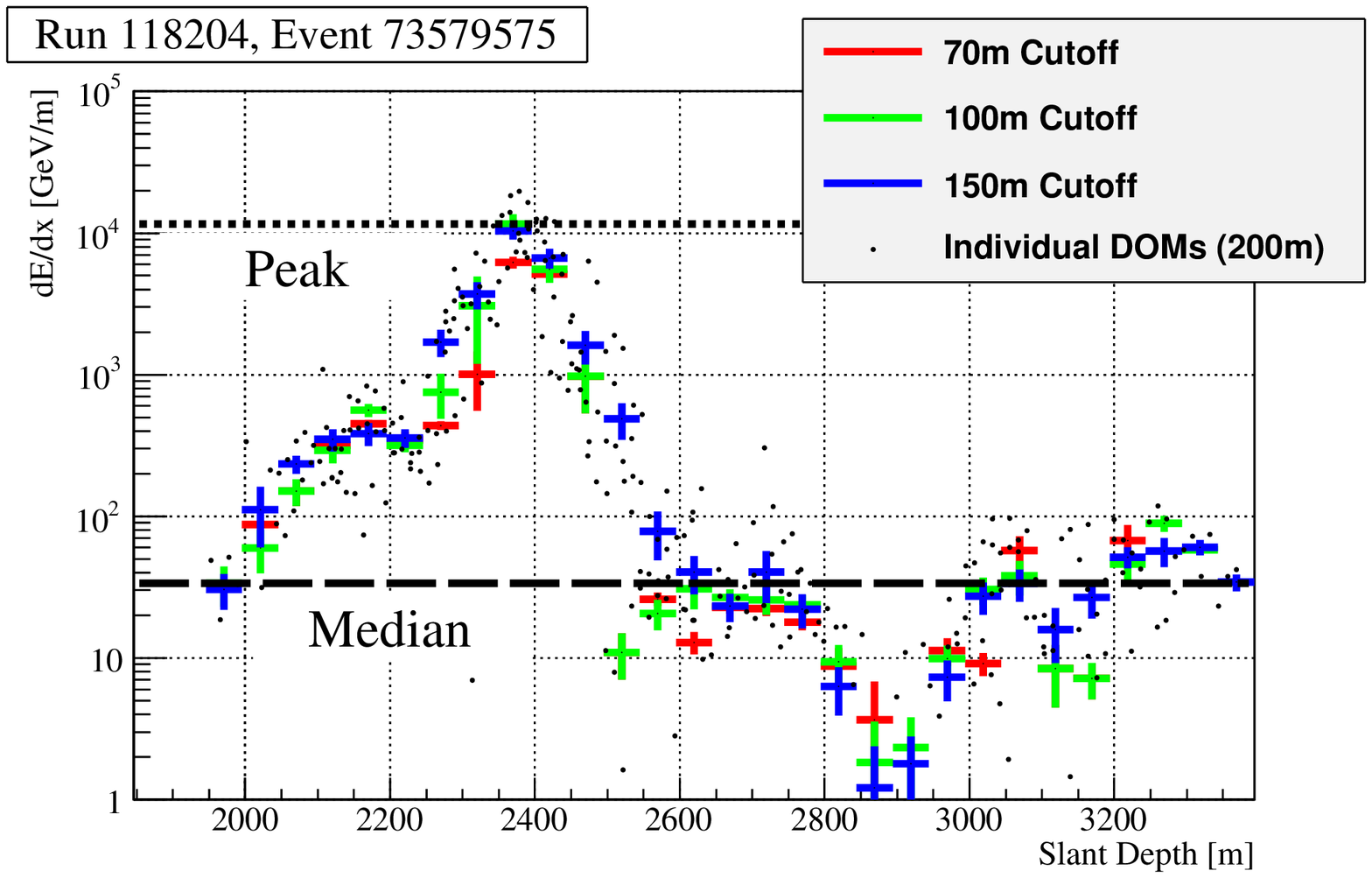}
  \includegraphics[width=180pt]{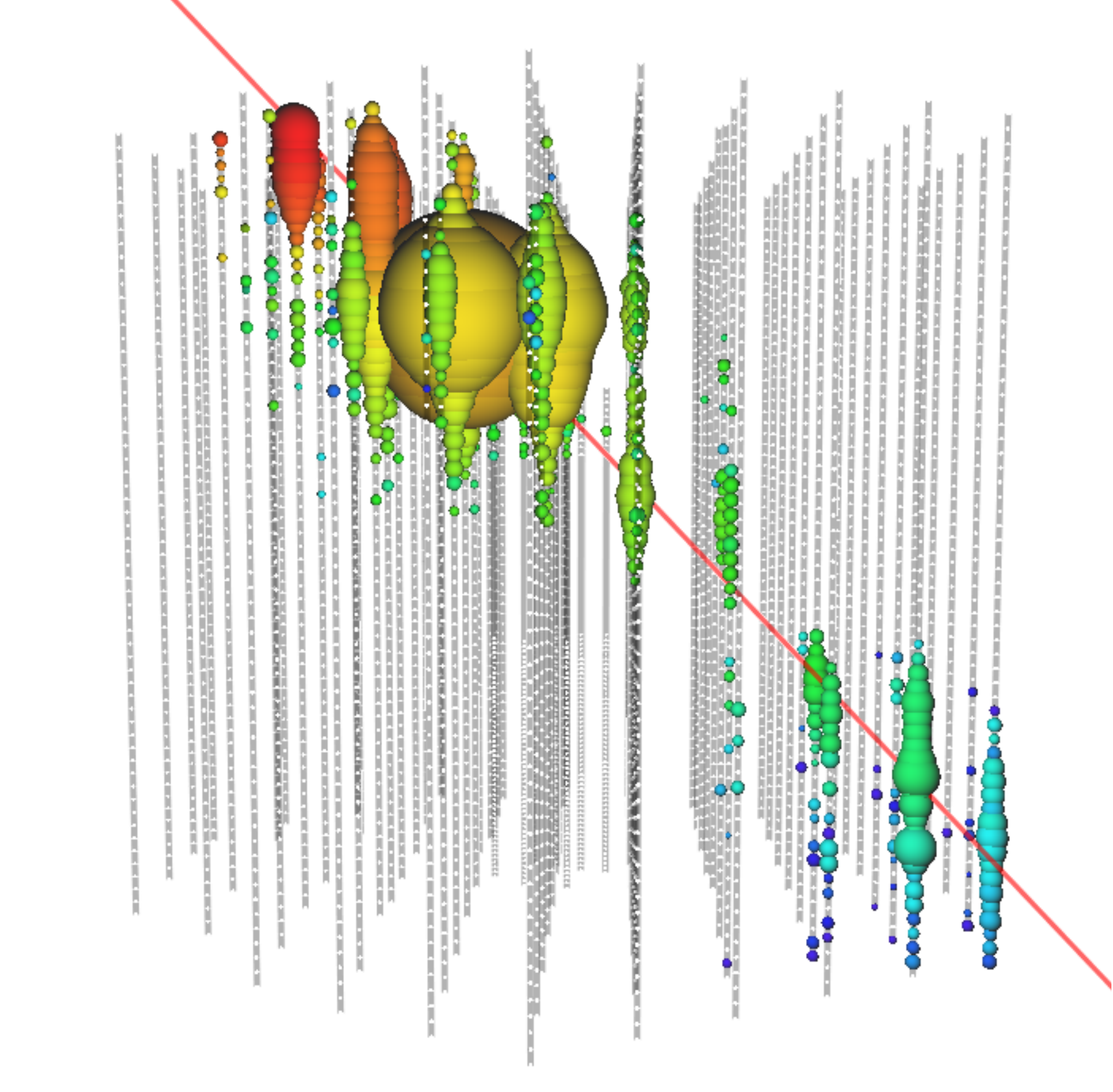}
  \caption{Example for peak to median energy loss ratio in high-energy muon candidate event found in experimental data. Top: Reconstructed differential energy loss in dependence of distance to surface, measured along the reconstructed track. Details of the method are described in \ref{sec:ddddr}. Bottom: Image of the event. The volume of each sphere is proportional to the signal registered by a given DOM. The color scheme corresponds to the arrival time of the first photon (red: earliest, blue: latest). 
Reconstructed event parameters are: $E_{\rm{loss}} = 550^{+220}_{-160} \textrm{ TeV}$, $E_{\rm{\mu,surf}} = 1.03^{+0.62}_{-0.39} \textrm{ PeV}$, $\theta_{\rm{zen}} = 45.1 \pm 0.2^{\circ}$
%Reconstructed event parameters are: $E_{\rm{loss}} = 554 \textrm{ TeV} \pm 0.15 \cdot \log_{\rm{10}}E $, $E_{\rm{\mu,surf}} = 1.034 \textrm{ PeV} \pm 0.20 \cdot \log_{\rm{10}}E $, $\theta_{\rm{zen}} = 45.1 \pm 0.2^{\circ}$
}
  \label{fig-pkmedrat}
\end{figure}

The selection of muon events with exceptional stochastic energy losses is primarily based on reconstructing the differential energy deposition and selecting tracks according to the ratio of peak to median energy loss as illustrated in Fig. \ref{fig-pkmedrat}. All other criteria are ancillary, and are only applied to minimize a possible contribution from misreconstructed tracks. An overview of the selection is given in Table \ref{hemu_cut_table}.

\begin{table*}[!ht]
  \small
  \begin{center}
    \begin{tabular}{|c|c|c|c|}
      \hline
      Quality Level  & Events ($\times10^{6}$) & Rate [$s^{-1}$] & Comment\\
      \hline
      All $Q_{\rm{tot}}>1,000$p.e. & 38.28 & 1.334 & Base Sample (86-String Configuration)\\
      $\cos\textrm{ }\theta_{\rm{zen}} > 0.1$ & 37.99 & 1.324 & Track zenith angle\\
      $q_{\rm{max}}/Q_{\rm{tot}}<0.5$ & 34.46 & 1.201 & Brightness dominated by single DOM\\
      $L_{\rm{dir}}>800\rm{m}$ & 27.55 & 0.960 & Track length in detector\\
      $N_{\rm{DOM, 150m}} > 40$ & 24.71 & 0.861 & Stochastic loss containment \\
      peak/median $dE/dx > 10$ & 2.795 & 0.0974 & \textbf{Exceptional energy loss along track}\\
      median $dE/dx > 0.2 \rm{GeV/m}$ & 2.769 & 0.0965 & Exclude dim tracks\\
      $E_{\rm{casc}} >$ 5 TeV  & 0.769 & 0.0268 & \textit{Exclude threshold region} \\
      \hline
    \end{tabular}	
    \caption{High-energy muon selection criteria and passing rates.}
    \label{hemu_cut_table}
  \end{center}
\end{table*}

A special case is the exclusion of events with a reconstructed shower energy of less than 5 TeV. This requirement was added to reduce uncertainties in the threshold region, which may not be well described by current understanding of systematic detector effects. 
The reason to choose a value of 5 TeV is that a typical electromagnetic shower of that energy will produce a signal of about 1,000 photo-electrons, coinciding with the base sample selection.

\subsection {Energy Estimator Construction}

\begin{figure}[ht!]
\centering
\includegraphics[width=220pt]{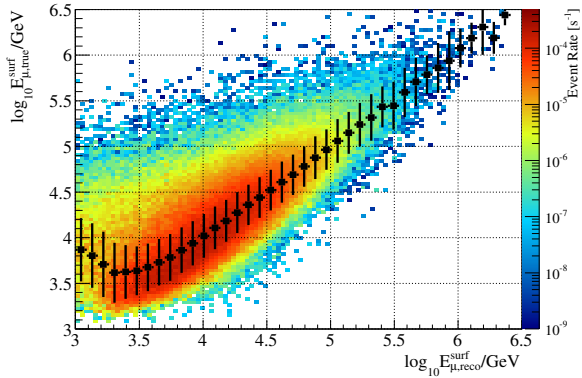}
\caption{Relation between reconstructed and true surface energy for simulated atmospheric muon data before excluding events with reconstructed shower energy of less than 5 TeV. The primary particle flux in the simulation was weighted according to a power law of the form $E^{-2.7}$. Also shown are mean and spread of the distribution.}
\label{fig-hemu_eneres}
\end{figure}

The energy reconstruction is based on the deterministic reconstruction method discussed in \ref{sec:ddddr}, which was designed specifically for this purpose. Subsequently developed likelihood methods \cite{Aartsen:2013vja} were evaluated, but gave no improvement in resolution while introducing a tail of substantially overestimated energies. 

In the first step, the energy $E_{\rm{casc,reco}}$ of the strongest loss (``cascade'') along the track was determined. The exact value is almost identical to the raw reconstructed energy $E_{\rm{casc,raw}}$ from the DDDDR algorithm, except for a minor correction factor of the form:

\begin{equation}\label{eq-cascest-atmu}
%\small
\log_{\rm{10}}E_{\rm{casc,reco}}/\rm{GeV} = 1.6888\cdot e^{0.214\cdot \log_{\rm{10}}E_{\rm{casc,raw}}/\rm{GeV}}
\end{equation}

In the energy region between 5 TeV and 1 PeV, the difference between raw and final value is smaller than 0.1 in $\log_{\rm{10}}E$.

The stochastic energy loss $E_{\rm{casc,reco}}$ was then used to estimate the most likely energy of the leading muon at the surface $E_{\rm{\mu,true}}^{\rm{surf}}$ in dependence of zenith angle $\theta_{\rm{zen}}$ and slant depth $d_{\rm{slant}}=z_{\rm{vert}}/\cos\textrm{ }\theta_{\rm{zen}}$, where $z_{\rm{vert}}$ is the vertical distance to the surface at the point of closest approach to the center of the detector.

\begin{figure}[ht!]
  \centering
  \includegraphics[width=220pt]{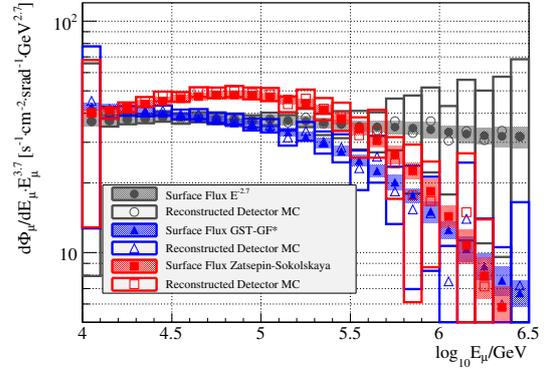}
  \caption{All-Sky surface flux predictions \cite{Fedynitch:2012fs} for three different cosmic ray models and spectrum extracted from full IceCube detector simulation with same primary weight. The error bars on the measured spectrum are the consequence of limited statistics.}
  \label{fig-hemu-fullcirc}
\end{figure}

The parametrized form of the measured muon surface energy is:

\begin{equation}
%\tiny
\label{musurf_param}
\begin{split}
\log_{\rm{10}}E_{\rm{\mu,reco}}^{\rm{surf}}/\rm{GeV}= 0.554+0.884\cdot \\ 
\left(\log_{\rm{10}}(3.44\cdot E_{\rm{casc,reco}}/\rm{GeV}) + \textit{f}_{\rm{corr}}(\cos \textrm{ }\theta_{\rm{zen}}, \textit{d}_{\rm{slant}}) \right)
%+\frac{0.242}{1+e^{-9.63\cdot \log_{10} (d_{slant}/m)-3.88}})
\end{split}
\end{equation}
where $f_{\rm{corr}}(\cos \textrm{ }\theta_{\rm{zen}}, d_{\rm{slant}})$ is a fifth-order polynomial. This relation represents a purely empirical parametrization based on the interpolation of detector-specific simulated data.

%The complexity is somewhat detrimental to the aesthetic value of the description, yet necessary to reduce biases due to higher-order effects. In future analyses, the entire apparatus will likely be hidden behind the veil of a multi-dimensional unfolding procedure.

The relation between the experimental muon surface energy estimator defined in Eq. \ref{musurf_param} and the true energy of the leading muon at the surface is shown in Fig. \ref{fig-hemu_eneres}. It is important to note that the definition is only valid for spectra reasonably close to that used in the construction.

\begin{figure*}[ht!]
  \centering
  \includegraphics[width=220pt]{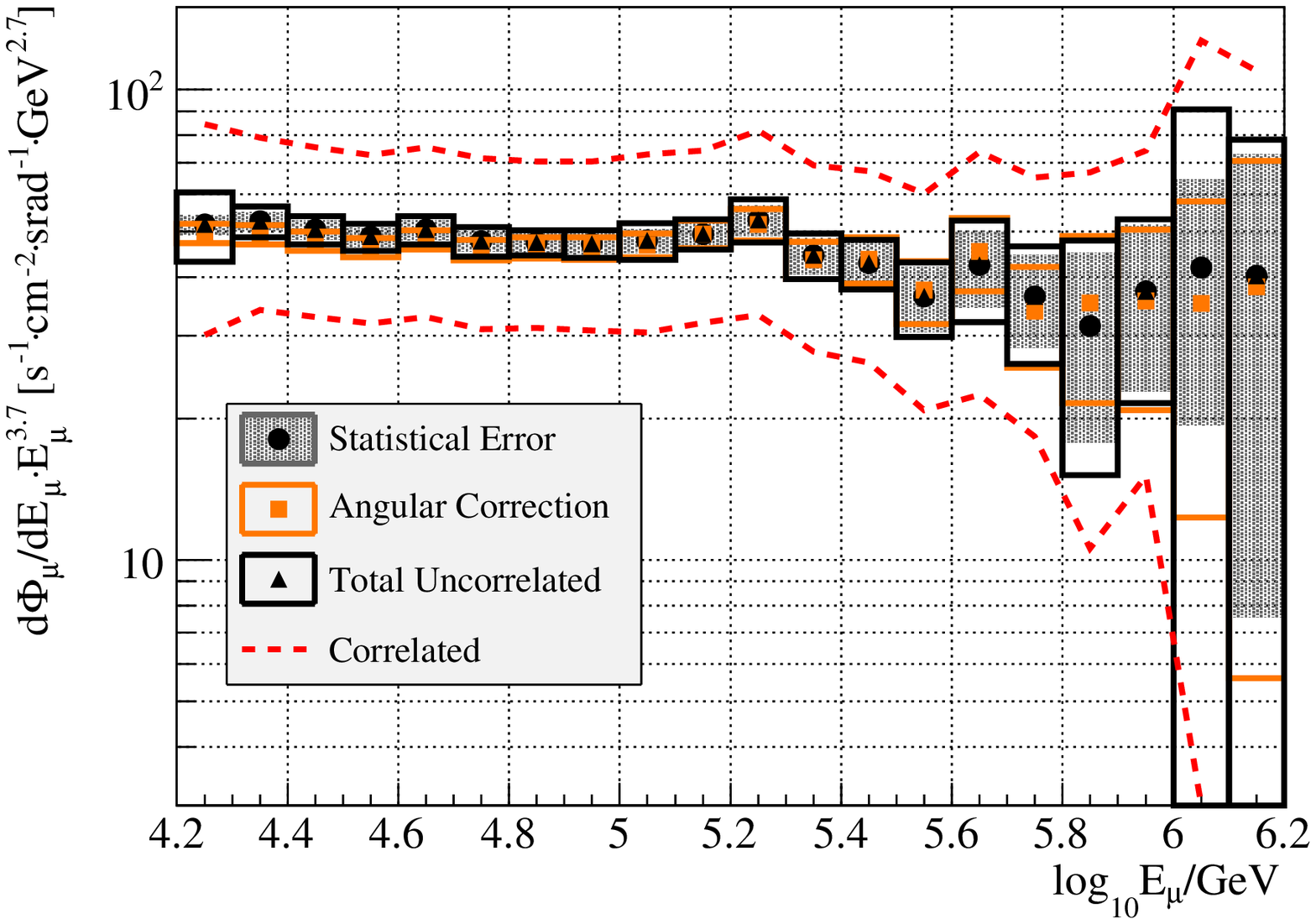}
  \includegraphics[width=220pt]{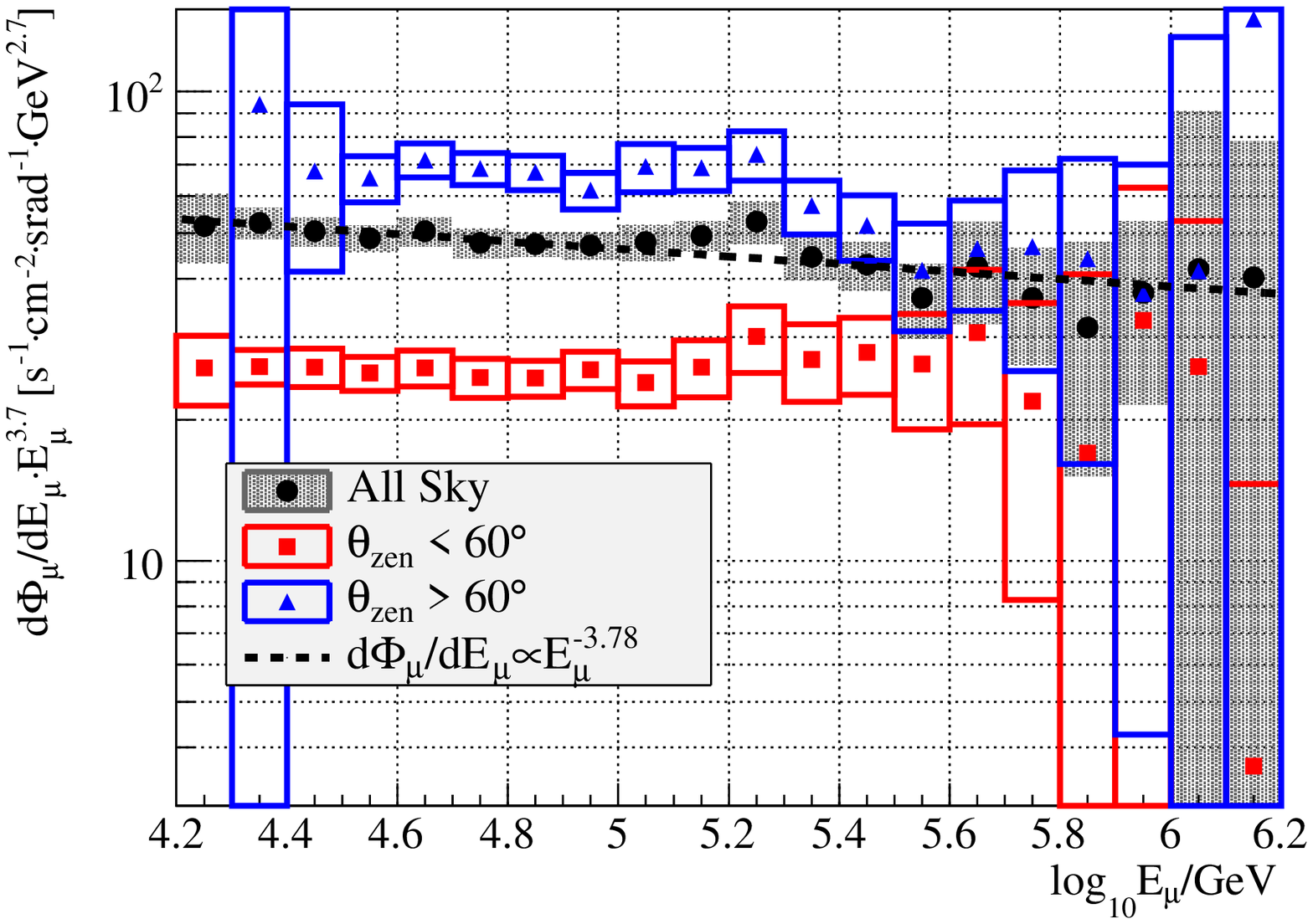}
  \caption{Experimentally measured spectrum of high-energy muons using one year of IceCube data. Left: All-Sky flux with bin-by-bin and correlated error margin. Right: All-Sky spectrum compared to flux above and below 60 degrees ($\cos\textrm{ }\theta_{\rm{zen}}=0.5$). Only bin-wise errors are shown. Between 15 TeV and 1.5 PeV, the all-sky spectrum is consistent with a power law of index $\gamma_{\mu} = -3.78$, illustrated by the dashed line.}
  \label{fig-hemuflux-trip}
\end{figure*}

\begin{table*}[!ht]
%\tiny
%\scriptsize
%\footnotesize
\small 
  \begin{center}
    \begin{tabular}{|c|c|c|c|c|}
      \hline
      Source & Type & Variation & Effect & Comment \\
      \hline
      Composition & uncorrelated & Fe, protons & variable & Negligible Above 25 TeV \\
      Angular Acceptance & uncorrelated & $0.2\cdot (\cos\textrm{ }\theta_{\rm{zen}}-0.5)$ & See Text & Unknown Cause \\
%      \hline
      DOM Efficiency & correlated & $\pm 10\%$ & $\pm 10\%$ Energy Shift & Effective light yield \\
      Optical Ice & correlated & 10\% Scattering, Absorption & $ \pm 10\% $ Energy Shift & Global variations \\
      Seasonal Variations & correlated & Summer vs. Winter & $\pm 5\%$ Flux Scaling & Prompt Invariant \\
      Muon Energy Loss & correlated & Theoretical uncertainty \cite{Koehne:2013gpa} & $\pm 1\%$ & Official IceCube Value \\
%      \hline
%      Hadronic Model & none & discrete & n/a & No impact on spectrum \\
      \hline
    \end{tabular}	
    \caption{Systematic uncertainties in the calculation of the high-energy muon energy spectrum.}
    \label{he_syst_table}
  \end{center}
\end{table*}

\subsection{Energy Spectrum}\label{sec-hemu-espec}

The final muon energy spectrum was calculated by dividing the histogrammed number of measured events $N_{\rm{data}}$ by a generic prediction from a full detector simulation $N_{\rm{detMC}}$, and then multiplying the ratio with the corresponding flux $\Phi_{\rm{surfMC}}$ at the surface. Specifically, IceCube detector simulation and external surface data set \cite{Fedynitch:2012fs} were weighted according to a power law of the form $E^{-2.7}$:

\begin{equation}
\frac{d\Phi_{\rm{\mu,exp}}}{dE_{\mu}}=\frac{\Delta N_{\rm{data}}}{\Delta E_{\rm{\mu,reco}}^{\rm{surf}}}\cdot \left( \frac{\Delta N_{\rm{detMC2.7}}}{\Delta E_{\rm{\mu,reco}}^{\rm{surf}}} \right )^{-1}\cdot \frac{d\Phi_{\rm{\mu,surfMC2.7}}}{dE_{\mu,\rm{true}}^{surf}}
\end{equation}

Figure \ref{fig-hemu-fullcirc} demonstrates the validity of the analysis procedure, and the robustness of the energy estimator construction against small spectral variations. The surface flux for different primary model assumptions can be extracted accurately from simulated experimental data. While a full unfolding would be preferable, the currently available simulated data statistics do not allow for the implementation of such a procedure.

\begin{figure*}[ht!]
  \centering
  \includegraphics[width=220pt]{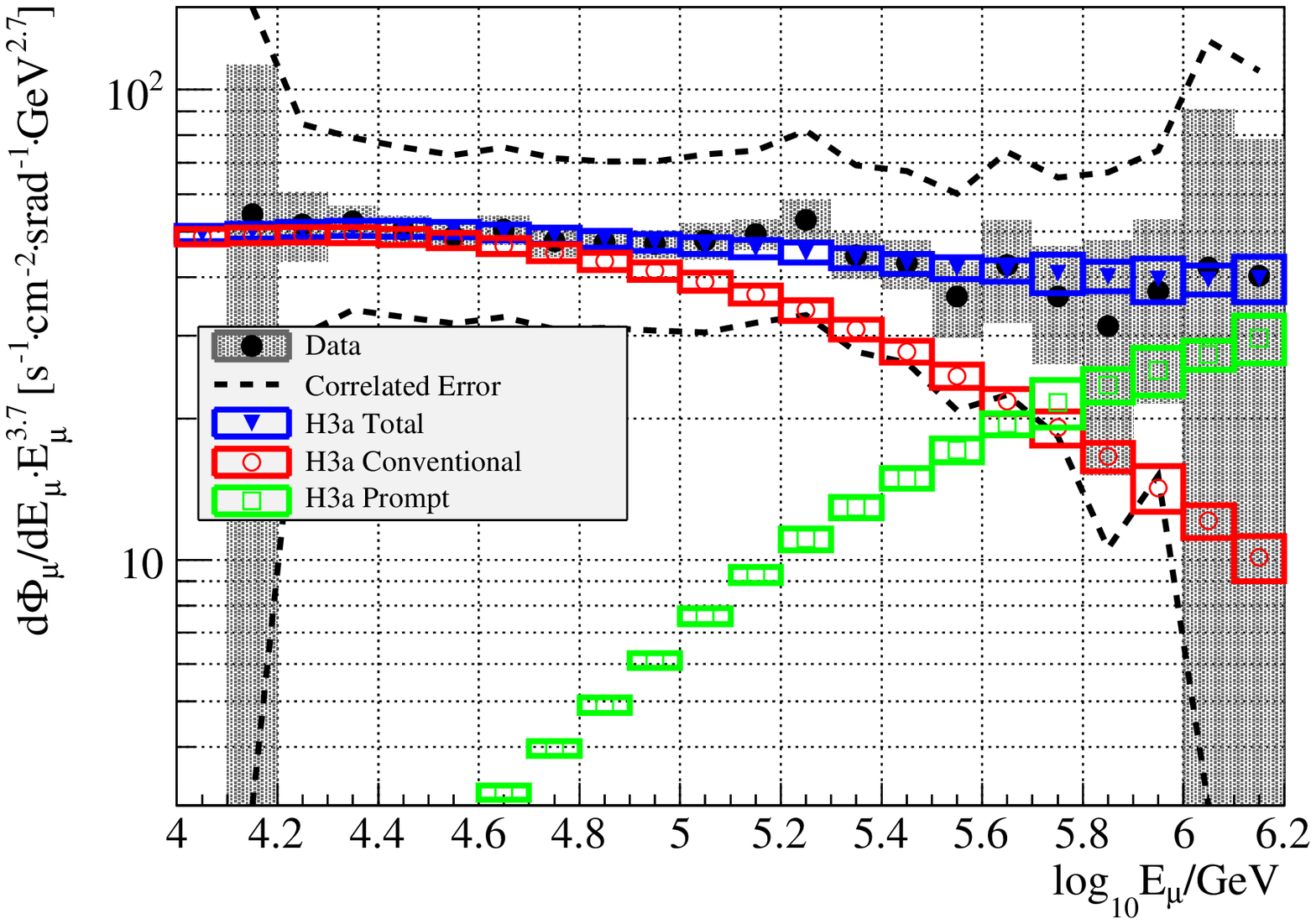}
  \includegraphics[width=220pt]{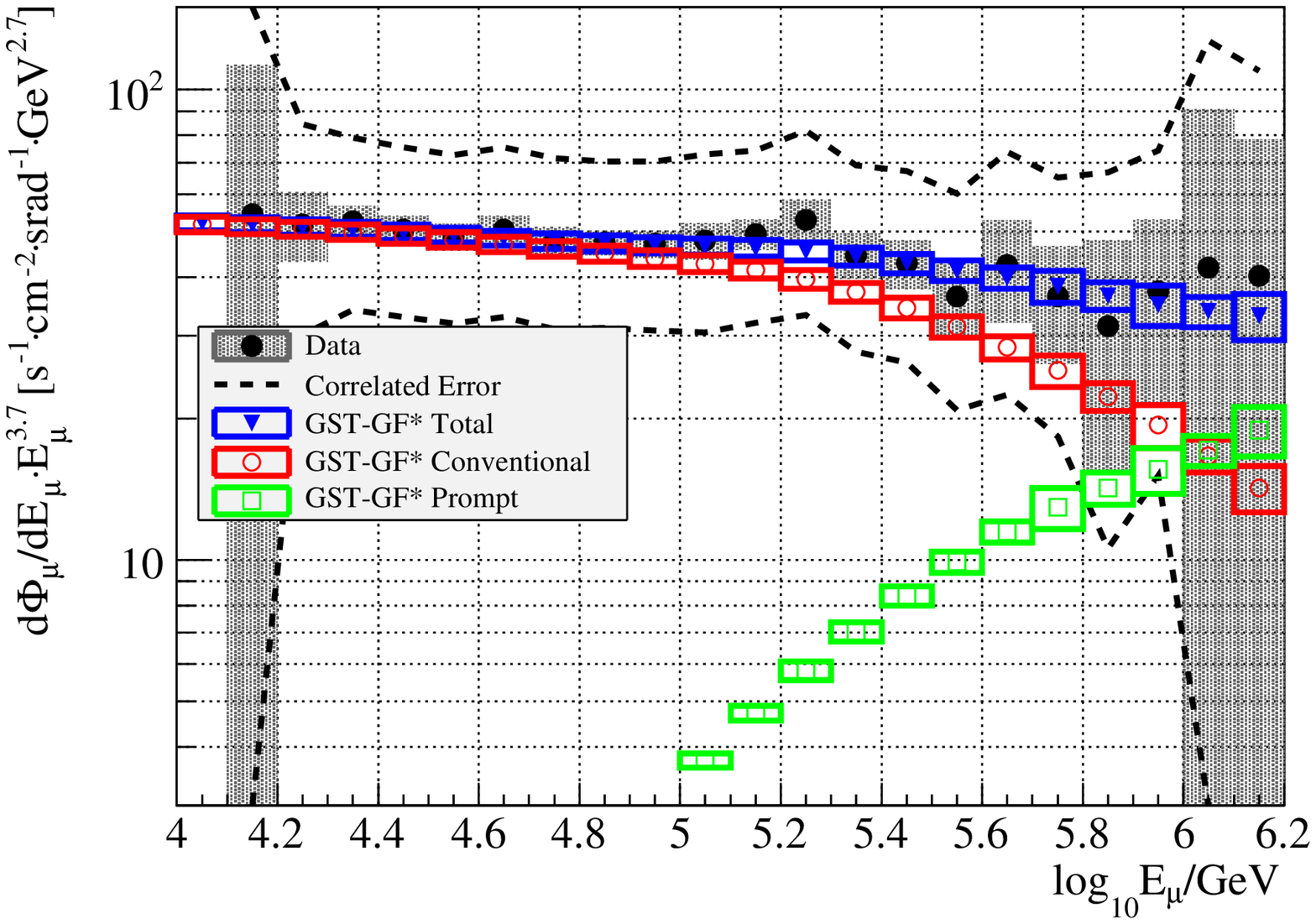}
  \caption{All-sky muon energy spectrum and predictions based on H3a (left) and Global Fit model (right) \cite{Gaisser:2013bla}. Best fit parameters are listed Table \ref{modeldep_table}.}
  \label{fig-hemuflux-allsky}
\end{figure*}

\begin{table*}[!ht]
  \begin{center}
%    \footnotesize
    \begin{tabular}{|c|c|c|c|c|c|}
      \hline
      CR Model & Best Fit (ERS) & $\chi^{2}$/dof & 1$\sigma$ Interval (90\% CL) & Pull ($\Delta\gamma$)  & $\sigma(\Phi_{\rm{Prompt}} > 0$)\\
      \hline
      GST-Global Fit \cite{Gaisser:2013bla} & 2.14 & 7.96/9 & 1.27 - 3.35 (0.77 - 4.30) & 0.01 & 2.64 \\
      H3a \cite{Gaisser:2013bla} & 4.75 & 9.09/9 & 3.17 - 7.16 (2.33 - 9.34) & -0.03 & 3.97 \\
      Zats.-Sok. \cite{Zatsepin:2006ci} & 6.23 & 13.98/9 & 4.55 - 8.70 (3.59 - 10.68) & -0.23 & 5.24 \\
      PG Constant $\Delta\gamma$ \cite{Hoerandel:2002yg} & 0.94 & 9.07/9 & 0.36 - 1.63 ($< 2.15$) & 0.03 & 1.52\\
      PG Rigidity \cite{Hoerandel:2002yg} & 6.97 & 5.86/9 & 4.73 - 10.61 (3.53 - 13.83) & -0.06 & 4.35 \\
%      Power Law Toy Model & 0 & $<0.16$ ($<0.37$) & n/a & 0 \\
      \hline
    \end{tabular}	
    \caption{Result of model-dependent fit to all-sky muon energy spectrum. Note that for muons, the prompt flux is expected to include a substantial contribution from electromagnetic decays of light vector mesons, which is not present in neutrino spectra \cite{Fedynitch:2015zma}.}
    \label{modeldep_table}
  \end{center}
\end{table*}

In the derivation of the experimental result, the systematic uncertainties listed in Table \ref{he_syst_table} were applied. The classification according to correlation is the same as in Section \ref{sec:multresult}. Except for a small effect due to primary composition near threshold, all experimental uncertainties lead to correlated errors. A special case is the angular acceptance. In light of the low-energy muon and multiplicity spectrum studies described in Sec. \ref{sec:lolev-result}, it is necessary to take into account the possibility of an unidentified error source distorting the distribution. This was done by calculating the energy spectrum once for the default angular acceptance and once with simulated events re-weighted by an additional factor $w_{\rm{corr}} = \alpha\cdot(\cos\textrm{ }\theta_{\rm{zen}}-0.5)$, where $\alpha$ corresponds to an ad-hoc linear correction parameter. The value $\alpha=0.2$, corresponding to the variation of $\pm 10\%$ seen in the other analyses, reflects the assumption that the effect is independent of the event sample.

The experimentally measured muon energy spectrum is shown in Fig. \ref{fig-hemuflux-trip}. Distortion due to possible angular effects are small compared to the statistical uncertainty. Within the present accuracy, the average all-sky flux above 15 TeV can be approximated by a simple power law:

\begin{equation}
\label{muflux_pl}
\begin{split}
\frac{d\Phi_{\mu}}{dE_{\mu}}=1.06^{+0.42}_{-0.32}\times10^{-10}\textrm{s}^{-1}\textrm{cm}^{-2}\textrm{srad}^{-1}\textrm{TeV}^{-1}\\
\cdot\left(\frac{E_{\mu}}{\textrm{10 TeV}}\right)^{-3.78\pm0.02(stat.)\pm0.03(syst.)}
\end{split}
\end{equation}

The translation to a vertical flux as commonly used in the literature is not trivial, since the angular dependence of the contribution from prompt hadron decays is different from that of light mesons, and its magnitude a priori unknown. 

The almost featureless shape of the measured spectrum might appear as a striking contradiction to the naive expectation of seeing a clear signature of the sharp cutoff of the primary nucleon spectrum at the knee. However, closer examination reveals that this is very likely a simple coincidence resulting from the fact the the prompt contribution approximately compensates for the effect of the knee if the flux is averaged over the whole sky.

\begin{figure*}[ht!]
  \centering
  \includegraphics[width=180pt]{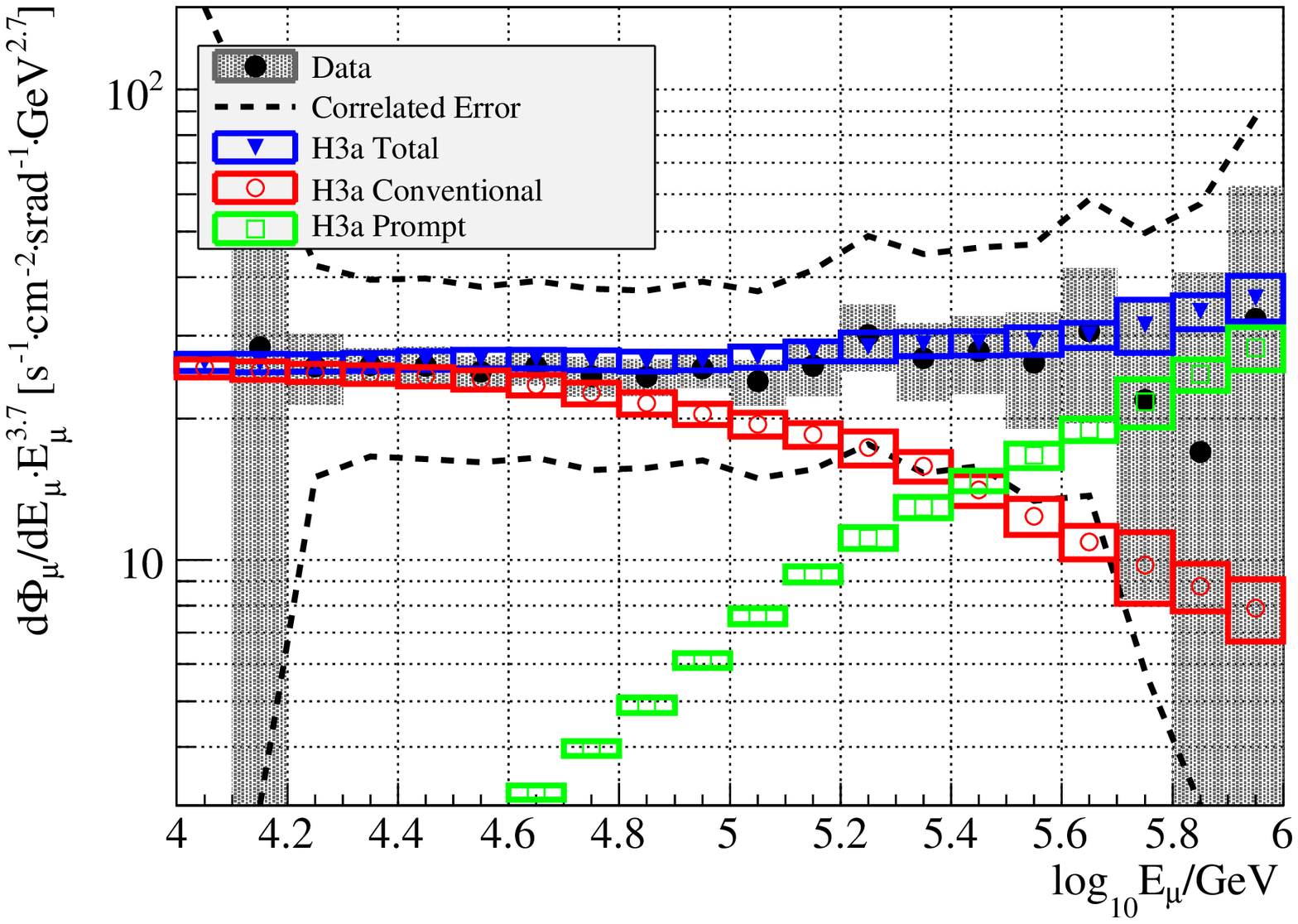}
  \includegraphics[width=180pt]{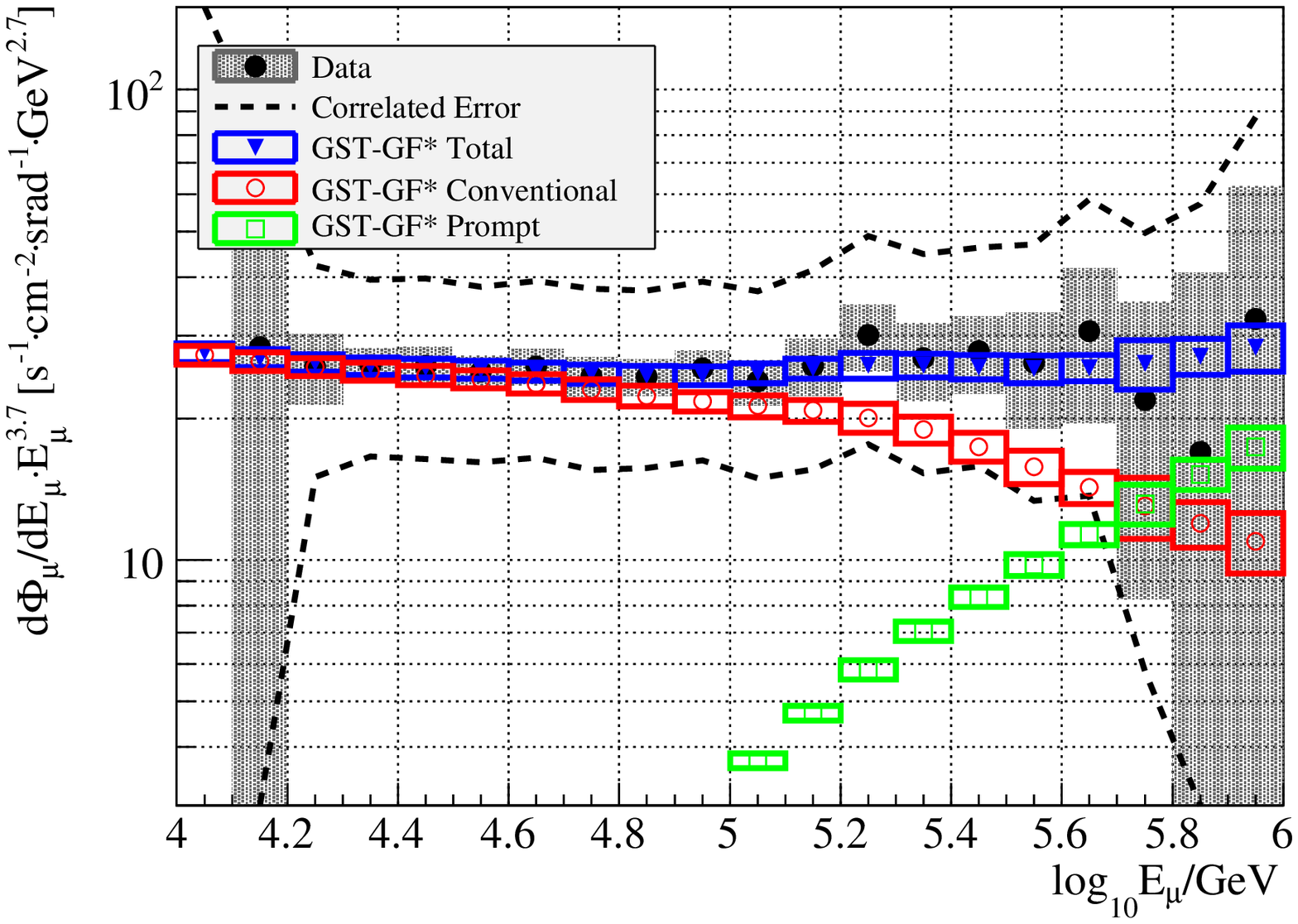}
  \includegraphics[width=180pt]{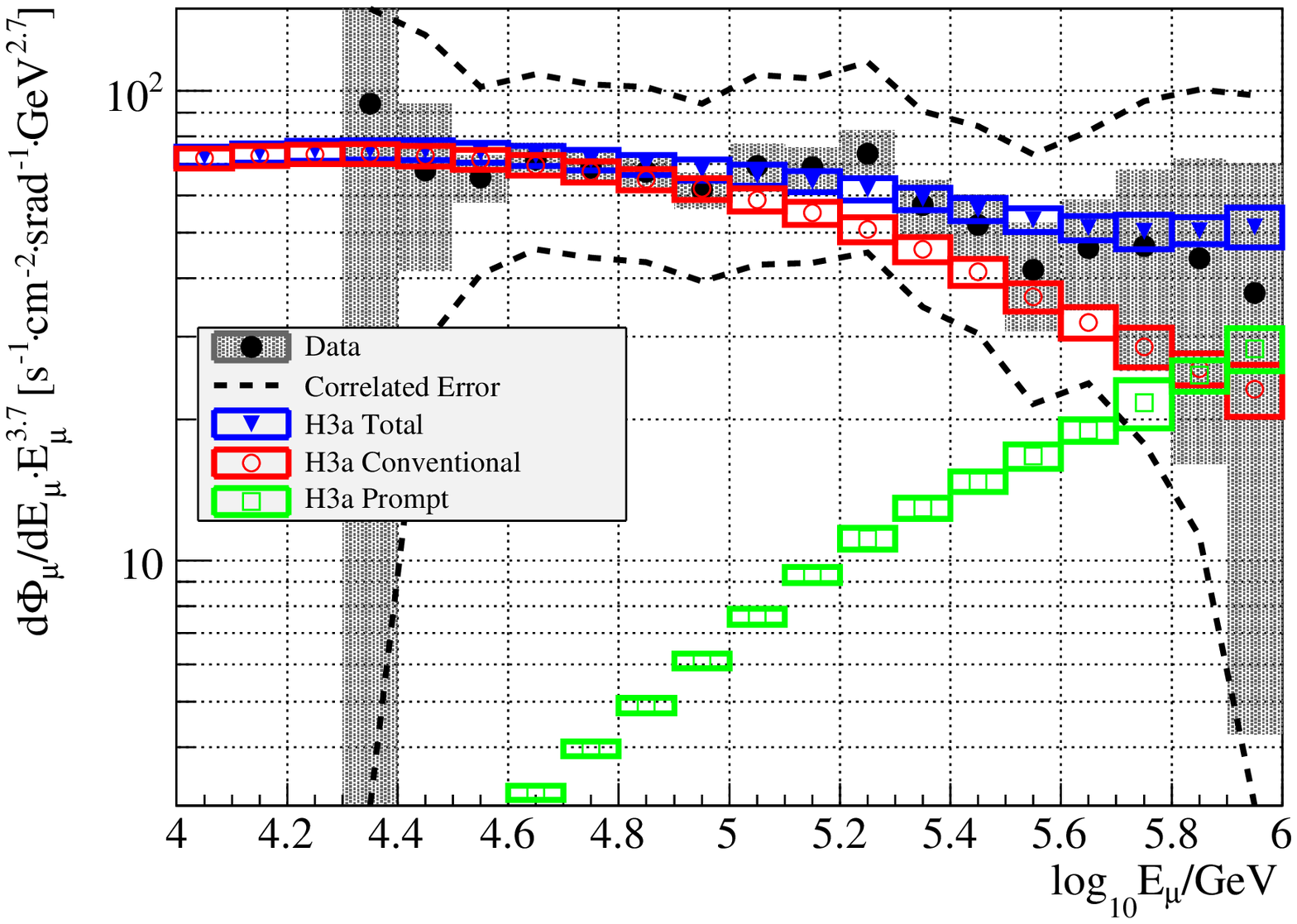}
  \includegraphics[width=180pt]{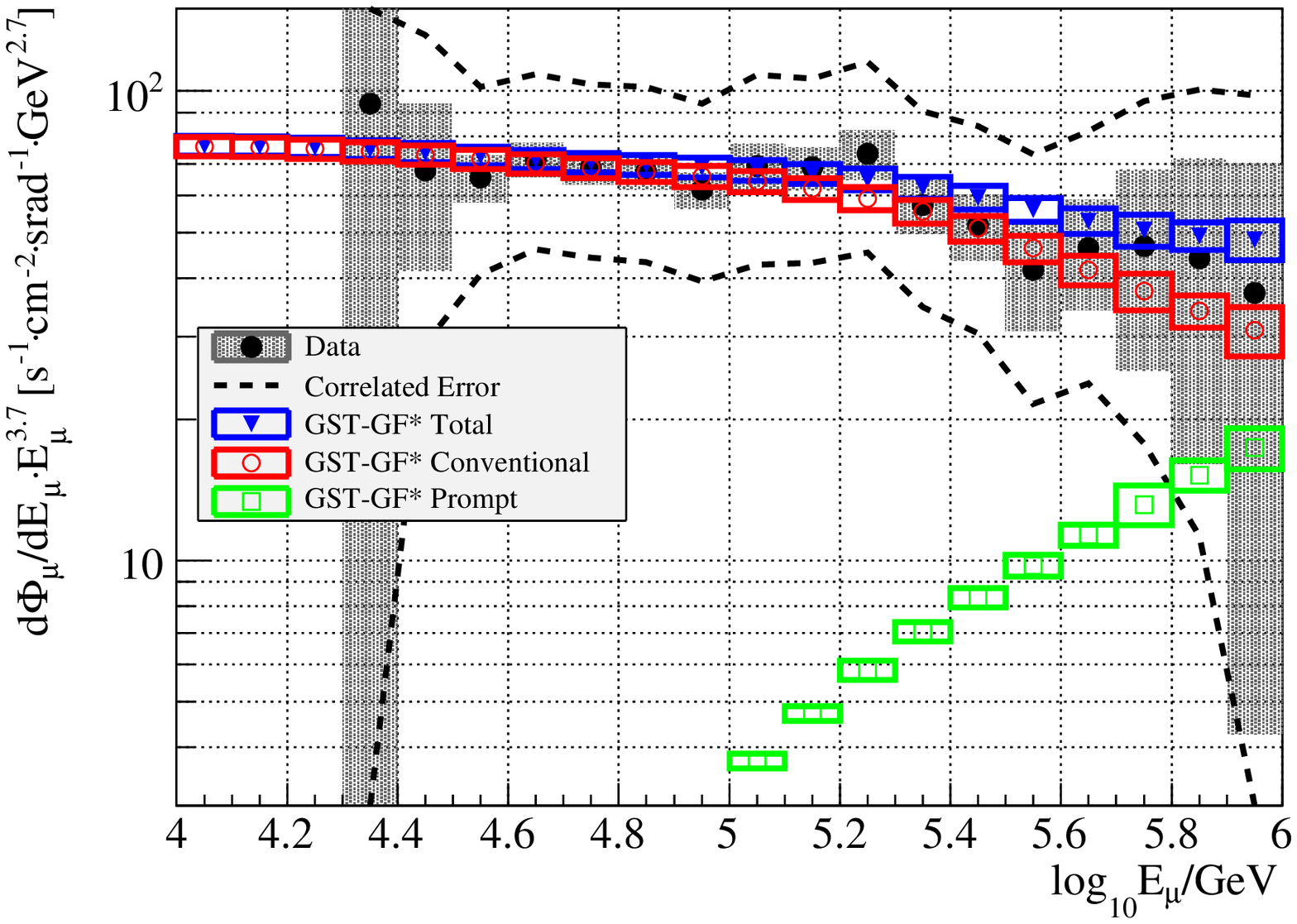}
  \caption{Horizontal and vertical muon energy spectra compared to prediction using best fit values to all-sky spectrum as listed in Table \ref{modeldep_table}. Top row: vertical (0-60 degrees from zenith), bottom row: horizontal (60-84 degrees from zenith). Left: H3a, Right: Global Fit Model.}
  \label{fig-hemuflux-region}
\end{figure*}

Calculating the spectra separately for angles above and below 60 degrees from zenith shows the expected increase of the muon flux toward the horizon. Beyond approximately 300 TeV, the two curves appear to converge, consistent with the emergence of an isotropic prompt component. A quantitative discussion of the angular distribution is given in the following section.

%\begin{figure}
%\centering
%\includegraphics[width=220pt]{modfit_k2_gampri.eps}
%\caption{modfit\_k2}
%\label{fig-modelfit-k2}
%\end{figure}

The final all-sky spectrum was then fitted to a combination of ``conventional'' light meson and prompt components, with a Gaussian prior of $\Delta\gamma = 0.1$ applied to the spectral index. The result in the case of H3a and GST-GF models is illustrated in Fig. \ref{fig-hemuflux-allsky}. The difference between the two measurements is due to the presence of a spectral component in the GST-GF model with a power-law index of -2.3 to -2.4 compared to about -2.6 in H3a. Even though the exponential cutoff energy of 4 PeV is identical in both cases, the influence of the steepening at the knee is effectively reduced in the harder spectrum.

The best fit values for the prompt contribution are listed in the second column of Table \ref{modeldep_table} relative to the ERS flux \cite{Enberg:2008te}. Note that unlike the theoretical prediction, which applies specifically to neutrinos from charm, the experimental result presented here is the sum of heavy quark and light vector meson decays. A detailed discussion can be found in \ref{sec:simple-prompt}.

Since only the energy spectrum is used here, the partial degeneracy between the behavior of the all-nucleon flux at the knee and the prompt contribution is preserved. Consequently, the magnitude of the prompt component strongly depends on the primary model. Except for the proposal by Zatsepin and Sokolskaya \cite{Zatsepin:2006ci}, each of the flux assumptions can be reconciled with the data without a major spectral adjustement.

\subsection{Angular Distribution}

The ambiguity between nucleon flux and prompt contribution can be resolved by the addition of angular information. Figure \ref{fig-hemuflux-region} shows the best fit results from the previous section compared to data separately for angles above and below 60 degrees from zenith. While neither of the two models shown here is obviously favored, it is clear that a substantial prompt contribution is needed in either case to explain the difference between the two regions.

A quantitative treatment can be derived from the different behavior of light meson and prompt components. The prompt flux is isotropic, whereas the contribution from light meson decays is in good approximation inversely proportional to $\cos\textrm{ }\theta_{\rm{zen}}$ \cite{Illana:2010gh}. Using the prompt flux description derived in \ref{sec:simple-prompt}, the experimentally measured fraction of prompt muons as a function of muon energy and zenith angle is:

\begin{equation}
\label{eq-promptzen}
\begin{split}
f_{\rm{prompt}}(E_{\mu},\cos\textrm{ }\theta)\equiv\frac{\Phi_{\rm{prompt}}(E_{\mu},\cos\textrm{ }\theta)}{\Phi_{\rm{total}}(E_{\mu},\cos\textrm{ }\theta)}\\
\simeq\left(1+\frac{E_{\rm{1/2}}\cdot\cos\textrm{ }\theta}{E_{\mu}\cdot f_{\rm{corr}}(E_{\mu})}\right)^{-1}
\end{split}
\end{equation}

In this approximation, the prompt contribution is described independent of the muon flux $\Phi_{\mu}(E_{\mu})$. The repartition between the two components at a given energy can therefore be measured from the angular distribution alone. The effect of higher order terms, such as departure of the angular distribution from a pure $\sec\theta_{\rm{zen}}$ dependence due to the curvature of the Earth and deviations of the nucleon spectrum from a simple power law, have been estimated as less than 10\% using a full DPMJET \cite{Berghaus:2007hp} simulation of the prompt component.

%The experimentally measured prompt fraction is then:

%\begin{equation}
%\footnotesize
%\label{eq-promptzenex}
%f_{prompt,exp}(E_{\mu})=\frac{\int^{1}_{\cos\theta_{max}}f_{prompt}(E_{\mu},\cos\theta)\cdot\eta_{det}(E_{\mu},\cos\theta)\textrm{d}\cos\theta}{\int^{1}_{\cos\theta_{max}}\eta_{det}(E_{\mu},\cos\theta)\textrm{d}\cos\theta}
%\end{equation}

\begin{figure}
\centering
\includegraphics[width=220pt]{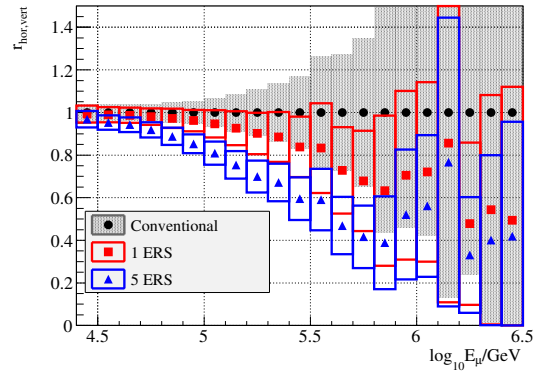}
\caption{Ratio parameter $r_{\rm{hor,vert}}$ expressing deviation of angular distribution from purely conventional flux for various prompt levels in simulation. The size of the error bars corresponds to the statistical uncertainty due to limited availability of simulated data.}
\label{fig-hemu-rhorvert-illu}
\end{figure}

In this study, the measurement of the prompt flux was based on splitting the event sample into two separate sets according to the reconstructed zenith angle. The ratios between experimental data and Monte-Carlo simulation were then combined into a single parameter defined as:

\begin{equation}
\label{eq-rhorvert}
r_{\rm{hor,vert}}=\frac{N_{\rm{\mu,data}}(\theta_{\rm{zen}}>60^{\circ})}{N_{\rm{\mu,MC}}(\theta_{\rm{zen}}>60^{\circ})}\cdot\left(\frac{N_{\rm{\mu,data}}(\theta_{\rm{zen}}<60^{\circ})}{N_{\rm{\mu,MC}}(\theta_{\rm{zen}}<60^{\circ})}\right)^{-1}
\end{equation}

The variation as a function of muon energy is illustrated in Fig. \ref{fig-hemu-rhorvert-illu}, where $N_{\rm{\mu,MC}}$ represents the purely conventional flux, and $N_{\rm{\mu,data}}$ is derived from simulation weighted according to two assumptions about the prompt flux level. Using two discrete samples is not the most statistically powerful way to exploit the angular information, but minimizes fluctuations resulting from limited simulation availability.

\begin{figure}
\centering
\includegraphics[width=220pt]{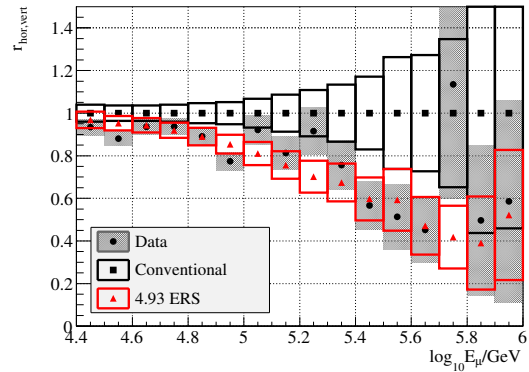}
\caption{Best angular prompt fit using default assumptions about systematic uncertainties. Expressed in multiples of the ERS flux \cite{Enberg:2008te}, the result is $4.9\pm0.9$, with $\chi^{2}$/dof=20.0/15.}
\label{fig-hemu-rhorvert}
\end{figure}

The experimental result is shown in Fig. \ref{fig-hemu-rhorvert}. The best estimate for the prompt flux is significantly higher than the theoretical prediction, but well within the margin permitted by the model-dependent fits to the energy spectrum discussed in the previous section.

\begin{figure}[ht]
\centering
\includegraphics[width=220pt]{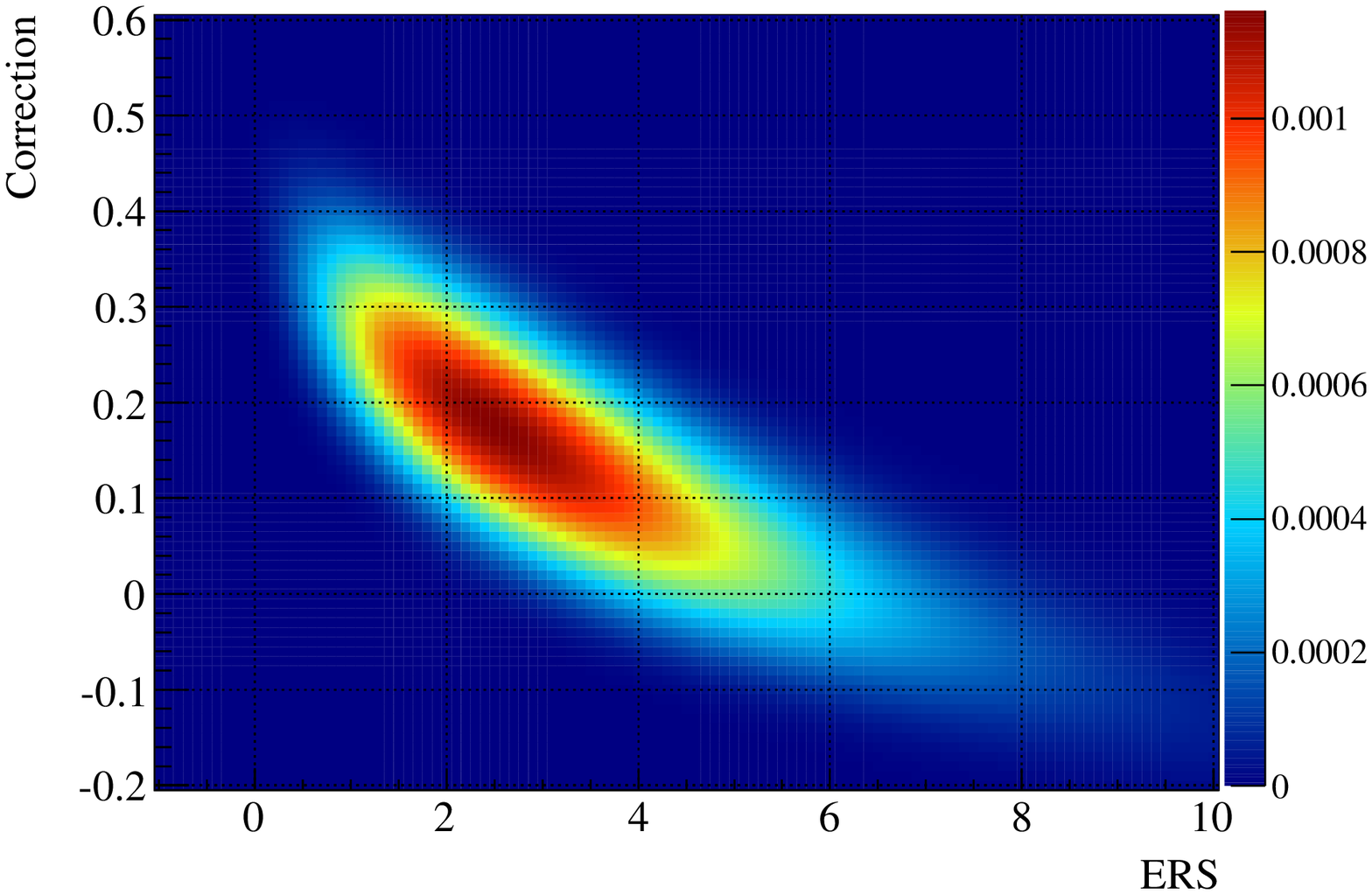}
\includegraphics[width=220pt]{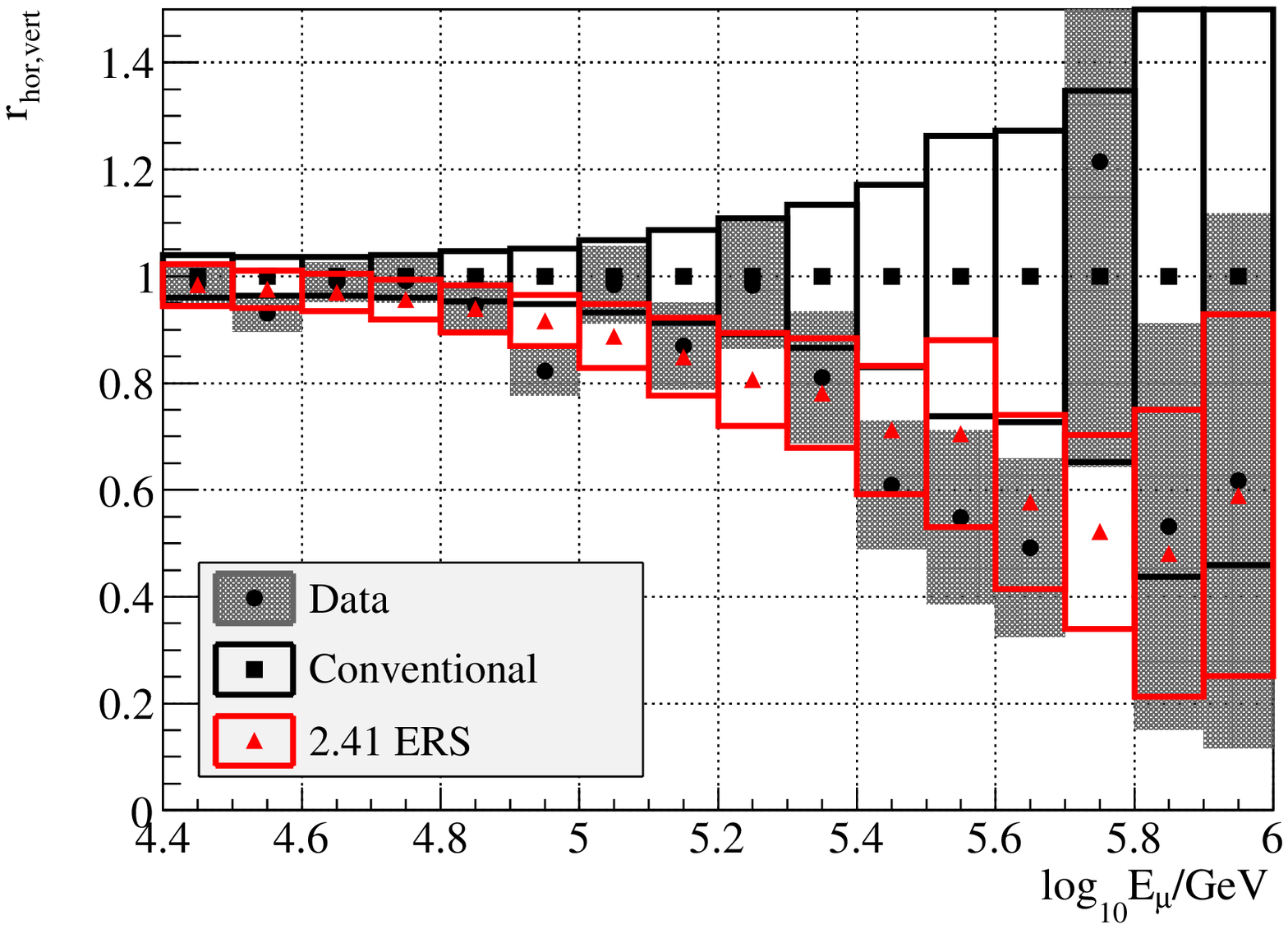}
\caption{Top: Two-dimensional probability distribution function of angular prompt fit results in the presence of an ad-hoc correction term as described in Section \ref{sec-hemu-espec}. The y-axis corresponds to the angular adjustment parameter $\alpha$. Bottom: Result for best overall fit with $\chi^{2}/$dof=14.9/15, located at $(2.41; 0.18)$.}
\label{fig-rhorvert-corr}
\end{figure}

Given the presence of an unknown systematic error in the low-level and high-multiplicity atmospheric muon samples as described in Sec. \ref{sec:lolev-result}, it is necessary to take into account the possibility that the angular distribution might be distorted. As the source of the effect is still unknown, the only choice is to evaluate the influence on the measurement by applying a generic correction term. 

Figure \ref{fig-rhorvert-corr} shows the consequence of re-weighting the simulated data by a linear term of the form $1+\alpha\cdot(\cos\textrm{ }\theta_{\rm{zen}})$. The two-dimensional distribution demonstrates that an imbalance between horizontal and vertical tracks with a magnitude of 18\% describes the data best. This value is suggestively close to the distortions observed in Sec. \ref{sec:le-muons} and \ref{sec:bundles}, although the limited statistical significance does not permit a firm conclusion.

\begin{table*}[!ht]
  \begin{center}
%    \footnotesize
    \begin{tabular}{|c|c|c|c|}
      \hline
      Sample & Best Fit (ERS) & 1$\sigma$ Interval (90\% CL) & $\sigma(\Phi_{\rm{prompt}} > 0$)\\
      \hline
      Uncorrected & 4.93 & 4.05-5.87 (3.55-6.56) & 9.43 \\
 %     Best $\chi^{2}$ Ang. Corr. & 2.41 & 1.83-2.99 (1.51-3.43) & 6.13 \\
      Marginalized Ang. Corr. & 3.19 & 1.64-5.48 (0.98-7.26) & 3.46 \\
      \hline
    \end{tabular}	
    \caption{Result of Angular Prompt Fit.}
    \label{angleprompt_table}
  \end{center}
\end{table*}

\begin{figure}[ht]
\centering
\includegraphics[width=220pt]{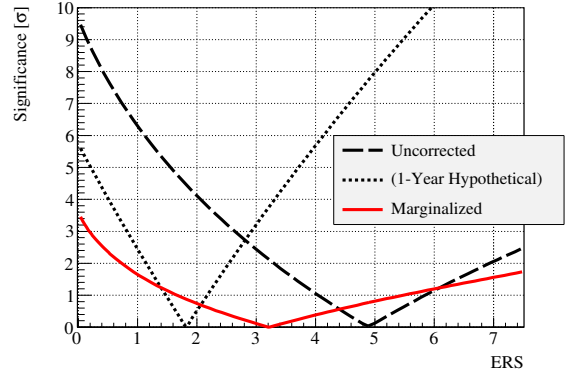}
\caption{Significance of prompt flux measurement based on angular information. The individual curves correspond to different assumptions about systematic effects as described in the text. Also shown is the hypothetical result which could be achieved with one year of experimental data given unlimited availability of simulated events, assuming a best fit value of 1.8 ERS consistent with theoretical predictions for inclusive prompt muon flux.}
\label{fig-hemu-potential}
\end{figure}

\subsection{Discussion}

A definite measurement of the prompt flux is not yet possible. Depending on which assumption is chosen for the systematic error, the final result varies considerably. Figure \ref{fig-hemu-potential} shows the significance levels for default assumption and full marginalization over the linear correction factor. Best fit values and confidence intervals for each case are listed in Table \ref{angleprompt_table}.

At present, the best neutrino-derived limit for the atmospheric prompt flux is 2.11 ERS at 90\% confidence level \cite{Aartsen:2015knd}. This result was derived by a likelihood fit combining four independent measurements from IceCube, and includes both track-like ($\nu_{\mu}$ charged current) and shower-like ($\nu_{\rm{e}}$ and $\nu_{\tau}$ charged current, all-flavor neutral current) neutrino event topologies. For comparisons it is important to keep in mind that the atmospheric muon measurement result represents the inclusive prompt flux, potentially including a substantial contribution from electromagnetic decays of unflavored vector mesons \cite{Fedynitch:2015zma}. It is also worth noting that recent studies show that the uncertainty of theoretical models for atmospheric lepton production in charm decays are larger than previously assumed \cite{Garzelli:2015psa}.

None of the model fluxes selected for the fit to the muon energy spectrum requires a prompt flux in disagreement with the neutrino measurement, with the exception of the proposal by Zatsepin and Sokolskaya. The rigidity-dependent poly-gonato model lacks an extragalactic component whose inclusion would lead to a higher nucleon flux and therefore a lower estimate for the prompt contribution.

The result based on the angular distribution alone is almost independent of the nucleon flux and would even at the present stage be statistically powerful enough to constrain competing primary nucleon flux models around the knee. Unfortunately this possibility is precluded by the likely presence of an unidentified systematic error source. Both uncorrected and ad-hoc corrected measurements could be reconciled with different predictions based on data from air shower arrays, notably the H3a and Global Fit models \cite{Gaisser:2013bla}. At present, the angular measurement is also fully consistent with constraints derived from neutrino data.

\section{Conclusion and Outlook}

The influence of cosmic rays on IceCube data is significant and varied. Given the presence of several energy regions where external measurements by direct detection or air shower arrays are sparse, it is necessary to develop a comprehensive picture including neutrinos, muons and surface measurements. Atmospheric muons play a privileged role, as they cover the largest energy range and provide the highest statistics. A consistent description of all experimental results will be an important contribution for the understanding of cosmic rays in general.

The studies presented in this paper have outlined the opportunities to extract meaningful results from atmospheric muon data in a large-volume underground particle detector. Once systematic effects are fully understood and controlled, it will be possible to measure the muon energy spectrum from 1 TeV to beyond 1 PeV by combining measurements based on angular distribution and catastrophic losses. Agreement between the two methods can then be verified in the overlap region around 10-20 TeV. 

There is a strong indication for the presence of a component from prompt hadron decays in the muon energy specrum, with best fit values generally falling at the higher side of theoretical predictions. In the future, it will be possible for the IceCube detector to precisely measure the prompt contribution and to constrain the all-nucleon primary flux before and around the knee. With more data accumulating, independent verification of the prompt measurement based on seasonal variations of the muon flux \cite{Desiati:2010wt} will soon become feasible as well.

The muon multiplicity spectrum provides access to the cosmic ray energy region beyond the knee. Even though a direct translation of the result to primary energy and average mass is impossible, combination with results from surface detectors or comparisons to model predictions provide valuable insights. In coming years, the measurement can be extended further into the transition region around the ankle. A possible contribution from heavy elements to the cosmic ray flux at EeV energies should then be discernible.

An important goal of this study was to verify the current understanding of systematic uncertainties. An unexplained effect was demonstrated using low-level data, and appears to be present in the other analysis samples as well. In order to improve the quality of future atmospheric muon measurements with IceCube, it will be essential to determine whether the observed discrepancy requires better understanding of the detector, or of the production mechanisms of muons in air showers.

Comparisons with measurements from the upcoming water-based KM3NeT detector \cite{Margiotta:2014eaa} will be invaluable to decide whether the inconsistencies seen in IceCube data are due to the particular detector setup, or represent unexplained physics effects.

\section{Acknowledgements}

We acknowledge the support from the following agencies: U.S. National Science Foundation-Office of Polar Programs, U.S. National Science Foundation-Physics Division, University of Wisconsin Alumni Research Foundation, the Grid Laboratory Of Wisconsin (GLOW) grid infrastructure at the University of Wisconsin - Madison, the Open Science Grid (OSG) grid infrastructure; U.S. Department of Energy, and National Energy Research Scientific Computing Center, the Louisiana Optical Network Initiative (LONI) grid computing resources; Natural Sciences and Engineering Research Council of Canada, WestGrid and Compute/Calcul Canada; Swedish Research Council, Swedish Polar Research Secretariat, Swedish National Infrastructure for Computing (SNIC), and Knut and Alice Wallenberg Foundation, Sweden; German Ministry for Education and Research (BMBF), Deutsche Forschungsgemeinschaft (DFG), Helmholtz Alliance for Astroparticle Physics (HAP), Research Department of Plasmas with Complex Interactions (Bochum), Germany; Fund for Scientific Research (FNRS-FWO), FWO Odysseus programme, Flanders Institute to encourage scientific and technological research in industry (IWT), Belgian Federal Science Policy Office (Belspo); University of Oxford, United Kingdom; Marsden Fund, New Zealand; Australian Research Council; Japan Society for Promotion of Science (JSPS); the Swiss National Science Foundation (SNSF), Switzerland; National Research Foundation of Korea (NRF); Danish National Research Foundation, Denmark (DNRF) 


\newpage
\appendix

\section{Data-Derived Deterministic Differential Deposition Reconstruction (DDDDR)}\label{sec:ddddr}
\setcounter{figure}{0}

\subsection{Concept}

The energy deposition of muons at TeV energies passing through matter is not continuous and uniform, but primarily a series of discrete catastrophic losses. In order to exploit the information contained in the stochasticity of muon events, it is necessary to reconstruct the differential energy loss along their tracks. 
The study presented in this paper requires a robust method for identification and energy measurement of major stochastic losses. Its principle is to use muon bundles in experimental data to characterize photon propagation in the detector and apply the result to the construction of a deterministic energy estimator.

\begin{figure}[ht]
\centering
\includegraphics[width=220pt]{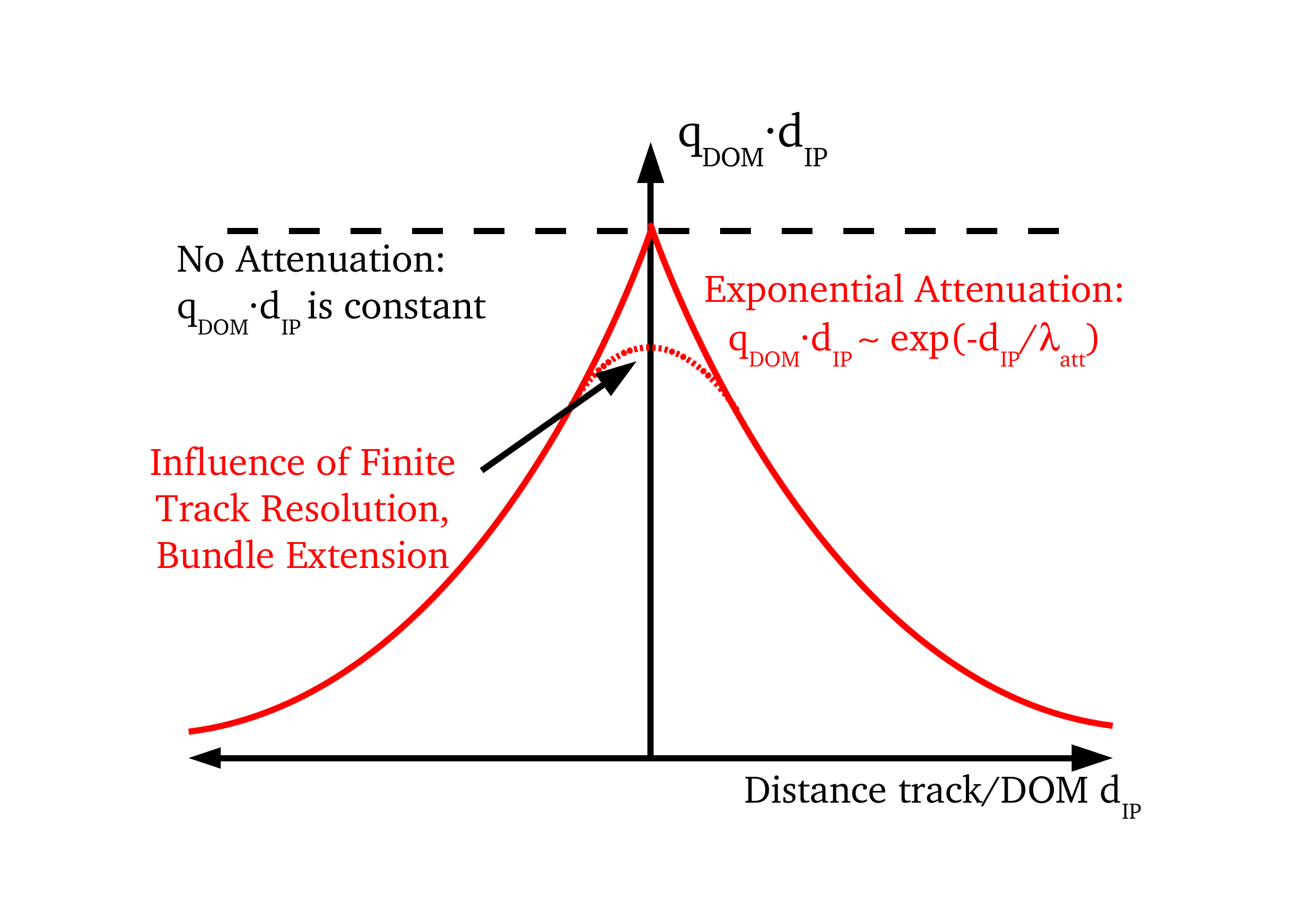}
\caption{Sketch of light attenuation around muon track in ice.}
\label{fig-d4r-lightatten}
\end{figure}

Figure \ref{fig-d4r-lightatten} shows a sketch of the photon intensity distribution around the reconstructed track of a muon bundle. In the ideal case of a perfectly transparent homogeneous medium and a precisely defined infinite one-dimensional track of arbitrarily high brightness, the light intensity would fall off as $1/d_{\rm{IP}}$, where the impact parameter $d_{\rm{IP}}$ is defined as the perpendicular distance to the track. Assuming the measured charge $q_{\rm{DOM}}$ in a given DOM to be proportional to the light density, and the emitted number of photons $N_{\rm{phot}}$ to be proportional to the energy deposition $\Delta E_{\mu}$, the relation between muon energy deposition and measurement then takes the form:

\begin{equation}\label{d4r_ideal}
\Delta E_{\mu}/\Delta x \sim N_{\rm{phot}} \sim q_{\rm{DOM}} \cdot d_{\rm{IP}}
\end{equation}

In reality, scattering and absorption in the detector medium require the addition of an exponential attenuation term $\exp(-d_{\rm{IP}}/\lambda_{\rm{att}})$:

\begin{equation}\label{d4r_att}
N_{\rm{phot}} \sim q_{\rm{DOM}} \cdot d_{\rm{IP}} \cdot exp(d_{\rm{IP}}/\lambda_{\rm{att}})
\end{equation}

where the attenuation length $\lambda_{\rm{att}}$ depends on the local optical properties in a given part of the detector. Approximating the structure of individual ice layers as purely horizontal, $\lambda_{\rm{att}}$ is simply a function of the vertical depth $z_{\rm{vert}}$.

\begin{figure}[ht!]
\centering
\includegraphics[width=220pt]{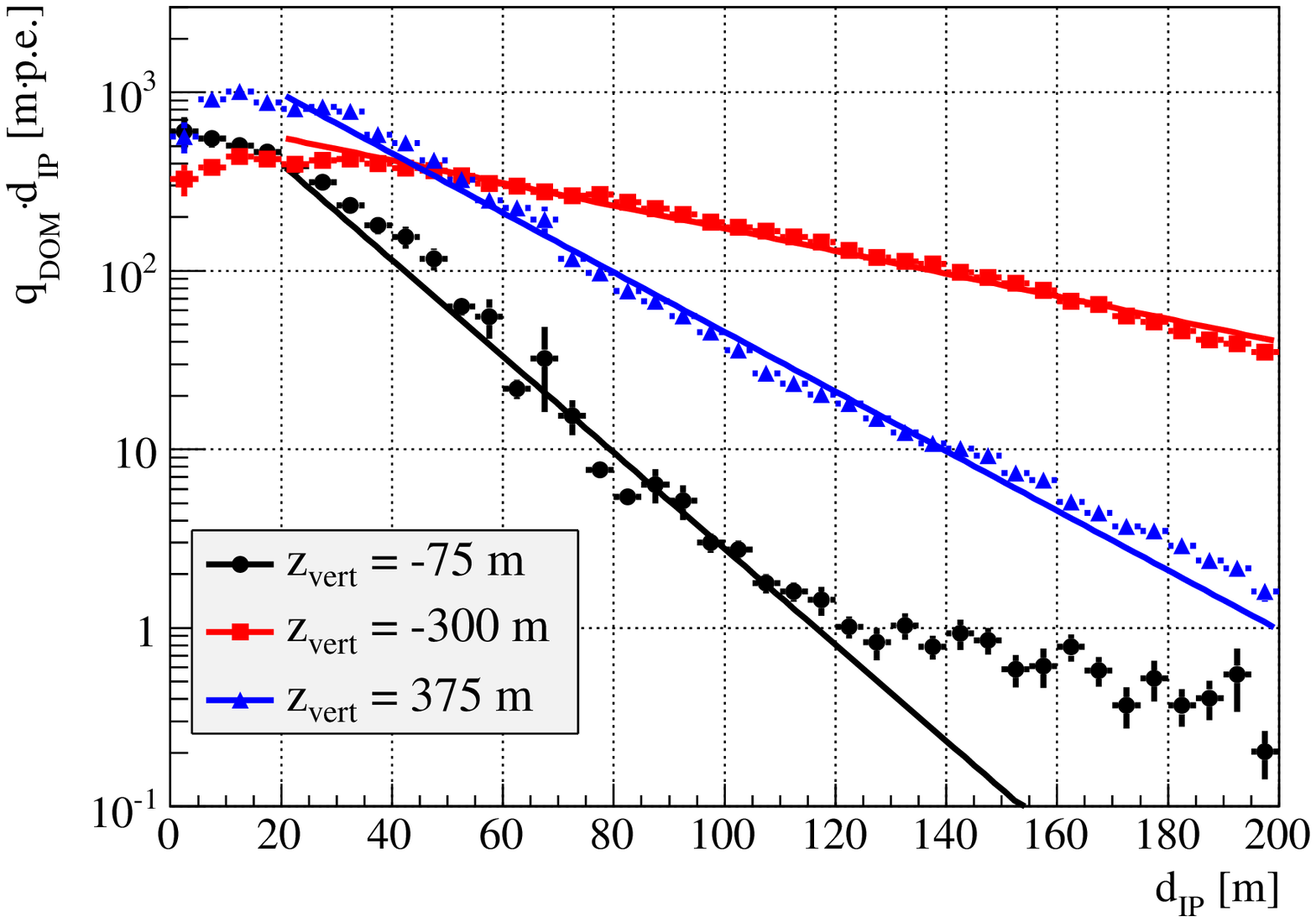}
\includegraphics[width=220pt]{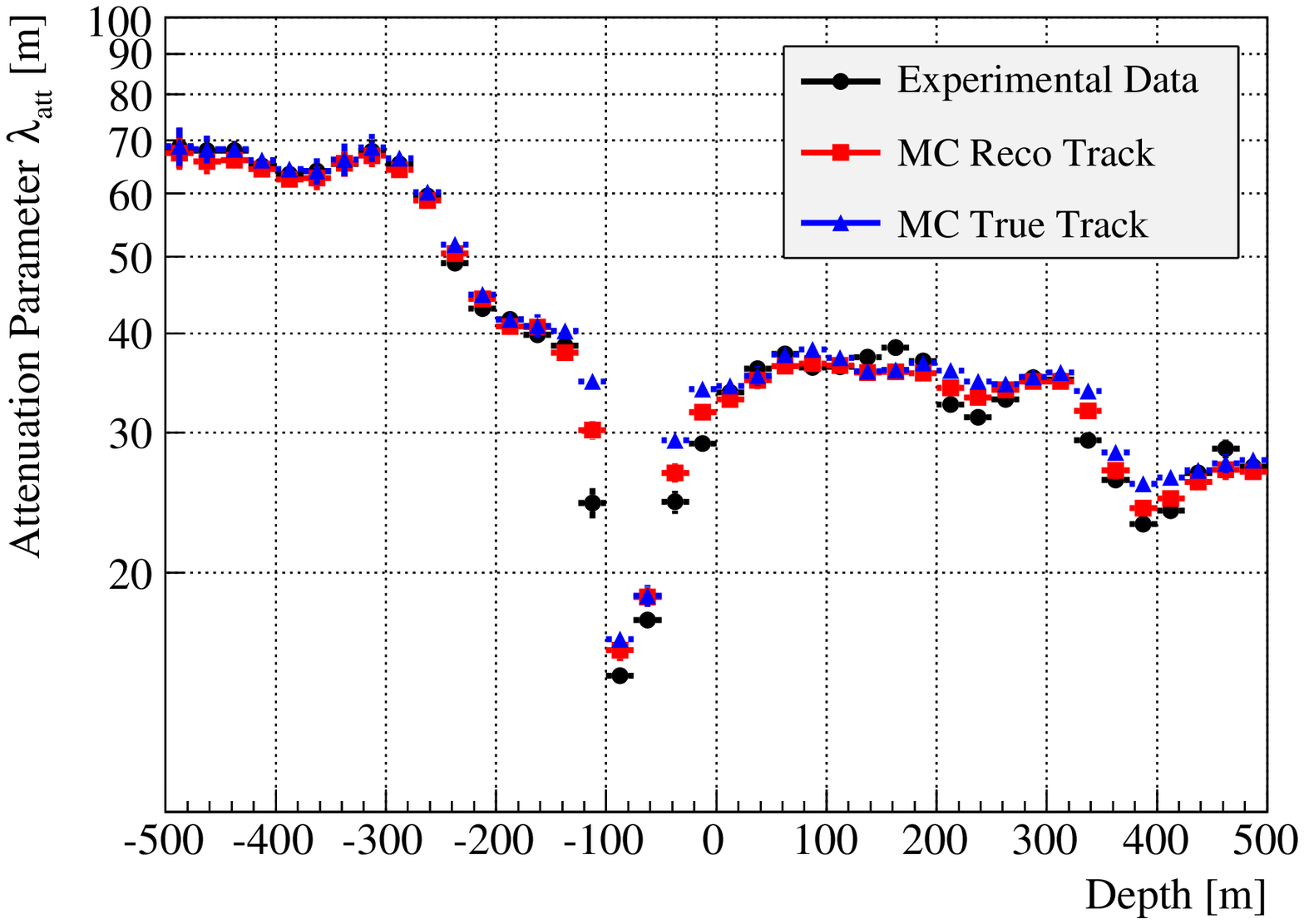}
\caption{Top: Lateral attenuation of photon intensity along muon bundle tracks in experimental data. The vertical depth ranges, corresponding to DOM position relative to the center of the detector 1949 m below the surface, were chosen to illustrate the strongly varying optical properties of the ice. Bottom: Effective attenuation parameter $\lambda_{\rm{att}}$ derived from exponential fit to the data distribution. Experimental values are compared to Monte-Carlo simulation using reconstructed and true track parameters for calculation of the impact parameter $d_{\rm{IP}}$.}
\label{fig-d4r-lightatt}
\end{figure}

The validity of this hypothesis is demonstrated in Fig. \ref{fig-d4r-lightatt}. A sample of bright downgoing tracks with $Q_{\rm{tot}}>1,000 pe$ was selected to obtain an unbiased data set fully covered by the online event filters. For each DOM within a given vertical depth range, the quantity 

\begin{equation}\label{d4r_dommeas}
\tilde{n}_{\rm{phot,ideal}}=\epsilon_{\rm{DOM}}^{-1}\cdot q_{\rm{DOM}}\cdot d_{\rm{IP}}
\end{equation}

is calculated, corresponding to the photon yield adjusted for the distance from the track and relative quantum efficiency $\epsilon_{\rm{DOM}}$ of the PMT, which is 1 in standard and about 1.35 for high-efficiency DeepCore DOMs. The curves are averaged over the entire event sample and include DOMs that did not register a signal. The solid lines shows the result of a fit to the function

\begin{equation}\label{d4r_fdip}
f(d_{\rm{IP}}) = c\cdot exp(-d_{\rm{IP}}/\lambda_{\rm{att}})
\end{equation}

with the effective attenuation length $\lambda_{\rm{att}}$ and the data sample-dependent normalization constant c as free fit parameters. Exponential attenuation as a function of the impact parameter is a valid assumption over a wide range, breaking down only for very close distances and in the layer with high dust concentration at $z_{\rm{vert}}\approx -100 \textrm{ m}$, where the vertical gradient of the optical ice properties is exceptionally steep.

The experimental result is well reproduced by the simulation, as illustrated in the lower plot. The very small difference between the curves using true and reconstructed track parameters means that track reconstruction inaccuracies can be neglected.

\subsection{Construction of Energy Observable}

Once the effective attenuation length has been determined, it can be used to construct a simple differential energy loss parameter. For each DOM within a given distance from the reconstructed track, an approximation for the photon yield corrected for PMT efficiency and ice attenuation can be calculated. The actual differential energy loss at the position of the DOM projected is related to the experimental observable by:

%\begin{equation}
%\label{d4r_nphotexp}
%\tilde{n}_{phot} \equiv \epsilon_{DOM}^{-1} \cdot q_{DOM} \cdot d_{IP} \cdot exp(d_{IP}/\lambda_{att})
%\end{equation}

\begin{equation}
  \label{d4r_dedx}
  \begin{split}
  \left (\frac{dE_{\mu}}{dx}\right)_{\rm{reco}}= \epsilon_{\rm{DOM}}^{-1}\cdot q_{\rm{DOM}}\cdot \\
  f_{\rm{scale}} \cdot
  \begin{cases}
%    d_{0}\cdot e^{d_{0}/\lambda_{\rm{att}}(z)} ,& d_{\rm{IP}} < d_{0} \\ 
    d_{0} ,& d_{\rm{IP}} < d_{0} \\ 
    d_{\rm{IP}}\cdot e^{(d_{\rm{track}}-d_{0})/\lambda_{\rm{att}}(z)} ,& d_{\rm{IP}} > d_{0}
  \end{cases}
  \end{split}
\end{equation}

where $f_{\rm{scale}} \simeq 0.020 \textrm{GeV}\cdot(\textrm{p.e}\cdot m^{2})^{-1}$ is a simple scaling factor that can be derived from a Monte Carlo simulation and $d_{0}(z) = 19 m+0.01\cdot z$ expresses the mild depth dependence of the point of transition from flat to exponential behavior. The vertical coordinate z is measured from the center of the detector at 1949 meters below the surface.

\begin{figure}[ht]
  \centering
  \includegraphics[width=200pt]{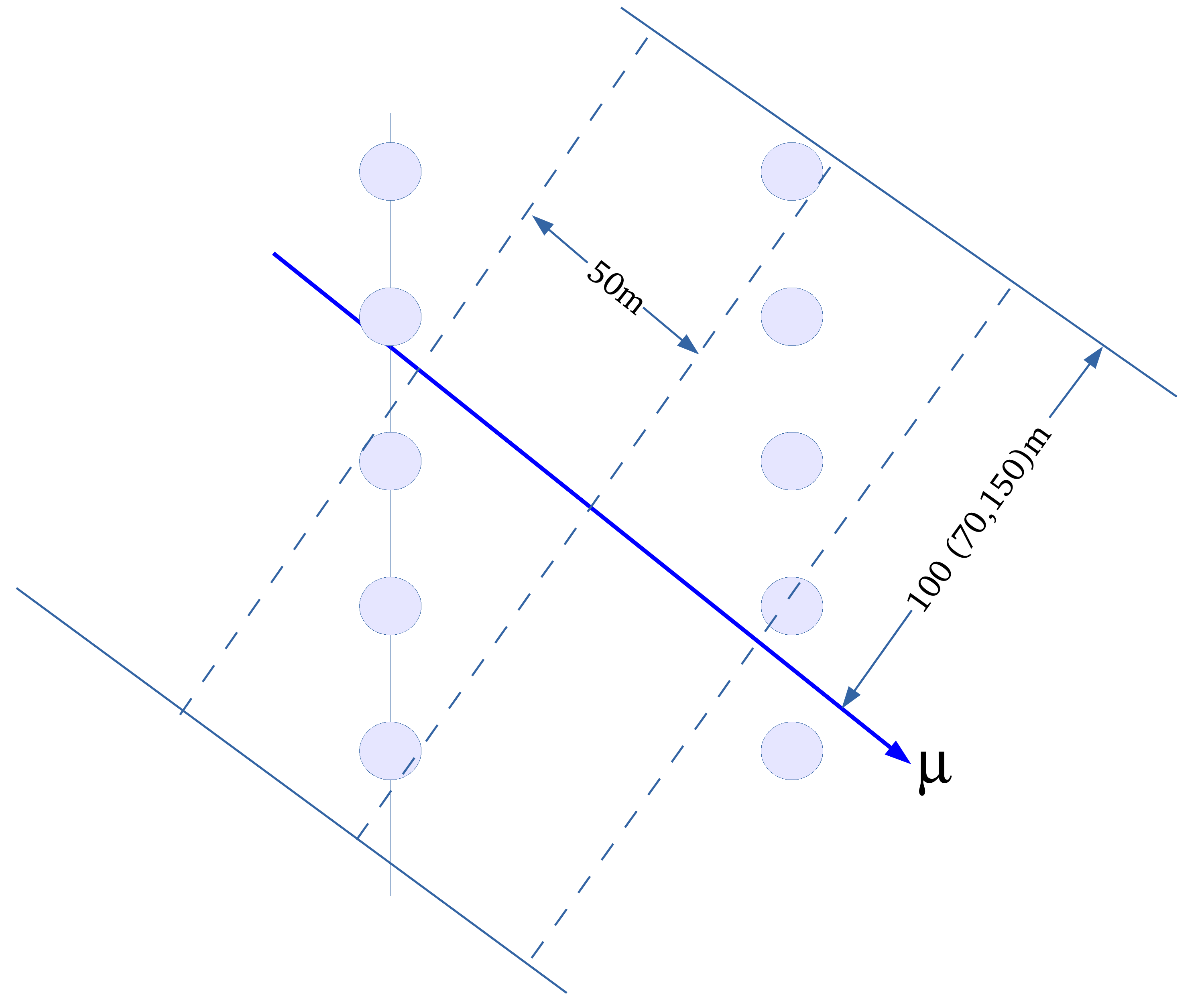}
  \includegraphics[width=220pt]{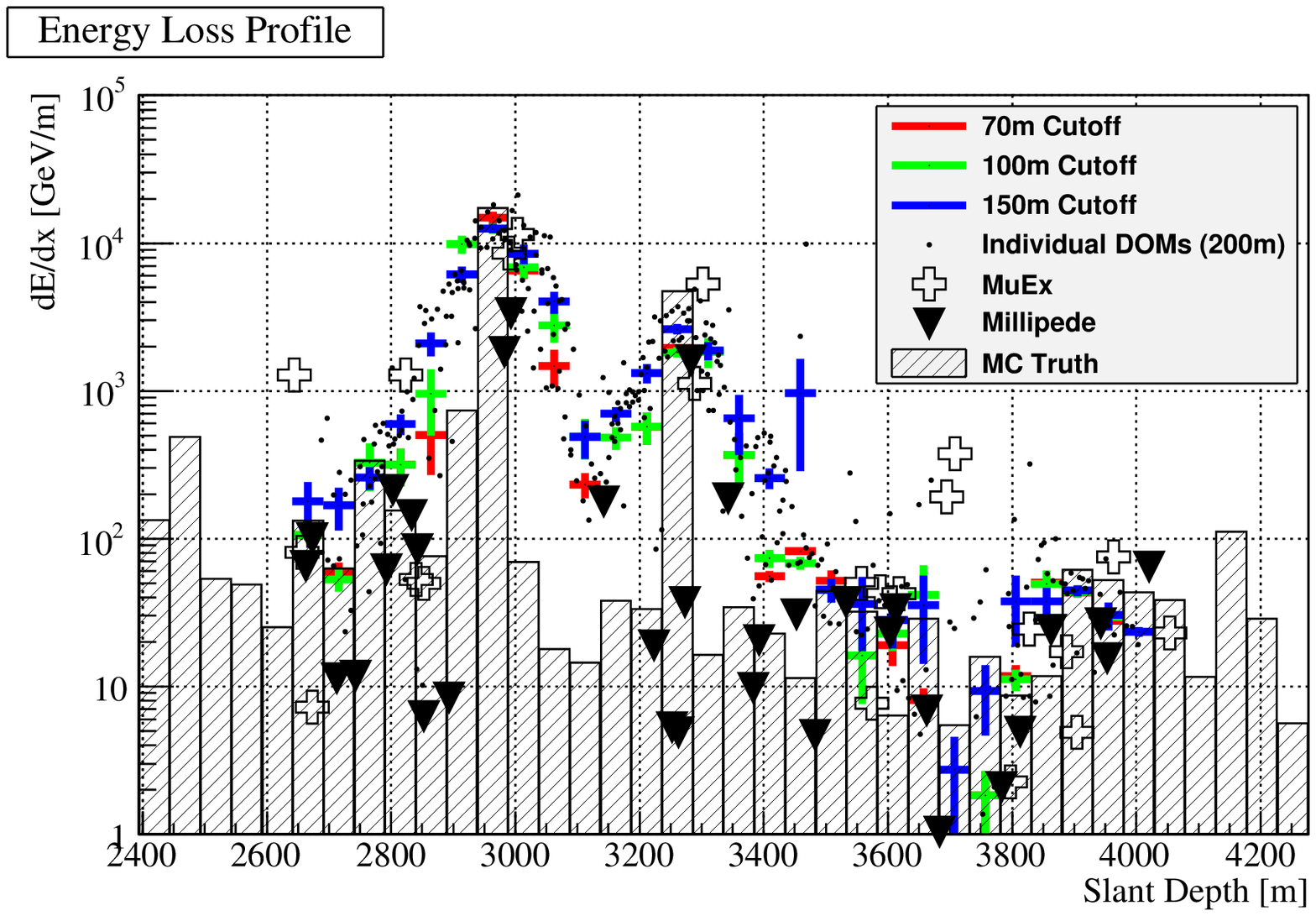}
  \caption{Top: Construction of differential energy deposition estimator. DOMs are represented by circles. The maximum lateral distance from the track up to which individual data points are included in the reconstruction can be varied depending on specific requirements. Bottom: Comparison between true and reconstructed energy loss in simulated event with parameters: $E_{\rm{shower,reco}}$ = 1165 TeV (True Value: 852 TeV), $\cos\textrm{ }\theta_{\rm{zen,reco}}$ = 0.556 (True Value: 0.551) $E_{\rm{\mu,reco}}$ = 2493 TeV (True Value: 1854 TeV). The shower energy corresponds to the highest single stochastic loss at approximately 3000 m slant depth. Reconstructions using two different likelihood methods  \cite{Aartsen:2013vja} are shown for comparison.}
  \label{fig-d4r-paracons}
\end{figure}

To account for fluctuations affecting individual measurements and DOMs that did not register a signal, the track is subdivided into longitudinal bins with a width of 50 meters, over which the measured parameter is averaged. The lateral limit for the inclusion of DOMs can be adjusted to find a compromise between sufficient statistics and adequate longitudinal resolution. The principle is illustrated in Fig. \ref{fig-d4r-paracons}.

Note that the exact value of dE/dx is only calculated for demonstration purposes and should be considered approximate. The measured observables, like any energy-dependent observable, are in practical applications directly related to physical parameters such as shower energy and muon multiplicity, where the exact conversion depends on the spectrum of the data distribution.

The energy of the strongest stochastic loss in the event could be derived immediately from the highest bin value in the profile. However, this estimate is often imprecise. Better results can be achieved by a dedicated reconstruction for the individual loss energy. The origin of the shower is assumed to coincide with the position of the DOM with the highest $dE/dx$ value projected on the track. Its energy is then calculated in a similar way as for the track, except that the photon emission is assumed to be point-like and isotropic. Instead of falling off linearly, the light intensity falls off quadratically as a function of distance, and the energy estimate becomes:

\begin{equation}
\label{d4r_lossen}
\begin{split}
E_{\rm{loss,reco}}= \epsilon_{\rm{DOM}}^{-1}\cdot q_{\rm{DOM}}\cdot \\f_{\rm{scale}} \cdot
 \begin{cases}
% r_{0}^{2}\cdot e^{r_{0}/\lambda_{\rm{att}}(z)} ,& r_{\rm{loss}} < r_{0} \\ 
   r_{0}^{2} ,& r_{\rm{loss}} < r_{0} \\ 
   r_{\rm{loss}}^{2}\cdot e^{(r_{\rm{loss}}-r_{0})/\lambda_{\rm{att}}(z)} ,& r_{\rm{loss}} > r_{0}
 \end{cases}
\end{split}
\end{equation}

\begin{figure}[ht!]
  \centering
  \includegraphics[width=220pt]{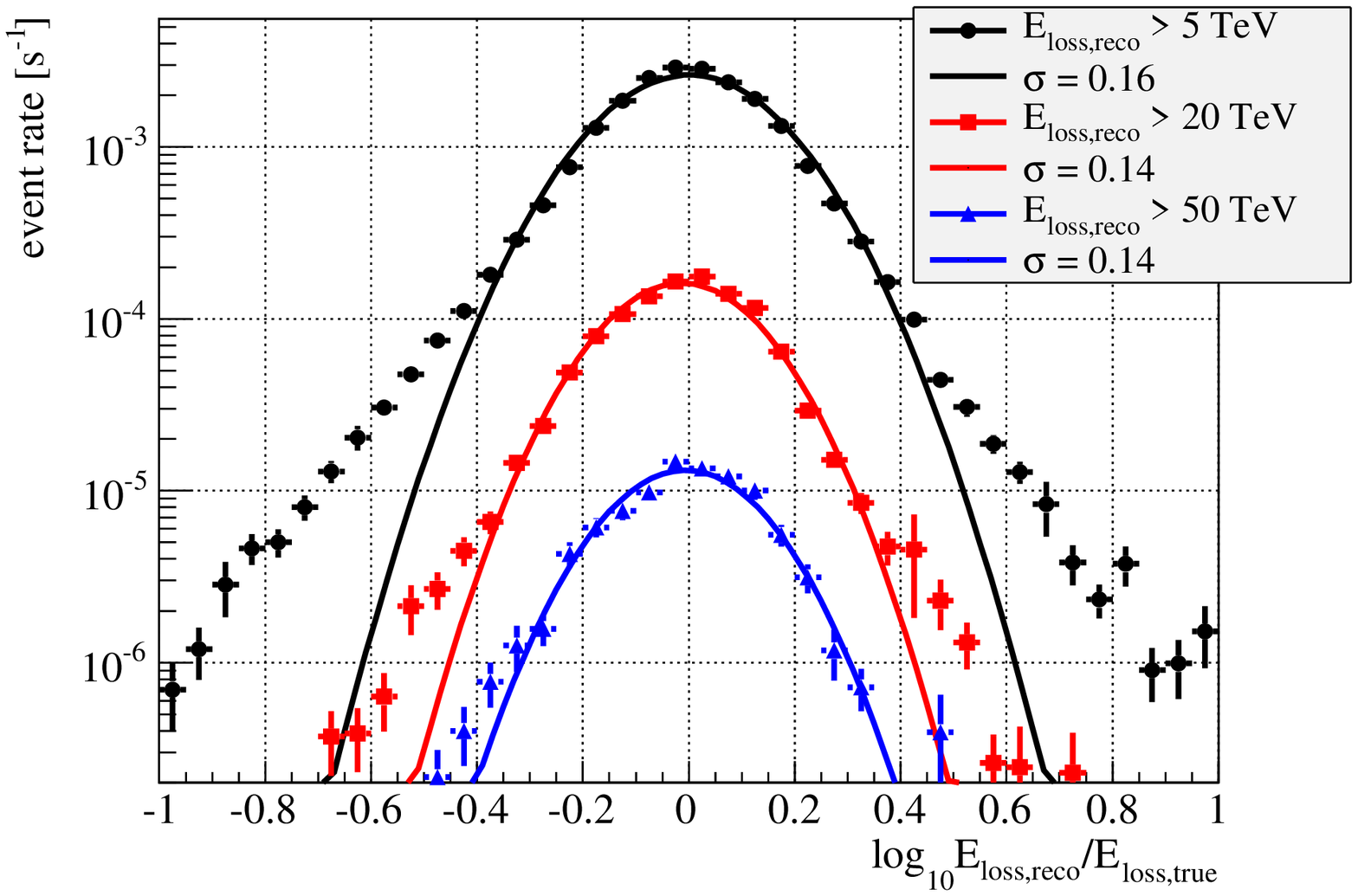}
  \caption{Ratio between reconstructed and true shower energy for simulated events weighted to an $E^{-2.7}$ power-law primary cosmic ray flux spectrum. Around the peak the distribution can be closely approximated by a Gaussian distribution with a width varying between approximately 0.16 and 0.14.}
  \label{fig-cascest_resol}
\end{figure}

The shower energy can then be determined by calculating the mean of the values for the individual DOMs. The energy resolution for events selected by the method described in Section \ref{sec:hemu} is shown in Fig. \ref{fig-cascest_resol}.

\section{Prompt Flux Calculation}\label{sec:simple-prompt}
\setcounter{figure}{0}
\setcounter{table}{0}

\subsection{Prompt Muon Flux Approximation}

The characteristics of the atmospheric muon energy spectrum at energies beyond 100 TeV are influenced by prompt hadron decays. In neutrino analyses, these can be taken into account by applying a simple weighting function to simulated data. Muons, on the other hand, are always part of a bundle, and in principle it would be necessary to generate a full air shower simulation including prompt lepton production.

The hadronic interaction generators integrated into the CORSIKA simulation package as of version 7.4 are not adequate for a prompt muon simulation mass production. QGSJET and DPMJET \cite{Berghaus:2007hp} are slow, and charm production in QGSJET is very small compared to theoretical predictions. The core CORSIKA propagator does not handle re-interaction effects for heavy hadrons, which become important at energies approaching 10 PeV. 

A version of SIBYLL that includes charm is at the development stage \cite{Engel:2015dxa}. The updated code also takes into account production and decay of unflavored light mesons, which form an important part of the prompt muon flux \cite{Illana:2010gh}. First published simulated prompt atmospheric muon spectra indicate consistency with the ERS model for charmed mesons, and an unflavored component of approximately equal magnitude \cite{Fedynitch:2015zma}.

In this paper, the prompt flux is expressed in dependence of the ``conventional' flux from light meson decays. In this way it can be modeled using simulated events from the standard IceCube CORSIKA mass production, including detector simulation and information about the primary cosmic ray composition.

\begin{figure}[ht!]
  \centering
  \includegraphics[width=220pt]{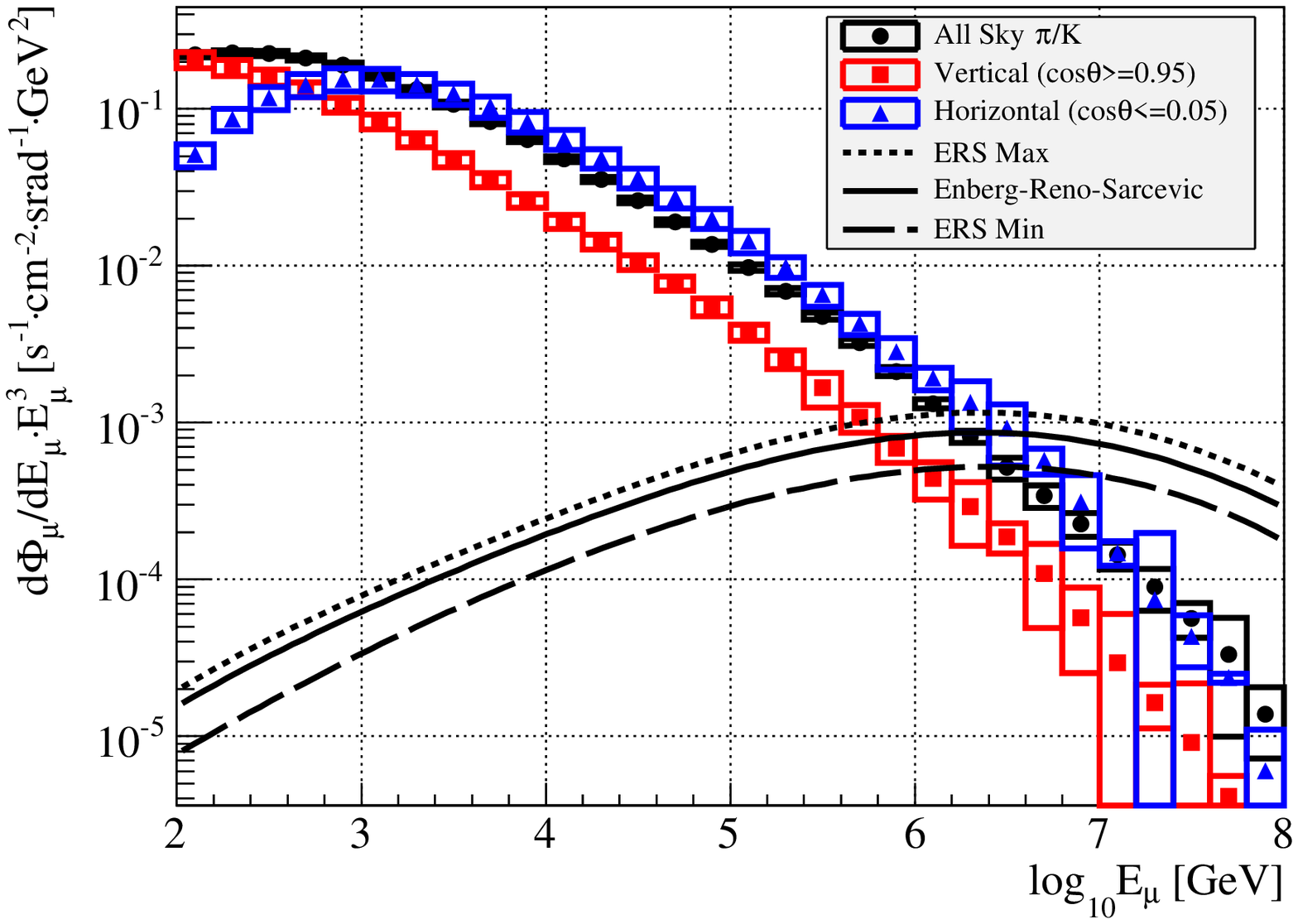}
  \caption{Muon flux predictions from full shower CORSIKA simulation \cite{Fedynitch:2012fs} and parametrization of theoretical calculation \cite{Enberg:2008te}.}
  \label{fig-pseudoprompt}
\end{figure}

Construction of the simulated prompt flux is based on the following assumptions:

\begin{itemize}
\item The spectral index of the prompt component $\gamma_{\rm{prompt}}$ is related to the conventional index $\gamma_{\rm{conv}}$ as $\gamma_{\rm{prompt}}=\gamma_{\rm{conv}}+1$. Higher-order effects, such as the varying cross section of charm production and re-interaction in the atmosphere, can be accounted for by a corrective term $f_{\rm{corr}}(E_{\mu})$.
\item The prompt flux is isotropic, the conventional flux increases proportional to $\sec\theta_{\rm{zen}}$ in the analysis region above $\cos\textrm{ }\theta_{\rm{zen}}=0.1$. Variations due to the curvature of the Earth \cite{Illana:2010gh} are neglected.
\item The influence of changes in the nucleon spectrum on the prompt flux is the same as on the conventional flux. Based on estimates using prompt muons simulated with DPMJET, this assumption is valid within 10\% for spectra with an exponential cutoff at the knee.
\item The contribution from light vector meson di-muon decays is small compared to that from heavy hadrons and/or has the same energy spectrum. For prompt muon fluxes simulated with the newest development version of SIBYLL, charm and unflavored spectra are almost identical in shape between 10 TeV and 1 PeV \cite{Fedynitch:2015zma}.
\end{itemize}

The approximated prompt flux is then:

\begin{equation}
\label{eq-pseudoprompt}
\begin{split}
\Phi_{\rm{\mu,prompt}} (E_{\mu},\theta_{\rm{zen}}) \simeq \Phi_{\rm{\mu,conv}}(E_{\mu},\theta_{\rm{zen}}) \\ \cdot \frac{E_{\mu}\cdot \cos\textrm{ }\theta_{\rm{zen}}}{E_{\rm{1/2}}}\cdot f_{\rm{corr}}(E_{\mu})
\end{split}
\end{equation}

The relative flux normalization is expressed in terms of $E_{\rm{1/2}}$, the crossover energy for prompt and conventional fluxes in vertical air showers. This parameter provides a simple and intuitively clear way to express the magnitude of the prompt flux, and can easily be estimated.

\begin{figure}[ht]
  \centering
  \includegraphics[width=220pt]{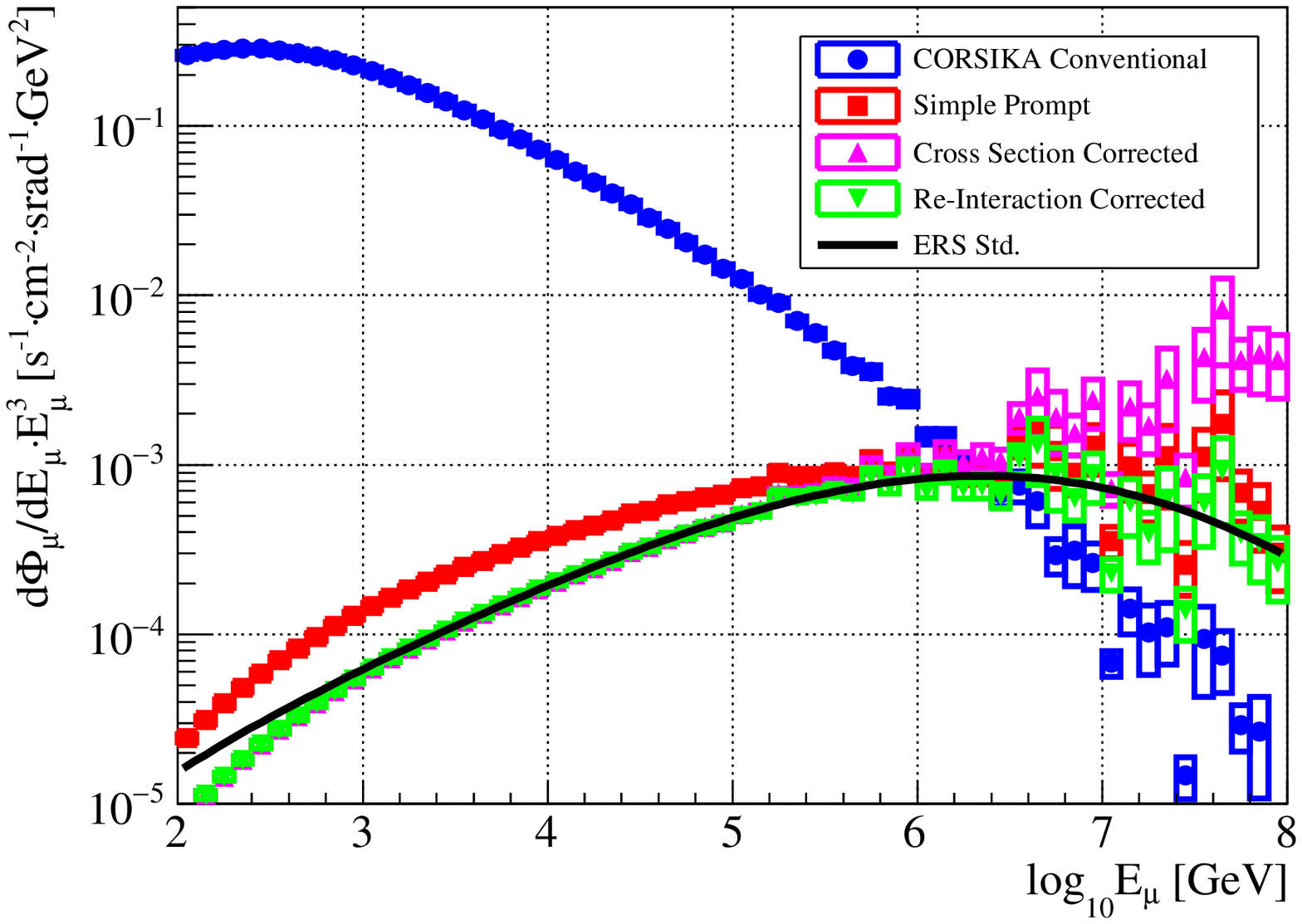}
  \caption{Effect of higher-order prompt flux correction factor on all-sky muon flux derived from simulation using CORSIKA. The separation into cross section and re-interaction correction should be considered approximate.}
  \label{fig-pseudoprompt_corr}
\end{figure}

To calculate the crossover energy $E_{\rm{1/2}}$ for a specific prediction, it is sufficient to compare conventional muon simulations with a prompt flux parametrization, as illustrated in Fig. \ref{fig-pseudoprompt}. The crossover energy can then be determined in a straightforward way by a fit to their ratio. Note that here the primary nucleon spectrum corresponds to the na\"ive TIG model \cite{Gondolo:1995fq} used in the theoretical calculation.

Since the full air shower simulation only needs to provide an estimate for the conventional flux, this procedure can be repeated for any interaction model. In this study, as in most IceCube analyses, the prompt prediction is based on the calculation by Enberg, Reno and Sarcevic \cite{Enberg:2008te}. The corresponding values are listed in Table \ref{e12_erstable}.

\begin{table}[!h]
  \footnotesize
  \begin{center}
    \begin{tabular}{|c|c|c|c|}
      \hline
      Hadronic Model & ERS (max) & ERS (default) & ERS (min) \\
      \hline
      SIBYLL & $5.71 \pm 0.02$ & $5.82 \pm 0.03$ & $5.99 \pm 0.03$ \\
      QGSJET-II & $5.62 \pm 0.02$ & $5.72 \pm 0.03 $ & $5.90 \pm 0.03$ \\
      QGSJET-01c & $5.65 \pm 0.02$ & $5.75 \pm 0.03 $ & $5.93 \pm 0.03$ \\
      \hline
    \end{tabular}	
    \caption{Vertical crossover energy $\log_{\rm{10}}E_{\rm{1/2}}/\textrm{GeV}$ for ERS flux and CORSIKA non-prompt muon simulation.}
    \label{e12_erstable}
  \end{center}
\end{table}

%The expected crossover energy for atmospheric muons, under the assumption that $E_{1/2}=2 PeV$ for the unflavored component, would consequently lie in between approximately 600 TeV for QGSJET-II+ERS (max) and 1 PeV for SIBYLL+ERS (min). Note that this value is in first approximation independent of the shape of the nucleon flux, which is very poorly constrained at the relevant energies.

Detailed features of a theoretical model are taken into account by a higher-order correction. In particular, those are the increase of the prompt production cross section as a function of primary energy and the appearence of re-interaction effects at energies of several PeV. Since the latter is negligible in the range covered by the study in this paper, its angular dependence was omitted. 

The parametrized form of the correction factor is:

\begin{equation}
\label{pseudoprompt-hecorr}
\begin{split}
f_{corr}(E_{\mu}) = f_{\rm{corr}}(c.s.)\cdot f_{\rm{corr}}(int.) =\\ \left[(3.74-0.461 \cdot \log_{\rm{10}}E_{\mu}/\rm{GeV})\cdot(1+e^{2.13 \cdot \log_{\rm{10}}E_{\mu}/4.9\rm{PeV}})\right]^{-1}
\end{split}
\end{equation}

After application of the correction, simulation-based flux prediction and theoretical model agree well, as illustrated in Fig. \ref{fig-pseudoprompt_corr}.

\subsection{Translation to Neutrino Flux}

Prompt muon and neutrino fluxes are not strictly identical. In particular, muons can originate in electromagnetic di-muon decays of vector mesons. The muon-derived measurement is a combination of unflavored and heavy quark-induced fluxes:

\begin{equation}
  \label{frac12_total}
  \Phi_{\rm{prompt,\mu}}=\Phi_{\rm{\mu,heavy}}+\Phi_{\rm{unflav}}
\end{equation}

Whereas previous estimates based on theoretical calculations indicated an unflavored contribution of 0.3-0.4 times the ERS flux \cite{Illana:2010gh}, recent numerical simulations result in a higher value, almost approaching the flux from heavy hadron decays \cite{Fedynitch:2015zma}.

 The contribution from vector meson decays is partially compensated by a relative suppression of the muon flux with respect to neutrinos of 15-20\% originating in the physics of $c\to s$ decay \cite{Lipari:2013taa}, here represented by the conversion factor $\zeta_{\nu,\mu}$. The resulting neutrino flux is therefore: 

\begin{equation}
  \label{frac12_nu}
  \Phi_{\rm{prompt,\nu}}=\zeta_{\mu,\nu}\cdot(\Phi_{\rm{prompt},\mu}-\Phi_{\rm{unflav}})
\end{equation}

An exact translation requires precise determination of spectrum and magnitude of the unflavored contribution and evaluation of the weak matrix element responsible for $\zeta_{\nu,\mu}$. At the moment, the calculation of a reliable estimate for the prompt atmosperic neutrino flux is precluded by the substantial uncertainties on the experimental measurement.

%\iffalse

\section{Influence of Bundle in High-Energy Muon Events}\label{sec:he-bundle}
\setcounter{figure}{0}

\begin{figure}[ht!]
  \centering
  \includegraphics[width=180pt]{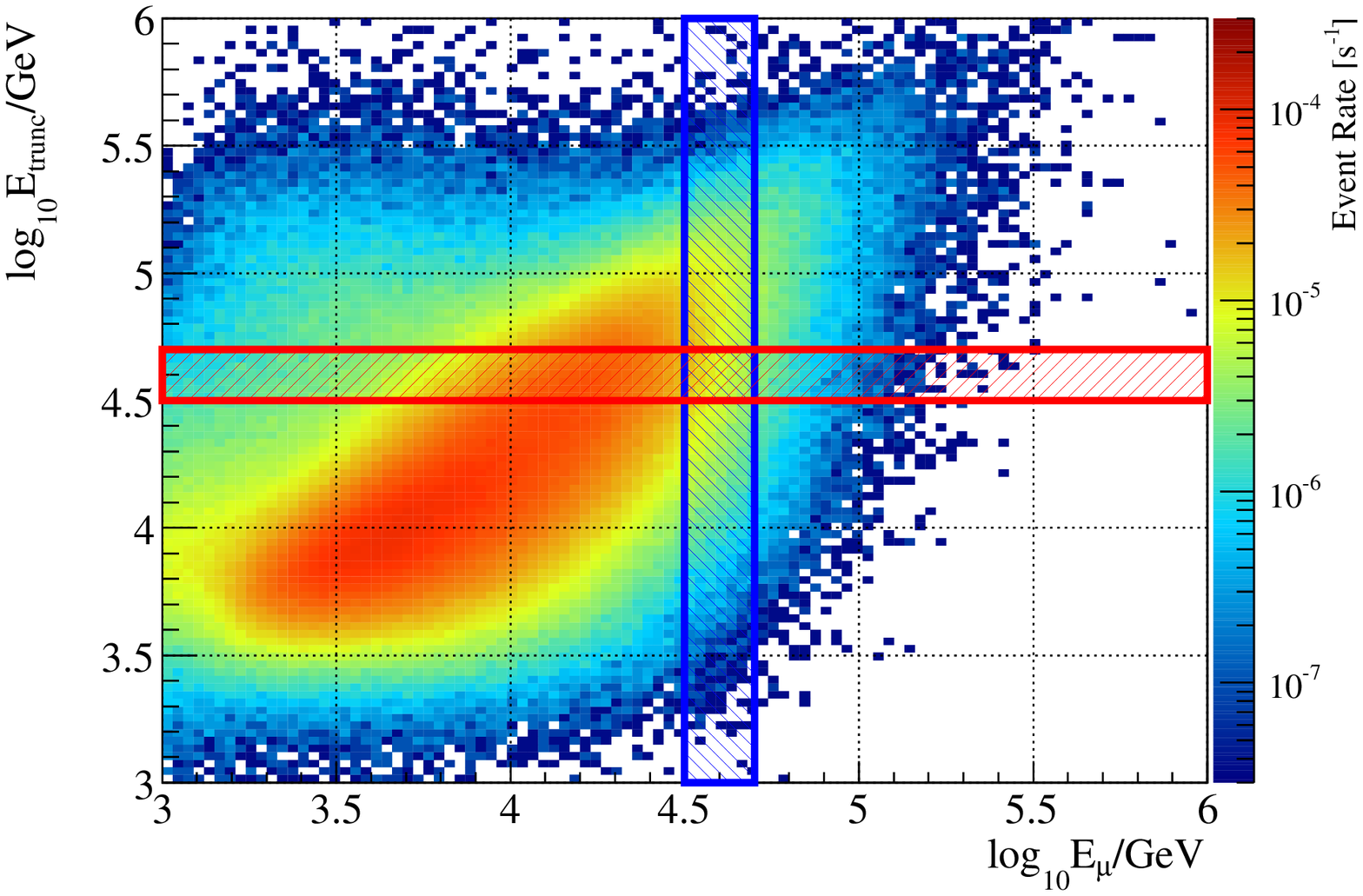}
  \includegraphics[width=180pt]{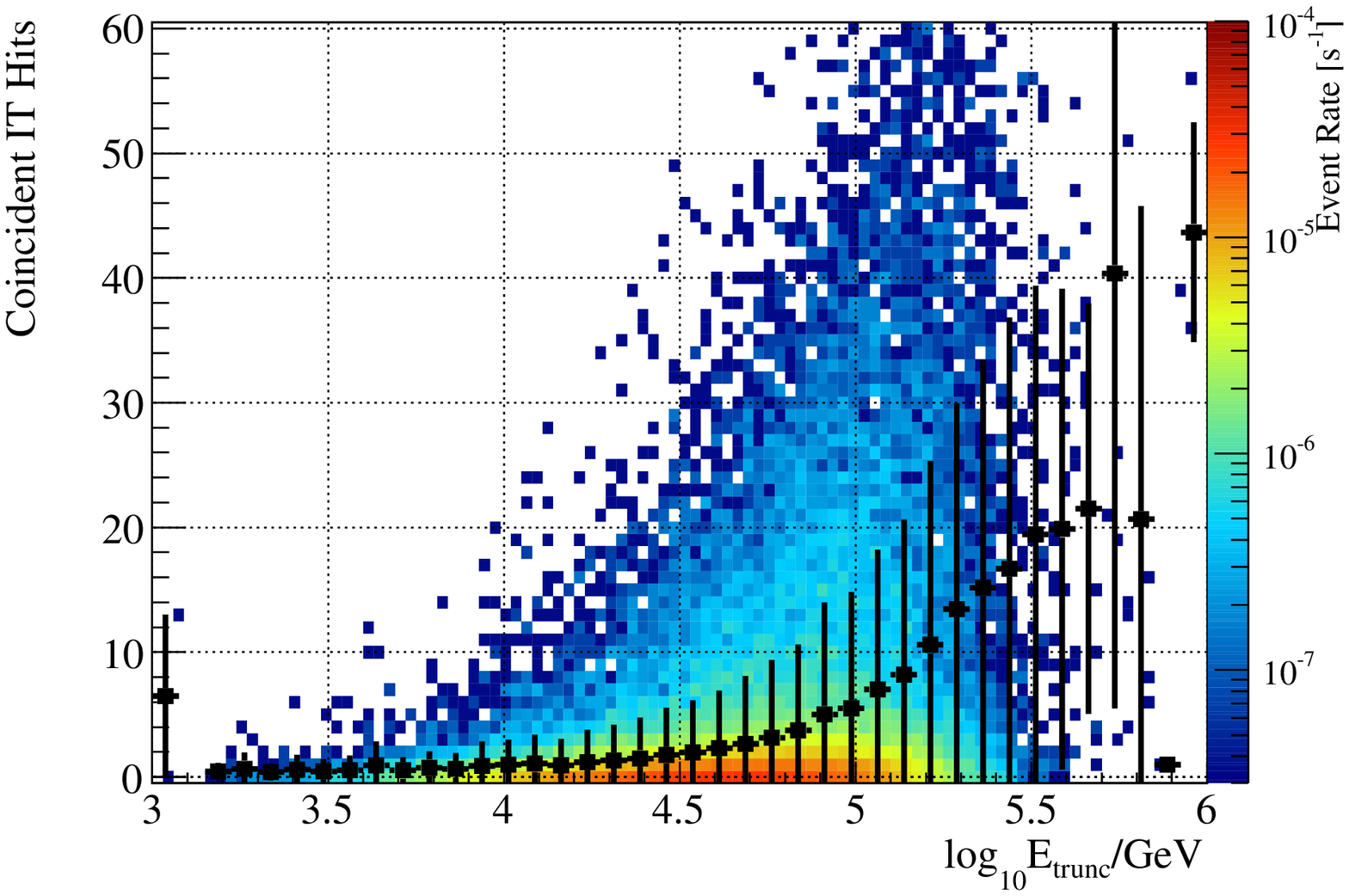}
  \includegraphics[width=180pt]{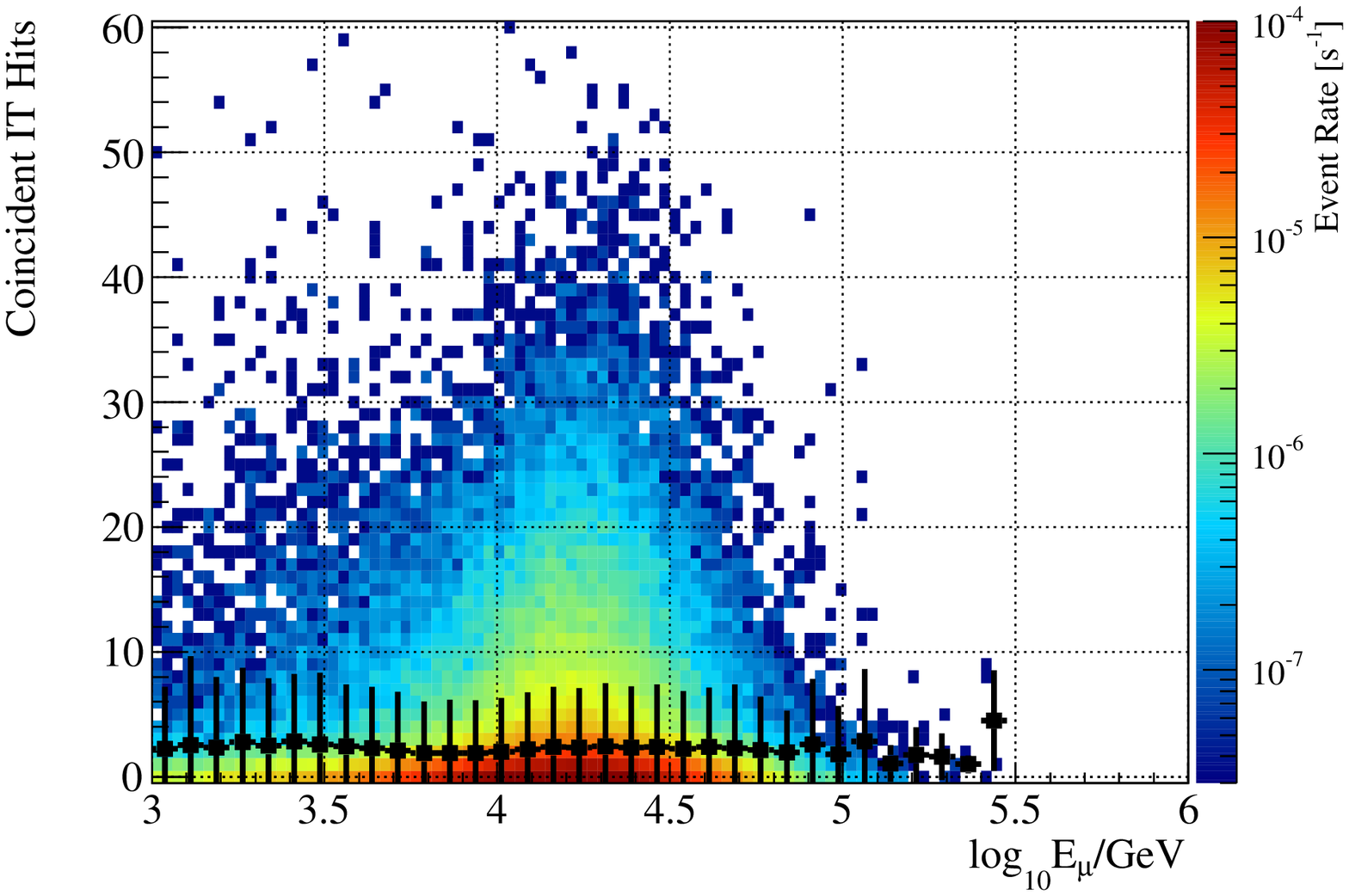}
  \caption{Top: Reconstructed muon surface energy and truncated mean \cite{Abbasi:2012wht} for experimental data. The sample corresponds to tracks with reconstructed angle within 37 degrees from zenith ($\cos\textrm{ }\theta_{\rm{zen}}>0.8$) in selection described in \ref{sec:hemu}, before exclusion of events with shower energies below 5 TeV. Red and blue boxes illustrate selection of data with approximately constant energy measurement. Middle: Number of IceTop tanks registering a signal in coincidence with muon track for fixed reconstructed muon surface energy (blue box). Bottom: Same for fixed truncated mean (red box). }
  \label{fig-bundtest-ithits}
\end{figure}

\begin{figure}[ht!]
  \centering
  \includegraphics[width=220pt]{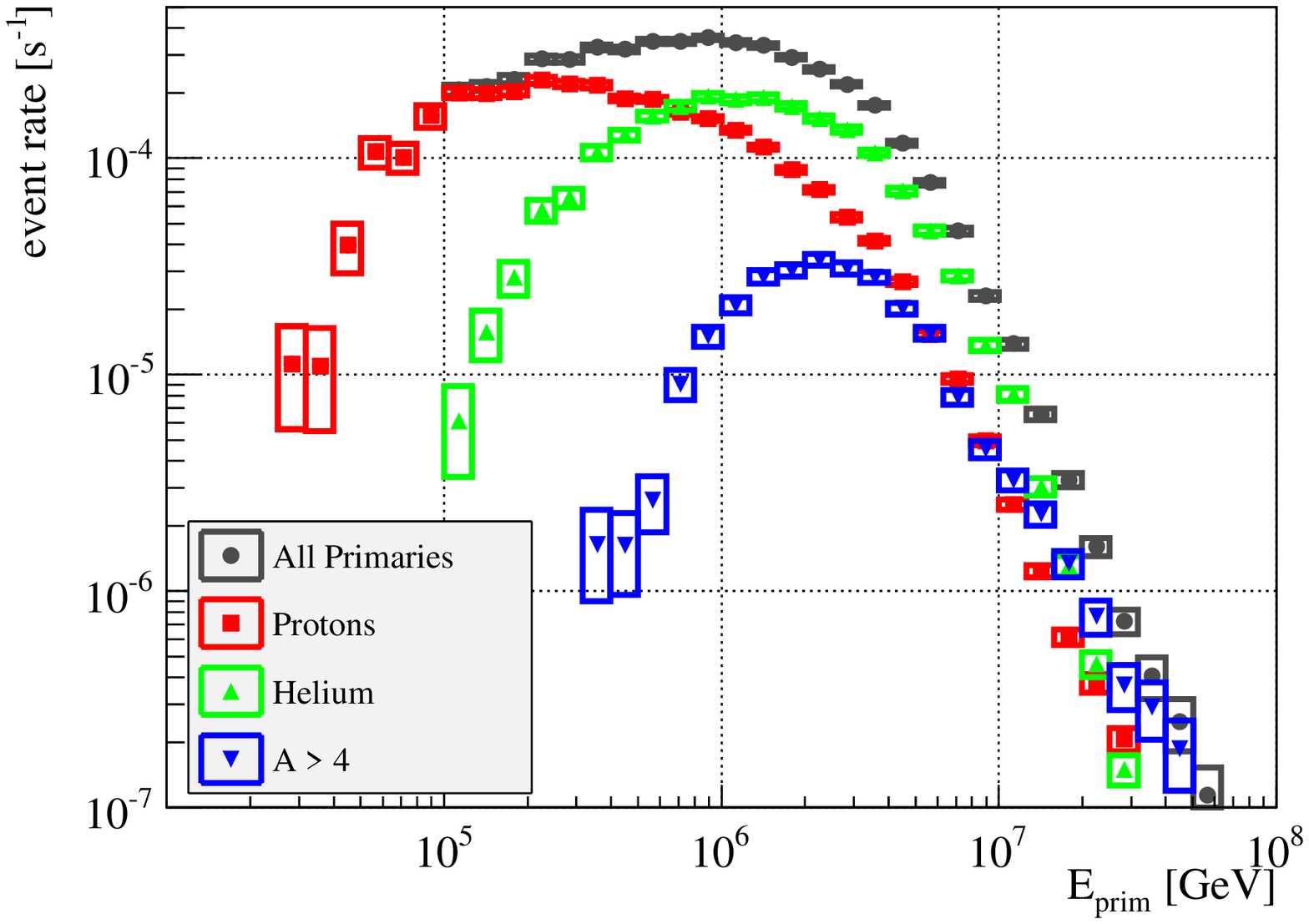}
  \includegraphics[width=220pt]{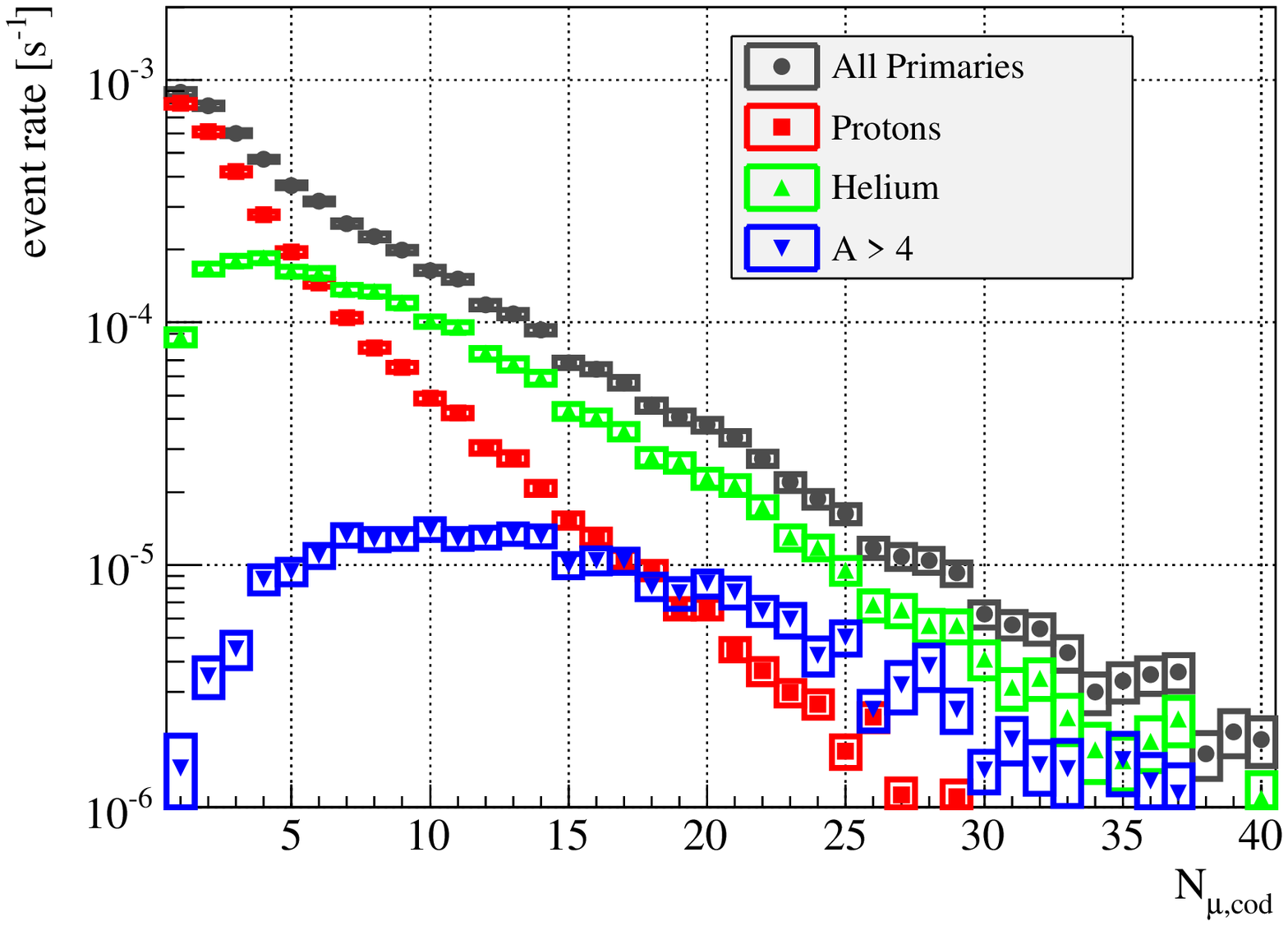}
  \caption{Parameter distributions separated by primary cosmic ray type for simulated high-energy muon events with reconstructed surface energies between 30 and 50 TeV. True primary energy (top) and muon bundle multiplicity at detector depth (bottom).}
  \label{fig-bundtest-separ}
\end{figure}

\begin{figure}[ht!]
  \centering
  \includegraphics[width=220pt]{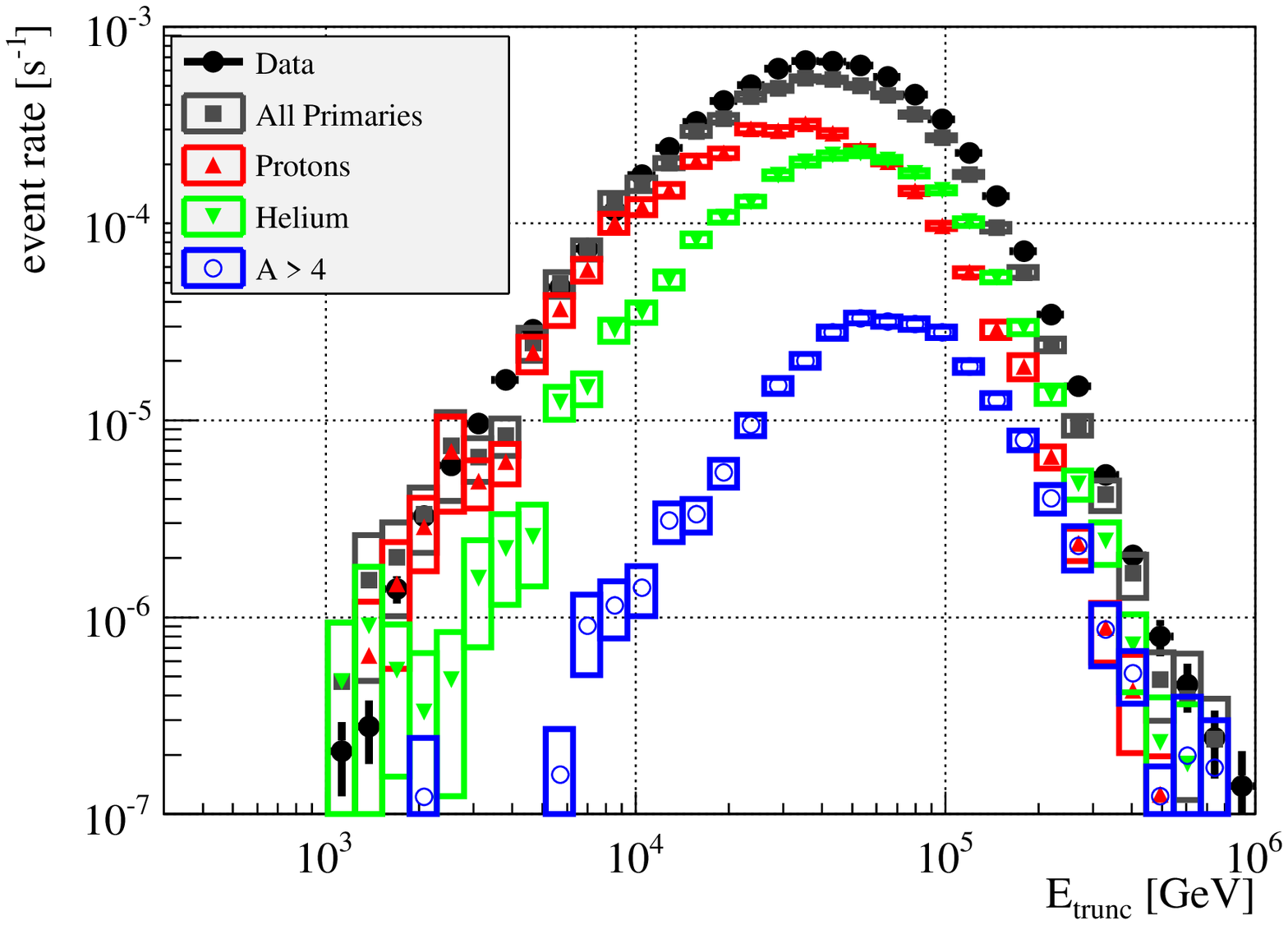}
  \includegraphics[width=220pt]{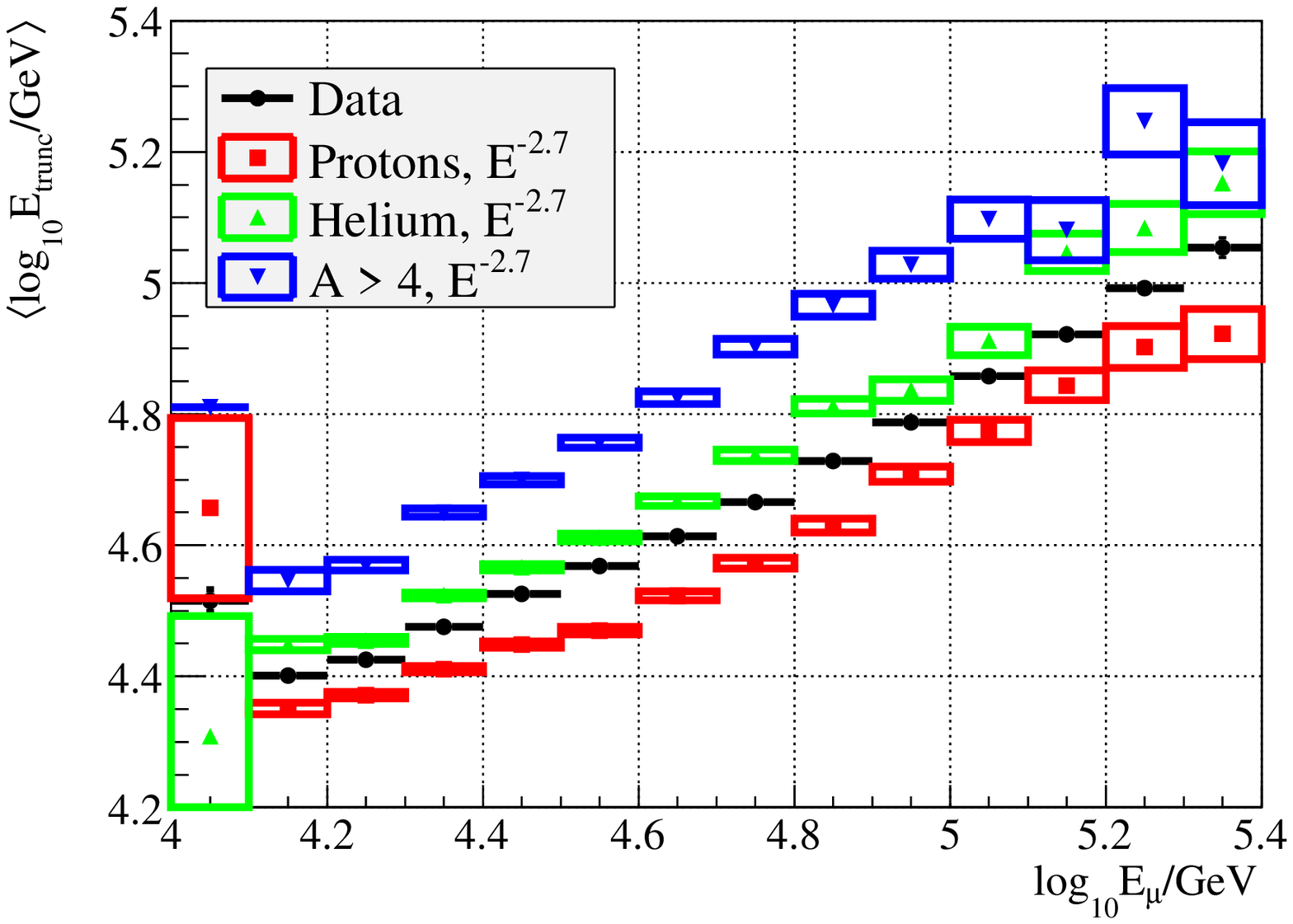}
  \caption{Top: Truncated Energy observable in CORSIKA simulation weighted to GST-Global Fit flux and experimental data. Event selection criteria are the same as in Fig. \ref{fig-bundtest-separ}. Bottom: Mean Truncated Energy observable in dependence of reconstructed leading muon surface energy for simulated and experimental data.}
  \label{fig-bundtest-compo}
\end{figure}

High-energy muon events rarely consist of single particles. Usually there is an accompanying bundle of low-energy muons, whose multiplicity depends on the primary type and energy. It is possible to demonstrate that the influence of secondary particles on the leading muon energy reconstruction is negligible, and that information about the cosmic ray primary can be extracted using an additional observable.

The accuracy of typical muon energy measurements can be increased by excluding exceptional catastrophic losses using the truncated mean of the energy deposition \cite{Abbasi:2012wht}. Since the high-energy muon energy estimate used in this paper relies only on the single strongest shower, the information used in the two reconstruction methods is fully independent.

The approximate orthogonality of the two observables can be demonstrated using only experimental data by including information from the surface array IceTop. Since the leading muon rarely takes away more than 10\% of the primary cosmic ray energy, its presence has almost no influence on the surface size of the air shower. The signal registered by IceTop should therefore only be correlated with the properties of the cosmic ray primary.

In Fig. \ref{fig-bundtest-ithits}, truncated mean and reconstructed muon surface energy are shown for the high-energy muon event sample as described in \ref{sec:hemu}. The lower two panels show the number of IceTop tanks registering a signal in coincidence with the air shower. The effect of varying the muon surface energy for a constant truncated mean is negligible, while in the inverse case a strong increase can be seen at the higher end. The result demonstrates that the total energy of the air shower, and consequently the size of the muon bundle, is not correlated with the measurement of the muon energy. On a qualitative level, it can also be seen that the truncated mean is related to the properties of the parent cosmic ray nucleus.

For the quantitative interpretation of the truncated mean measurement, it is necessary to rely on simulated data, as illustrated in Fig. \ref{fig-bundtest-separ}. The true primary energy distributions for proton and helium are clearly separated. For the same nucleon energy, helium nuclei are four times more energetic than protons. The consequence is a substantially larger bundle multiplicity in the detector. To be distinguishable in the truncated mean observable, the energy deposition from the muon bundle needs to be comparable to that from leading muon. The relation between muon multiplicity and truncated mean is therefore less clear than in the muon multiplicity measurement as described in Section \ref{sec:bundles}.

A comparison between simulation and experimental data is shown in Fig. \ref{fig-bundtest-compo}. The simulated curves are based on the simplified assumption of a straight power law primary spectrum. While a detailed analysis goes beyond the scope of this paper, the quantitative behavior of the experimental data conforms to the expectation that the average mass of the parent cosmic ray flux falls in between proton and helium.

%\fi


%% References


\bibliographystyle{elsarticle-num}



\end{document}